\documentclass[preprint,15pt]{elsarticle}

\usepackage{amssymb}
\usepackage{amsmath}
\usepackage{mathrsfs}
\usepackage{subfig}
\usepackage{graphicx} %use graph format
\usepackage{epstopdf}
\usepackage{latexsym}
\usepackage{color}
\usepackage{lineno} %line numbers
\usepackage{pstricks}
\usepackage{pst-all}

\usepackage{booktabs}
\usepackage{multirow}
\newtheorem{thm}{Theorem}[section]
\newtheorem{lem}[thm]{Lemma}
\newtheorem{definition}[thm]{Definition}
\newtheorem{prop}[thm]{Proposition}
\newdefinition{rmk}{Remark}
\newdefinition{example}{Example}
\newtheorem{coro}[thm]{Corollary}
\newdefinition{ini}{Initialization}
\newproof{pf}{Proof}
\newproof{pot}{Proof of Theorem \ref{thm2}}
\usepackage[ruled,vlined,linesnumbered]{algorithm2e}
\begin{document}
%\linenumbers
\begin{frontmatter}

\title{Space Mapping of Spline Spaces over Hierarchical T-meshes}%\thanks{Grants or other notes
\author[a]{Jingjing Liu}
\author[b]{Fang Deng}
\author[a]{Jiansong Deng\corref{cor1}}
\cortext[cor1]{
Corresponding author.}
\ead{dengjs@ustc.edu.cn}

\address[a]{School of Mathematical Sciences, University of Science and Technology of China, Hefei, Anhui 230026, P. R. China}
\address[b]{School of Mathematics and Statistics,
North China University of Water Resources and Electric Power, ZhengZhou, Henan, 450045, P. R. China}

\begin{abstract}
In this paper, we construct a bijective mapping between a biquadratic spline space over the hierarchical T-mesh
and the piecewise constant space over the corresponding crossing-vertex-relationship graph (CVR graph).
We propose a novel structure, by which we offer an effective and easy operative method for constructing the basis functions of the biquadratic spline space.
The mapping we construct is an isomorphism.
The basis functions of the biquadratic spline space hold the properties such as linearly independent, completeness and the property of partition of unity, which are the same with the properties for the basis functions of piecewise constant space over the CVR graph.
To demonstrate that the new basis functions are efficient, we apply the basis functions to fit some open surfaces.
\end{abstract}
\begin{keyword}
Spline spaces over T-meshes \sep Dimension  \sep CVR graph  \sep Space mapping \sep Basis functions
\end{keyword}
\end{frontmatter}
Mathematical Sciences Classification: 65D07

\section{Introduction}

Splines are useful tools for representing functions and surface models. Non-uniform Rational B-Splines (NURBS), which are defined on tensor product meshes, are the most popular splines in the industry. Due to the tensor product structure, however, the local refinement of NURBS is impossible; furthermore, NURBS models generally contain a large number of superfluous control points. Therefore, many splines that are defined on T-meshes are developed and can be adaptively locally refined.

There are four main types of splines that can be defined over T-meshes.
%Hierarchical B-splines \cite{HB} are polynomial splines that are defined over hierarchical T-meshes.
Hierarchical B-splines provides a classical approach to obtain local refinement in geometric modeling,
the construction of the basis guarantees nested spaces and linear independence of the basis functions.
The definition is improved as hierarchical B-splines with the partition of unity in \cite{AVISO}.
An increasing number of published papers \cite{JUTHB,JUCHB,TDHB}
discuss the completeness and partition of unity.
T-splines \cite{Sed03,Sed04} are defined over T-meshes, where T-junctions between axis aligned segments are allowed.
T-splines have been used efficiently in CAD applications, being able to produce watertight and locally refined models.
However, the use of the most general T-spline concept in IGA is limited by the risk of linear dependence of the resulting splines \cite{Buffa2010}.
Therefore, analysis-suitable T-splines are introduced in \cite{AST}.
Polynomial splines over hierarchical
T-meshes (PHT-splines) \cite{DengPHT} are developed directly from the spline spaces.
The basis functions of PHT-splines are linearly independent
and form a partition of unity.
An adaptive extended IGA (XIGA) approach based PHT-splines for modeling crack propagation
is presented in \cite{ISPHT}. More works are done for IGA in \cite{ISmeshfree, freeVIs}.
LR-splines \cite{LR} are also an important kind of the splines defined over T-meshes,
and their definition is inspired by the knot insertion refinement process of tensor B-splines,
they also proposed an efficient algorithm to seek and destroy linear dependence relations.
In practice, linear dependence of LR B-splines can be controlled and much knowledge
exists with respect to mesh configurations resulting in linear dependent LR B-splines.
In \cite{LRMins}, a first analysis on the necessary conditions for encountering
a linear dependence relation has been presented.
In \cite{PropertyLR}, different properties of the LR-splines are analyzed:
in particular the coefficients for polynomial representations and
their relation with other properties such as linear independence and the number of B-splines covering each element.

%Among these splines, one fundamental problem is to understand the spline space over a T-mesh;
%this spline space is constructed by piecewise polynomials of a given smoothness on a T-mesh.

To discuss the splines from the view of spline spaces,
\cite{Deng06} proposed the spline space over a T-mesh $\mathbf{S}(m,n,\alpha,\beta,\mathscr{T})$
which is a bi-degree $(m,n)$ piecewise polynomial
spline space over the T-mesh $\mathscr{T}$,
with the smoothness order $\alpha$ and $\beta$ in two directions.
When $m \geqslant 2\alpha+1$ and $n \geqslant 2\beta+1$,
a dimension formula is given in \cite{Deng06},
and the basis functions are constructed in \cite{DengPHT}.
%However, if we relax the constraints of $m \geqslant 2\alpha+1$ and $n \geqslant 2\beta+1$,
%a general dimension formula does not exist.
In 2011, \cite{XLbuwen} discovered that the dimension
of the associated spline space has instability over particular T-meshes, i.e.,
the dimension is associated not only
with the topological information of the T-mesh
but also with the geometric information of the T-mesh.
In addition, \cite{Dbbuwen}
gives two additional examples of $\mathbf{S}(5,5,3,3,\mathscr{T})$ and $\mathbf{S}(4,4,2,2,\mathscr{T})$ for the instability of dimensions.
To overcome the instability of dimensions, weighted T-meshes \cite{PT}, diagonalizable T-meshes \cite{XLScof},
and T-meshes for a hierarchical B-spline \cite{thb2013},
over which the dimensions are stable, are developed.
\cite{thb2013} addresses hierarchical T-meshes, which have a nature tree
structure and have existed in the finite element analysis community for a long time.
For a hierarchical T-mesh, \cite{Deng13} derives
a dimension formula for biquadratic $C^1$ spline spaces,
and \cite{w13} provides a dimension formulae for $\mathbf{S}(d,d,d-1,d-1,\mathscr{T})$
over a very special hierarchical T-mesh using the homological algebra technique.
Using tools from homological algebra,
\cite{Toshniwala2019} discusses the dimension of polynomial splines of mixed smoothness on T-meshes,
\cite{Toshniwala2020}
 provides combinatorial bounds on the dimension for polynomial spline spaces of non-uniform bi-degree on T-meshes.
\cite{Zeng15b} gives a dimension formula of
$\mathbf{S}(3,3,2,2,\mathscr{T})$  over a T-mesh that is more general than that in  \cite{w13}
but also a special hierarchical
T-mesh.
Using a corresponding crossing-vertex-relationship graph (CVR graph), \cite{Fang17} constructed
basis functions of $\mathbf{S}(2,2,1,1,\mathscr{T})$
over hierarchical T-meshes and all basis functions are B-spline basis functions.
However, the basis construction in \cite{Fang17}
need to obey the limitation that the level differences of the hierarchical T-meshes are not more than one.
In other words, the basis construction method is not considered on the general hierarchical T-meshes.
%is that the maximum level difference of the adjacent cells is one.
%In order to maintain this rule, some redundant cells are subdivided, the number of basis functions is increased,
%and the computation has also increased.

In this paper, we overcome the limitations in \cite{Fang17},
and we discuss the dimensions and construct the basis functions from a space mapping standpoint.
For the hierarchical T-mesh $\mathscr{T}$, we denote the corresponding CVR graph as $\mathscr{G}$,
we do the works as follows:

\begin{enumerate}
\item Without any additional restrictions over $\mathscr{T}$,
we give a bijective mapping between $\overline{\mathbf{S}}(2,2,1,1,\mathscr{T})$ and $\overline{\mathbf{S}}(0,0,-1,-1,\mathscr{G})$.
And $\overline{\mathbf{S}}(2,2,1,1,\mathscr{T})$  is isomorphic to $\overline{\mathbf{S}}(0,0,-1,-1,\mathscr{G})$.
\item By the tools which are called T-structures, we give a general method to construct each basis function
for $\mathbf{S}(2,2,1,1,\mathscr{T})$ when there is no limitation for level difference of $\mathscr{T}$.
\item By the isomorphic space, we prove that the basis functions of $\mathbf{S}(2,2,1,1,\mathscr{T})$
hold the properties of linearly independence, completeness and partition of unity.
\end{enumerate}

This paper is organized as follows.
In Section \ref{prelimi}, we recall some notations about hierarchical T-meshes,
spline spaces over hierarchical T-meshes and B-net method.
In Section \ref{UDANmapping}, we give a bijective mapping for univariate spline spaces.
It illuminates us for considering the mapping between the spline space over a hierarchical T-mesh and the piecewise constant space over the corresponding CVR graph
in Section \ref{mappingformulea111111}. To ensure the mapping is bijective,
we introduce some conclusions about T-structures, by which we describe a general method to construct the basis functions of $\overline{\mathbf{S}}(2,2,1,1,\mathscr{T})$ in Section \ref{basisfunctions}.
We discuss the properties of the mapping and the properties of the basis functions in Section \ref{Properties}.
%We compare our basis functions with some classical splines in Section \ref{comparing}.
In Section \ref{fitting}, the basis functions are applied to fit some open surfaces.
We end the paper with conclusions and future works in Section \ref{Conclusions}.
\section{ Hierarchical T-meshes,  spline spaces and B-net method}\label{prelimi}
%For spline spaces on hierarchical T-meshes,
%we want to construct a piecewise constant space, which is isomorphic to the biquadratic spline  space.
%Similar to section \ref{UDANmapping}, we need to construct the mapping between the two spaces.
In this section, we recall some notations about hierarchical T-meshes, spline spaces and B-net method.

\subsection{Hierarchical T-meshes and some notations for hierarchical T-meshes}

Instead of considering general T-meshes,
we focus our attention on $2 \times 2$ division hierarchical T-meshes\cite{DengPHT} as follows:

\begin{definition}\cite{DengPHT}
Given a tensor product mesh (level 0),
at least one cell of level $k$ be subdivided into $2 \times 2$ equal subcells,
which are cells at level $k+1$.The resulting T-mesh is called a \textbf{hierarchical T-mesh of $2 \times 2$ division}.
The maximal level number that appears is defined as the \textbf{level} of the hierarchical T-mesh, we denote the mesh of level $k$ as $\mathscr{T}^k$.
\end{definition}

\begin{figure}[!ht]
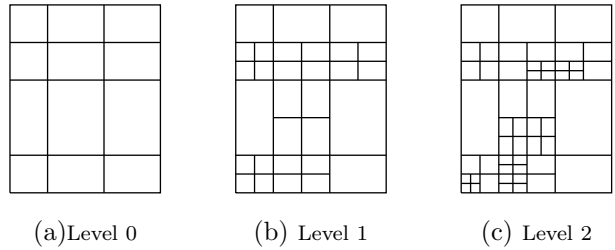

\captionsetup{font={small}}
\begin{center}
\psset{unit=0.5cm,linewidth=0.5pt}
\begin{tabular}{c@{\hspace*{1cm}}c@{\hspace*{1cm}}c}
\pspicture(0,0)(4,5) \psline(0,0)(4,0)(4,5)(0,5)(0,0)
\psline(0,1)(4,1)\psline(0,3)(4,3)\psline(0,4)(4,4)
\psline(1,0)(1,5)\psline(2.5,0)(2.5,5)
\endpspicture &
\pspicture(0,0)(4,5)
\psline(0,0)(4,0)(4,5)(0,5)(0,0)\psline(0,1)(4,1)\psline(0,3)(4,3)\psline(0,4)(4,4)
\psline(1,0)(1,5)\psline(2.5,0)(2.5,5)\psline(1.75,0)(1.75,4)
\psline(0,0.5)(2.5,0.5)\psline(0.5,0)(0.5,1)\psline(2.5,2)(1,2)\psline(0,3.5)(4,3.5)
\psline(0.5,3)(0.5,4)\psline(3.25,3)(3.25,4)
\endpspicture &\pspicture(0,0)(4,5) \psline(0,0)(4,0)(4,5)(0,5)(0,0)
\psline(0,1)(4,1)\psline(0,3)(4,3)\psline(0,4)(4,4)
\psline(1,0)(1,5)\psline(2.5,0)(2.5,5)\psline(1.75,0)(1.75,4)
\psline(0,0.5)(2.5,0.5)\psline(0.5,0)(0.5,1)\psline(2.5,2)(1,2)\psline(0,3.5)(4,3.5)
\psline(0.5,3)(0.5,4)\psline(3.25,3)(3.25,4)\psline(0,0.25)(0.5,0.25)
\psline(0.25,0)(0.25,0.5)\psline(1.375,0)(1.375,2)\psline(1,0.25)(1.75,0.25)
\psline(1,0.75)(1.75,0.75)\psline(1,1.5)(2.5,1.5)\psline(2.125,1)(2.125,2)
\psline(2.125,3)(2.125,3.5)\psline(1.75,3.25)(3.25,3.25)\psline(2.875,3)(2.875,3.5)
%\psset{linewidth=1pt,linecolor=green}\psline(2.125,1)(2.125,2)
\endpspicture \\[2mm]
(a)\footnotesize {Level 0}  & (b) \footnotesize{Level 1}  & (c) \footnotesize{Level 2}
\end{tabular}
\caption{A hierarchical T-mesh. \label{HTmesh}}
\end{center}
\end{figure}
\begin{figure}[!ht]
\captionsetup{font={small}}
\begin{center}
\psset{unit=0.65cm,linewidth=0.5pt}
\begin{tabular}{c@{\hspace*{1cm}}c}
\begin{pspicture}(0,0)(4,4)
\psframe*[linecolor=blue](1.5,2)(2,2.5)\psframe*[linecolor=green](0,3)(1,4)
\psline(0,0)(0,4)(4,4)(4,0)(0,0)
\psline(0,1)(4,1)\psline(1,1.5)(2,1.5)\psline(0,2)(4,2)\psline(1,2.5)(3,2.5)\psline(0,3)(4,3)
\psline(1,0)(1,4)\psline(1.5,1)(1.5,3)\psline(2,0)(2,4)\psline(2.5,2)(2.5,3)\psline(3,0)(3,4)
\rput(1,1){\footnotesize${\bullet}$}\rput(0.75,0.7){\scriptsize${v_{2}}$}
\rput(2,1){\footnotesize${\bullet}$}\rput(2.25,0.7){\footnotesize${v_{3}}$}
\rput(1,1.5){\footnotesize${\bullet}$}\rput(0.7,1.5){\footnotesize${v_{0}}$}
\rput(2,1.5){\footnotesize${\bullet}$}\rput(2.3,1.5){\footnotesize${v_{1}}$}
\rput(1,2.5){\footnotesize${\bullet}$}\rput(0.7,2.5){\scriptsize${v_{4}}$}
\rput(3,2.5){\footnotesize${\bullet}$}\rput(3.3,2.5){\scriptsize${v_{5}}$}

\rput(0,2){\footnotesize${\bullet}$}\rput(-0.3,2){\footnotesize${b_0}$}
\rput(0,1){\footnotesize${\bullet}$}\rput(-0.3,1){\footnotesize${b_1}$}
\rput(0,0){\footnotesize${\bullet}$}\rput(-0.3,0){\footnotesize${b_2}$}
\rput(4,0){\footnotesize${\bullet}$}\rput(4.4,0){\footnotesize${b_3}$}
\rput(4,1){\footnotesize${\bullet}$}\rput(4.3,1){\footnotesize${b_4}$}
\rput(4,2){\footnotesize${\bullet}$}\rput(4.3,2){\footnotesize${b_5}$}
\end{pspicture}\\
\end{tabular}
\caption{The vertices, edges and cells.\label{aPHTmesh}}
\end{center}
\end{figure}
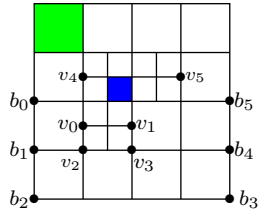

Fig. \ref{HTmesh} illustrates the process of generating a hierarchical T-mesh.
In the hierarchical T-mesh $\mathscr{T}$,  the definitions of vertex, edge and cell are the same as in \cite{Deng06},
we recall the notations of $\mathscr{T}$ as follows:

In the  hierarchical T-mesh $\mathscr{T}$, a grid point in $\mathscr{T}$ is also called a \textbf{vertex} of $\mathscr{T}$.
If a vertex is on the boundary grid line
of $\mathscr{T}$,  then it is called a \textbf{boundary-vertex}. Otherwise, it is called an \textbf{interior-vertex}.
There are two types of interior-vertices.
An interior-vertex of valence four is called a \textbf{crossing-vertex}.
An interior-vertex of valence three is called a \textbf{T-junction}.

The line segment connecting two adjacent vertices on a grid line is called an \textbf{edge} of $\mathscr{T}$.
If an edge is on the boundary of $\mathscr{T}$, then it is called a \textbf{boundary-edge};
otherwise it is called an \textbf{interior-edge}.
If an edge is the longest possible line segment whose two end points are either boundary vertices or T-junctions, we refer to the edge as a \textbf{l-edge}.
If an l-edge is comprised of some boundary-edges, then it is
called a \textbf{boundary-l-edge}; otherwise, it is called an \textbf{interior-l-edge}.
If the two end
points of an interior-l-edge are both T-junctions, the l-edge is called a \textbf{T-l-edge}\cite{Zeng15b}.
If an edge is the longest possible line segment whose inner-vertices
are T-junctions, the edge is referred to as a \textbf{c-edge}.

Each rectangular grid element is referred to as a \textbf{cell} of $\mathscr{T}$.
A cell is called an \textbf{interior-cell}  if all its edges are interior edges; otherwise, it is called a \textbf{boundary-cell}.

In Fig. \ref{aPHTmesh}, $b_i,i=0,...,5$ are boundary-vertices, while $v_i,i=0,...,5$ are interior-vertices,
$v_2$ is a crossing-vertex while $v_0$ is a T-junction.
$b_1v_2$ is an interior-edge while $b_1b_2$  is a boundary-edge,
$b_2b_3$ is a boundary-l-edge while $v_0v_1$ is an interior-l-edge,
$v_0v_1$  is also a T-l-edge, and $v_2v_3$ is a c-edge.
The blue cell is an interior-cell while the green cell is a boundary-cell.

\subsection{Spline spaces}

Given a T-mesh $\mathscr{T}$,
we use $\mathscr{F}$ to denote all of the cells in $\mathscr{T}$
and $\Omega$ to denote the region occupied by the cells in $\mathscr{T}$.
The spline spaces are defined as \cite{Deng06}:
\begin{equation}
\mathbf{S}(m,n,\alpha,\beta,\mathscr{T}):=\{f(x,y)\in
C^{\alpha,\beta}(\Omega): f(x,y)|_{\phi}\in \mathbb{P}_{mn},\forall
\phi\in\mathscr{F}\},
\end{equation}
where $\mathbb{P}_{mn}$ is the space of the polynomials with bi-degree $(m,n)$
and $C^{\alpha,\beta}$ is the space consisting of all of the bivariate
functions continuous in $\Omega$ with order $\alpha$ along the $x$-direction and $\beta$ along the
$y$-direction. It is obvious that $\mathbf{S}(m,n,\alpha,\beta,\mathscr{T})$ is a linear space.

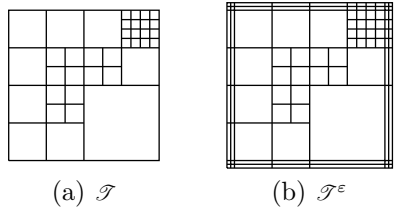
\begin{figure}[!ht]
\captionsetup{font={small}}
\begin{center}
\psset{unit=0.5cm,linewidth=0.5pt}
\begin{tabular}{c@{\hspace*{1cm}}c}
\begin{pspicture}(0,-0.2)(4,4)
\psline(0,0)(0,4)(4,4)(4,0)(0,0)
\psline(0,1)(2,1)\psline(1,1.5)(2,1.5)\psline(0,2)(4,2)\psline(1,2.5)(3,2.5)\psline(0,3)(4,3)
\psline(1,0)(1,4)\psline(1.5,1)(1.5,3)\psline(2,0)(2,4)\psline(2.5,2)(2.5,3)\psline(3,2)(3,4)
\psline(3,3.25)(4,3.25)\psline(3,3.5)(4,3.5)\psline(3,3.75)(4,3.75)
\psline(3.25,3)(3.25,4)\psline(3.5,3)(3.5,4)\psline(3.75,3)(3.75,4)
\end{pspicture}&
\begin{pspicture}(0,-0.2)(4,4)
\psline(0,0)(0,4)(4,4)(4,0)(0,0)\psline(-0.1,-0.1)(-0.1,4.1)(4.1,4.1)(4.1,-0.1)(-0.1,-0.1)
\psline(-0.2,-0.2)(-0.2,4.2)(4.2,4.2)(4.2,-0.2)(-0.2,-0.2)
\psline(-0.1,-0.2)(-0.1,-0.1)\psline(-0.2,-0.1)(-0.1,-0.1)
\psline(0,-0.2)(0,0)\psline(-0.2,0)(0,0)\psline(-0.2,4)(0,4)(0,4.2)
\psline(-0.2,4.1)(-0.1,4.1)(-0.1,4.2)\psline(4.2,4)(4,4)(4,4.2)\psline(4.2,4.1)(4.1,4.1)(4.1,4.2)
\psline(4,-0.2)(4,0)(4.2,0)\psline(4.1,-0.2)(4.1,-0.1)(4.2,-0.1)
\psline(0,1)(2,1)\psline(1,1.5)(2,1.5)\psline(0,2)(4,2)\psline(1,2.5)(3,2.5)\psline(0,3)(4,3)
\psline(1,0)(1,4)\psline(1.5,1)(1.5,3)\psline(2,0)(2,4)\psline(2.5,2)(2.5,3)\psline(3,2)(3,4)
\psline(3,3.25)(4,3.25)\psline(3,3.5)(4,3.5)\psline(3,3.75)(4,3.75)
\psline(3.25,3)(3.25,4)\psline(3.5,3)(3.5,4)\psline(3.75,3)(3.75,4)
\psline(1,0)(1,-0.2)\psline(2,-0.2)(2,0)\psline(4,2)(4.2,2)\psline(4,3)(4.2,3)
\psline(4,3.25)(4.2,3.25)\psline(4,3.5)(4.2,3.5)\psline(4,3.75)(4.2,3.75)
\psline(3.25,4)(3.25,4.2)\psline(3.5,4)(3.5,4.2)\psline(3.75,4)(3.75,4.2)
\psline(3,4)(3,4.2)\psline(2,4)(2,4.2)\psline(1,4)(1,4.2)\psline(-0.2,1)(0,1)
\psline(-0.2,3)(0,3)\psline(-0.2,2)(0,2)
\end{pspicture} \\
(a) \footnotesize{$\mathscr{T}$}  & (b) \footnotesize{$\mathscr{T}^\varepsilon$}
\end{tabular}
\caption{A T-mesh $\mathscr{T}$ and its extended T-mesh $\mathscr{T}^\varepsilon$ associated with $\mathbf{S}(2,2,1,1,\mathscr{T})$. \label{ETmesh}}
\end{center}
\end{figure}

For a T-mesh $\mathscr{T}$ of $\mathbf{S}(m,n,\alpha,\beta,\mathscr{T})$,
we can obtain
an extended T-mesh in the following fashion.
$m$  edges are added to the horizontal boundaries averagely,
$n$  edges are added to the vertical boundaries averagely,
and then connect the boundary-vertexes of $\mathscr{T}$ to the outermost edges.
The resulting mesh, which we denote as
$\mathscr{T}^\varepsilon$, is called the extended T-mesh
of $\mathscr{T}$ associated with $\mathbf{S}(m,n,\alpha,\beta,\mathscr{T})$.
$\mathscr{T}^\varepsilon$ is also called
an extended T-mesh.
Fig. \ref{ETmesh} shows an example of the extended T-mesh.

The corresponding biquadratic spline spaces over $\mathscr{T}$ with homogeneous
boundary conditions (HBC) were defined as follows \cite{Deng13}:
\begin{equation}
\overline{\mathbf{S}}(m,n,\alpha,\beta,\mathscr{T}):=\{f(x,y)\in
C^{\alpha,\beta}(\mathbb{R}^2): f(x,y)|_{\phi}\in
\mathbb{P}_{22},\forall \phi\in\mathscr{F}, \text{and} f|_{\mathbb{R}^2\setminus\Omega} \equiv 0 \}.
\end{equation}

One important observation in \cite{Deng13} is that the two spline spaces $\mathbf{S}(m,n,\alpha,\beta,\mathscr{T})$
and $\overline{\mathbf{S}}(m,n,\alpha,\beta,\mathscr{T}^{\varepsilon})$ are closely related.

\begin{thm}\cite{Deng13}
\label{thm2.1}
Given a T-mesh $\mathscr{T}$, assume that
$\mathscr{T}^{\varepsilon}$ is its extension associated with
$\mathbf{S}(m,n,\alpha,\beta,\mathscr{T})$ and that $\Omega$ is the region occupied by
the cells in $\mathscr{T}$. Then,
\begin{equation}
\label{qicikuochong1111}
\mathbf{S}(m,n,\alpha,\beta,\mathscr{T}) = \overline{\mathbf{S}}(m,n,\alpha,\beta,\mathscr{T}^{\varepsilon})|_{\Omega},
\end{equation}
\begin{equation}
\label{qicikuochong22222}
\dim \mathbf{S}(m,n,\alpha,\beta,\mathscr{T}) = \dim \overline{\mathbf{S}}(m,n,\alpha,\beta,\mathscr{T}^{\varepsilon}).
\end{equation}
\end{thm}

\subsection{B-net method}\label{Bnetmothod}
The B-net method is based on Bernstein-B$\acute{e}$zier representation of polynomials.
Refer to \cite{Deng06, Approximation} for details.

\begin{figure}[!ht]
\captionsetup{font={small}}
\begin{center}
\psset{unit=0.65cm,linewidth=0.5pt}
\begin{tabular}{c@{\hspace*{1cm}}c}
\begin{pspicture}(0,0)(4,3)
\psline(0,0)(4,0)(4,2)(0,2)(0,0)
\psline[linecolor=green](0.05,1)(4,1)\psline[linecolor=green](0.05,1.1)(4,1.1)\psline[linecolor=green](0.05,1.2)(4,1.2)\psline[linecolor=green](0.05,1.3)(4,1.3)
\psline[linecolor=green](0.05,1.4)(4,1.4)\psline[linecolor=green](0.05,1.5)(4,1.5)\psline[linecolor=green](0.05,1.6)(4,1.6)\psline[linecolor=green](0.05,1.7)(4,1.7)
\psline[linecolor=green](0.05,1.8)(4,1.8)\psline[linecolor=green](0.05,1.9)(4,1.9)

\psline[linecolor=green](0.1,1)(0.1,1.95)\psline[linecolor=green](0.2,1)(0.2,1.95)\psline[linecolor=green](0.3,1)(0.3,1.95)\psline[linecolor=green](0.4,1)(0.4,1.95)
\psline[linecolor=green](0.5,1)(0.5,1.95)\psline[linecolor=green](0.6,1)(0.6,1.95)\psline[linecolor=green](0.7,1)(0.7,1.95)\psline[linecolor=green](0.8,1)(0.8,1.95)
\psline[linecolor=green](0.9,1)(0.9,1.95)\psline[linecolor=green](1,1)(1,1.95)
\psline[linecolor=green](1.1,1)(1.1,1.95)\psline[linecolor=green](1.2,1)(1.2,1.95)\psline[linecolor=green](1.3,1)(1.3,1.95)\psline[linecolor=green](1.4,1)(1.4,1.95)
\psline[linecolor=green](1.5,1)(1.5,1.95)\psline[linecolor=green](1.6,1)(1.6,1.95)\psline[linecolor=green](1.7,1)(1.7,1.95)\psline[linecolor=green](1.8,1)(1.8,1.95)
\psline[linecolor=green](1.9,1)(1.9,1.95)\psline[linecolor=green](2,1)(2,1.95)
\psline[linecolor=green](2.1,1)(2.1,1.95)\psline[linecolor=green](2.2,1)(2.2,1.95)\psline[linecolor=green](2.3,1)(2.3,1.95)\psline[linecolor=green](2.4,1)(2.4,1.95)
\psline[linecolor=green](2.5,1)(2.5,1.95)\psline[linecolor=green](2.6,1)(2.6,1.95)\psline[linecolor=green](2.7,1)(2.7,1.95)\psline[linecolor=green](2.8,1)(2.8,1.95)
\psline[linecolor=green](2.9,1)(2.9,1.95)\psline[linecolor=green](3,1)(3,1.95)
\psline[linecolor=green](3.1,1)(3.1,1.95)\psline[linecolor=green](3.2,1)(3.2,1.95)\psline[linecolor=green](3.3,1)(3.3,1.95)\psline[linecolor=green](3.4,1)(3.4,1.95)
\psline[linecolor=green](3.5,1)(3.5,1.95)\psline[linecolor=green](3.6,1)(3.6,1.95)\psline[linecolor=green](3.7,1)(3.7,1.95)\psline[linecolor=green](3.8,1)(3.8,1.95)
\psline[linecolor=green](3.9,1)(3.9,1.95)

\psline(2,2)(2,3)(3,3)(3,2)
\psline[linecolor=yellow](2,2.5)(3,2.5)\psline[linecolor=yellow](2,2.4)(3,2.4)\psline[linecolor=yellow](2,2.3)(3,2.3)\psline[linecolor=yellow](2,2.2)(3,2.2)\psline[linecolor=yellow](2,2.1)(3,2.1)
\psline[linecolor=yellow](2.1,2)(2.1,2.5)\psline[linecolor=yellow](2.2,2)(2.2,2.5)\psline[linecolor=yellow](2.3,2)(2.3,2.5)\psline[linecolor=yellow](2.4,2)(2.4,2.5)
\psline[linecolor=yellow](2.5,2)(2.5,2.5)
\psline[linecolor=yellow](2.6,2)(2.6,2.5)\psline[linecolor=yellow](2.7,2)(2.7,2.5)\psline[linecolor=yellow](2.8,2)(2.8,2.5)\psline[linecolor=yellow](2.9,2)(2.9,2.5)
\rput(0.2,0.2){\footnotesize${\circ}$}\rput(2,0.2){\footnotesize${\circ}$}\rput(3.8,0.2){\footnotesize${\circ}$}
\rput(0.2,1){\footnotesize${\circ}$}\rput(2,1){\footnotesize${\circ}$}\rput(3.8,1){\footnotesize${\circ}$}
\rput(0.2,1.8){\footnotesize${\circ}$}\rput(2,1.8){\footnotesize${\circ}$}\rput(3.8,1.8){\footnotesize${\circ}$}

\rput(2.1,2.1){\tiny${\bullet}$}\rput(2.5,2.1){\tiny${\bullet}$}\rput(2.9,2.1){\tiny${\bullet}$}
\rput(2.1,2.5){\tiny${\bullet}$}\rput(2.5,2.5){\tiny${\bullet}$}\rput(2.9,2.5){\tiny${\bullet}$}
\rput(2.1,2.9){\tiny${\bullet}$}\rput(2.5,2.9){\tiny${\bullet}$}\rput(2.9,2.9){\tiny${\bullet}$}
\rput(0.1,-0.25){\footnotesize${x_0}$}\rput(4,-0.2){\footnotesize${x_3}$}\rput(2,3.25){\footnotesize${x_1}$}\rput(3,3.25){\footnotesize${x_2}$}
\rput(-0.25,0){\footnotesize${y_0}$}\rput(-0.2,2){\footnotesize${y_1}$}\rput(1.75,3){\footnotesize${y_2}$}
\rput(4.3,1){\footnotesize${C_1}$}\rput(3.35,2.5){\footnotesize${C_2}$}
\end{pspicture}\\
\end{tabular}
\caption{\footnotesize{The B-net method.}\label{b-net}}
\end{center}
\end{figure}
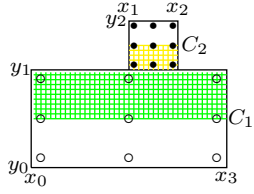
In Fig. \ref{b-net}, let $f_1(x,y)$ and $f_2(x,y)$ be two polynomials with bi-degree $(2, 2)$
defined over two adjacent cells
$C_1:[x_0,x_3] \times [y_0,y_1]$
and $C_2:[x_1,x_2] \times [y_0,y_1]$, respectively.
They can be expressed in the Bernstein - B$\acute{e}$zier forms:

\begin{equation}
f_1(x,y)= \sum_{j=0}^{2} \sum_{k=0}^{2} b^1_{j,k}B^2_j\left(\frac{x-x_0}{x_3-x_0}\right)B^2_k\left(\frac{y-y_0}{y_1-y_0}\right),
\label{Tbordi}
\end{equation}
\begin{equation}
f_2(x,y)= \sum_{j=0}^{2} \sum_{k=0}^{2} b^2_{j,k}B^2_j\left(\frac{x-x_1}{x_2-x_1}\right)B^2_k\left(\frac{y-y_1}{y_2-y_1}\right),
\end{equation}
where $B^2_j(t)$ and $B^2_k(t)$ are the Bernstein polynomials.
$b^1_{j,k}$ and $b^2_{j,k}$ are referred to as the \textbf{B$\acute{e}$zier-ordinates (B-ordinates)}
of $f_1(x,y)$ and $f_2(x,y)$, respectively.
$b^1_{j,k}$ corresponds to the point
$P^1_{j,k}:\left( \frac{(2-j)x_0+jx_3}{2}, \frac{(2-k)y_0+ky_1}{2}\right)$,
which is  referred to as the \textbf{domain-points} \cite{Zeng15c} associated with $C_1$.
$b^2_{j,k}$ corresponds to the point
$P^2_{j,k}:\left( \frac{(2-j)x_1+jx_2}{2}, \frac{(2-k)y_1+ky_2}{2}\right)$,
which is  referred to as the domain-points associated with $C_2$.
The domain-points of $C_1$ and $C_2$ are denoted by  ``$\circ$" and ``${\bullet}$" respectively.

%the smoothness conditions indicate that
As $f_1(x,y)$ and $f_2(x,y)$ are $C^1$ continuous across their common
boundary,
when $b^1_{j,k}, 0 \leqslant j \leqslant 2, 1 \leqslant k \leqslant 2$ are given,
$b^2_{j,k}, 0 \leqslant j \leqslant 2, 0 \leqslant k \leqslant 1$ are determined.
As shown in Fig. \ref{b-net}, if $f_1(x,y)$ and $f_2(x,y)$ are $C^1$ continuous across their common
boundary, when the two rows of
the B-ordinates in the green domain are given,
the  two rows of B-ordinates in the yellow domain are determined. When $C_1$ and $C_2$ are two vertical adjacent cells, we have similar conclusions.
We call the B-ordinates that correspond to the domain points on the cell $C$ as the the B-ordinates on $C$ for convenience.

By the preliminary knowledge above,
we mainly discuss the spline space $\mathbf{S}(2,2,1,1,\mathscr{T})$ in this paper.
\cite{Deng13} gives the conclusion as
\[ \dim \overline{\mathbf{S}}(2,2,1,1,\mathscr{T})=N_{\mathscr{G}}.\]
where $N_{\mathscr{G}}$ is the number of cells in $\mathscr{G}$.
The piecewise constant space on $\mathscr{G}$ is $\mathbf{S}(0,0,-1,-1,\mathscr{G})$,
and \[ \dim\overline{\mathbf{S}}(0,0,-1,-1,\mathscr{G}) = N_{\mathscr{G}},\]
we obtain
\begin{equation}
\label{relationshipbetweenTG}
\dim \overline{\mathbf{S}}(2,2,1,1,\mathscr{T})=\dim \overline{\mathbf{S}}(0,0,-1,-1,\mathscr{G}).
\end{equation}
By Theorem \ref{thm2.1}, to consider the spline space over a T-mesh, we only need to consider
the corresponding spline space with homogeneous boundary conditions over its extended T-mesh.
The mapping for univariate spline spaces can enlighten us well.

\section{Mapping for univariate spline spaces}\label{UDANmapping}

In this section, we construct a bijective mapping between
the univariate quadratic spline space and the corresponding univariate piecewise constant spline space.
By the massage of basis function of the univariate piecewise constant spline space,
a new method for constructing the basis functions of the quadratic spline space is given.

\subsection{The univariate spline space and some notations}

Given the knots $\mathrm{T}:t_0<t_1<...<t_n$,
we use $\mathscr{F}$ to denote all the intervals in $\mathrm{T}$,
$\phi$ is referred to as an element of $\mathscr{F}$,
and $\Omega$ to denote the range occupied by the intervals in $\mathrm{T}$.
All of the interior knots  $t_1<t_2<...<t_{n-1}$
are referred to as the \textbf{C-knots} of $\mathrm{T}$,
which is denoted as $\mathrm{G}$.

The quadratic spline spaces is defined as:
\begin{equation}
\mathbf{S}(2,1,\mathrm{T}):=\{p(t)\in
C^{1}(\Omega): p(t)|_{\phi}\in \mathbb{P}_{2},\forall
\phi\in\mathscr{F}\}.
\label{s}
\end{equation}

For the knot sequence $\mathrm{T}$ of $\mathbf{S}(2,1,\mathrm{T})$,
we can obtain an
extended knot sequences by inserting two knots at each end of $\mathrm{T}$.
The \textbf{extension}\cite{Dengbook} of $\mathrm{T}$  is referred to as
$\mathrm{T}^{\varepsilon}:t_{-2}<t_{-1}<t_0<t_1<...<t_n<t_{n+1}<t_{n+2}$.

The corresponding quadratic spline spaces over $\mathrm{T}$ with homogeneous
boundary conditions (HBC) can be defined as follows:
\begin{equation}
\overline{\mathbf{S}}(2,1,\mathrm{T}):=\{p(t)\in
C^1(\mathbb{R}): p(t)|_{\phi}\in
\mathbb{P}_{2},\forall \phi\in\mathscr{F},   p(t)|_{\mathbb{R}\setminus\Omega} \equiv 0 \}.
\label{s-}
\end{equation}

The two spline spaces in Equations (\ref{s})
and (\ref{s-}) are closely related as follows:

\begin{equation}
\mathbf{S}(2,1,\mathrm{T}) = \overline{\mathbf{S}}(2,1,\mathrm{T}^{\varepsilon})|_\Omega.
\label{usp}
\end{equation}
\begin{equation}
\dim \mathbf{S}(2,1,\mathrm{T}) = \dim \overline{\mathbf{S}}(2,1,\mathrm{T}^{\varepsilon}).
\label{uspdim}
\end{equation}

With Equation (\ref{usp}) and Equation (\ref{uspdim}), to consider the spline space over the  knots,
we need only to consider the corresponding spline space with homogeneous boundary conditions over its extended  knots.
We just need to construct the bijective mapping between $\overline{\mathbf{S}}(2,1,\mathrm{T})$ and $\overline{\mathbf{S}}(0,-1,\mathrm{G})$.
and apply the mapping to  $\overline{\mathbf{S}}(2,1,\mathrm{T}^{\varepsilon})$ and
$\overline{\mathbf{S}}(0,-1,\mathrm{G}^{\varepsilon})$, where $\mathrm{G}^{\varepsilon}$ is the C-knots of $\mathrm{T}^{\varepsilon}$.

\subsection{ The mapping between $\overline{\mathbf{S}}(2,1,\mathrm{T})$ and $\overline{\mathbf{S}}(0,-1,\mathrm{G})$}
In this subsection, we first define a mapping functional by the B-ordinates.
And then, we use the mapping functional to define the  mapping formulae between $\overline{\mathbf{S}}(2,1,\mathrm{T})$ and $\overline{\mathbf{S}}(0,-1,\mathrm{G})$.

Let $p(t)$ be a  polynomial with degree 2 defined over the interval $\mathcal{I}:[t_0,t_1]$.
It can be expressed in the Bernstein - B$\acute{e}$zier form:
\begin{equation}
p(t) = \sum_{j=0}^{2}b_jB^2_j(u),  u=\frac{t-t_0}{t_1-t_0},
\label{theunivaretaB-ordintes}
\end{equation}
where $B^2_j(u)$ is the quadratic Bernstein polynomial and $\sum_{j=0}^{2}B^2_j(u)=1$.
$b_j$ is referred to as the {B$\acute{e}$zier-ordinates(B-ordinates)}
corresponding to $t_j:\frac{(2-j)t_0+jt_1}{2}, j=0,1,2$.

With Equation \ref{theunivaretaB-ordintes}, $(t-t_0)^2$ is the factor of $B^2_0(u)$ and $(t-t_1)^2$ is the factor of $B^2_2(u)$ , together with $\sum_{j=0}^{2}B^2_j(u)=1$,
the mapping functional is defined as follows:
\begin{definition}
Given the quadratic polynomial function $p(t)$ and $\mathcal{I} = [t_0,t_1]$, let $u=\frac{t-t_0}{t_1-t_0}$,
\[p(t)=b_0B^2_0(u)+b_1B^2_1(u)+b_2B^2_2(u)=b_1+(b_0-b_1)B^2_0(u)+(b_2-b_1)B^2_2(u).\]
We define the mapping functional $\mathbf{\varphi}$, \[\mathbf{\varphi} (p(t): \mathcal{I})=b_1|_{\mathcal{I}},\]
where %$\mathbf{\varphi} (p(t): \mathcal{I})$ denotes the functional of $p(t)$ on $\mathcal{I}$,
$b_1|_{\mathcal{I}}$ denotes a piecewise constant function whose value is $b_1$ on $\mathcal{I}$.
\label{danF}
\end{definition}

We can define the mapping formulae via the mapping functional in Definition \ref{danF} as follows:
\begin{definition}
Given the knots $\mathrm{T}: t_0<t_1<...<t_n$,
the C-knots of $\mathrm{T}$ are denoted as $\mathrm{G}:t_1<t_2<...<t_{n-1}$.
Each interior interval of  $T$ is denoted as $\mathcal{I}$,
the corresponding interval of $I$ on $\mathrm{G}$ is denoted as $\mathcal{IG}$.
We give the mapping  formulae between $\overline{\mathbf{S}}(2,1,\mathrm{T})$ and $\overline{\mathbf{S}}(0,-1,\mathrm{G})$ as:
\begin{equation}
\label{dan11111111111111111111}
\mathbf{\Phi}: \overline{\mathbf{S}}(2,1,\mathrm{T}) \rightarrow    \overline{\mathbf{S}}(0,-1,\mathrm{G}),
\end{equation}
\[\mathbf{\varphi} (p(t)|_{\mathcal{I}} : \mathcal{I}) \to q(t)|_{\mathcal{IG}},\]
where $p(t) \in \overline{\mathbf{S}}(2,1,\mathrm{T})$,
$p(t)|_\mathcal{I}$ denotes the expression of $p(t)$ on $\mathcal{I}$,
$q(t) \in \overline{\mathbf{S}}(0,-1,\mathrm{G})$,
and $q(t)|_{\mathcal{IG}}$ denotes the expression of $q(t)$ on $\mathcal{IG}$.
\label{mapping1}
\end{definition}

\begin{figure}[htp!]
\captionsetup{font={small}}
  \centering
  \subfloat[\footnotesize{$p(t)$}]{\includegraphics[width=3.5cm]{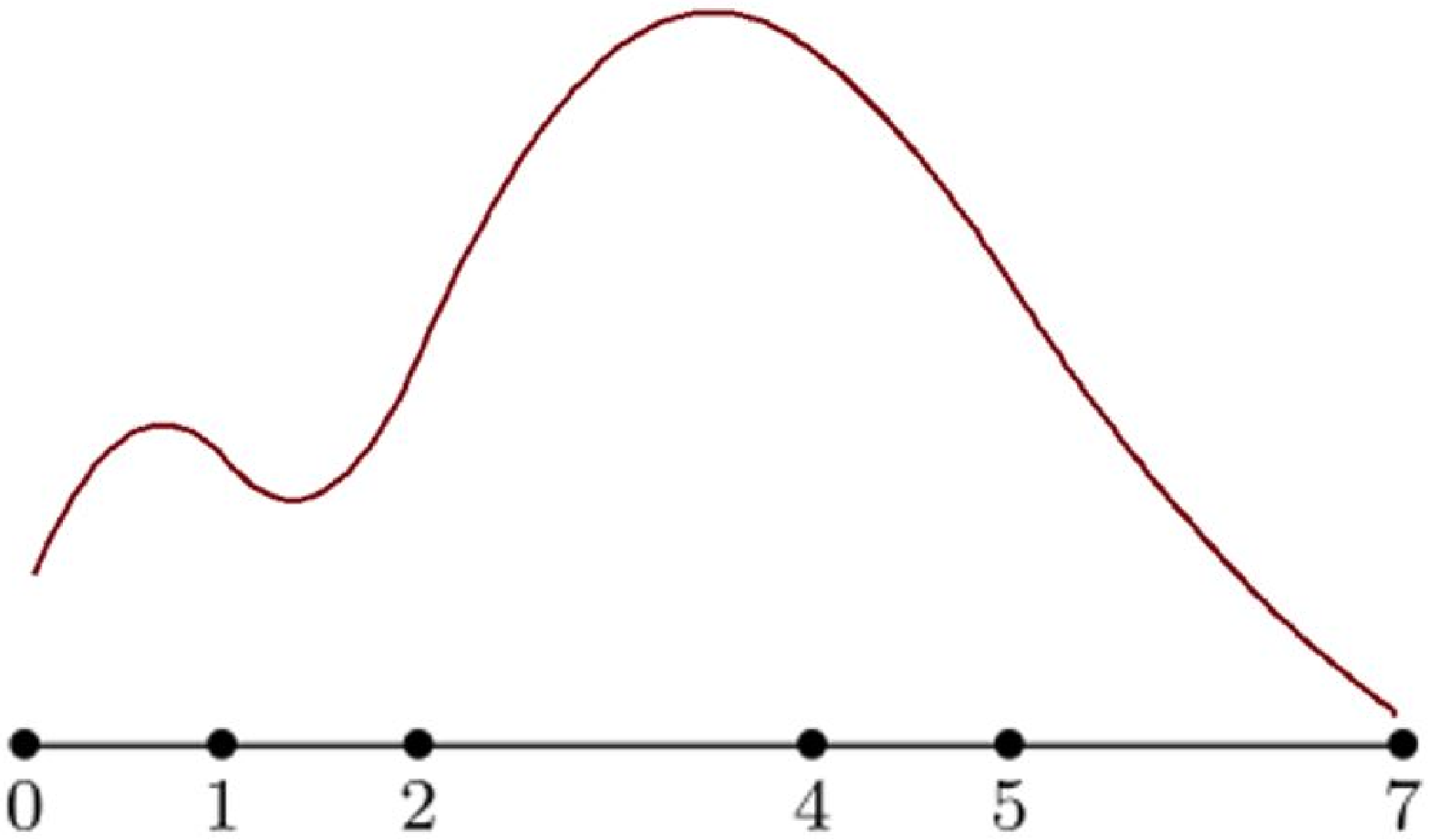}} \hspace{1cm}
   \subfloat[\footnotesize{B-ordinates of $p(t)$}]{\includegraphics[width=3.5cm]{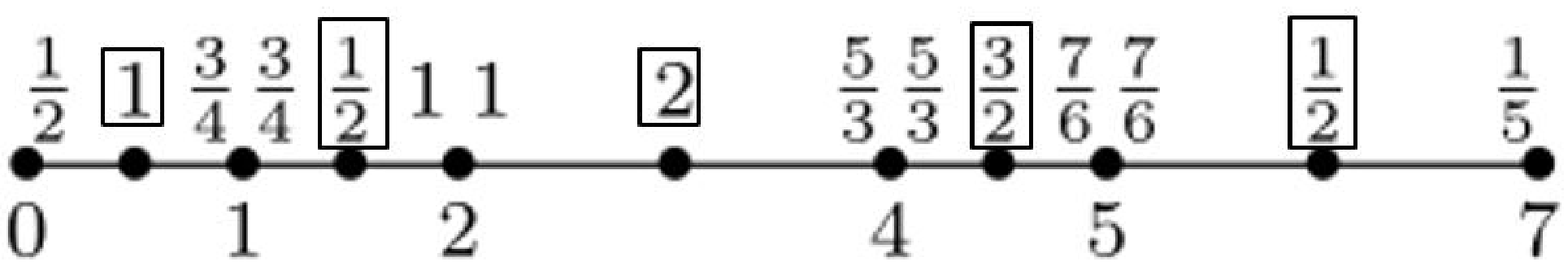}} \hspace{1cm}
  \subfloat[\footnotesize{$q(t)$}]{\includegraphics[width=3.5cm]{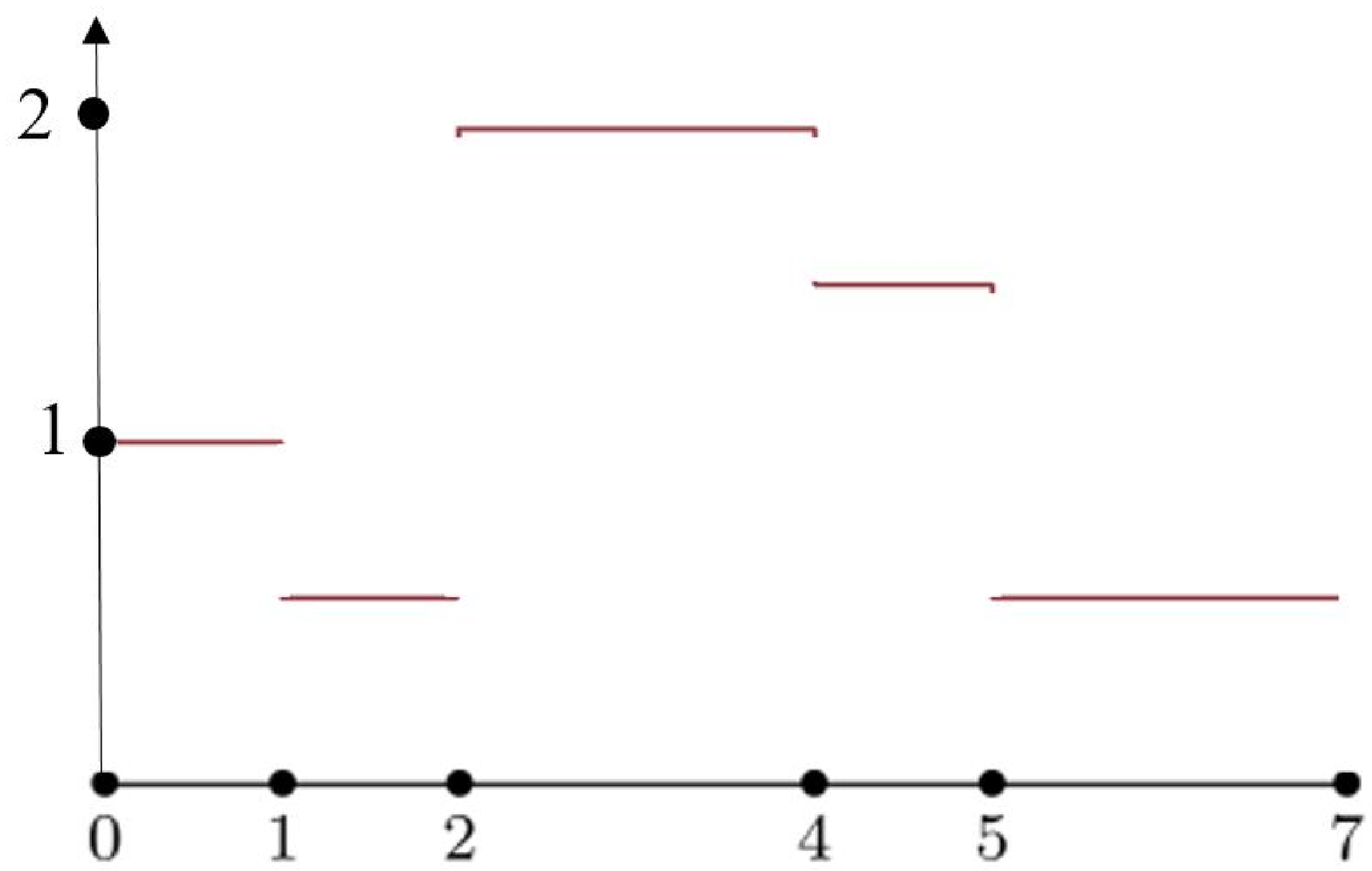}}
  \caption{The mapping of $p(t)$.}\label{univariatemapping}
\end{figure}

Fig. \ref{univariatemapping}(a) shows $p(t) \in \overline{\mathbf{S}}(2,1,\mathrm{T})$ on some interior intervals of $\mathrm{T}$,
Fig. \ref{univariatemapping}(c) shows the mapping result $q(t) \in \overline{\mathbf{S}}(0,-1,\mathrm{G})$ on the corresponding intervals.
%$q(t) \in \overline{\mathbf{S}}(0,-1,\mathrm{G})$ is the mapping result on the interior intervals.
In fact, the value of $q(t)$ on each interval in Fig. \ref{univariatemapping}(c)
is the B-ordinate on the centre of each interval, which is shown in Fig. \ref{univariatemapping}(b).

%\subsubsection{The injectivity property of the mapping}

%It suffices to show that
\begin{lem}
\label{injectiveuni1111111111111111}
The mapping defined in Definition \ref{mapping1} is injective.
\end{lem}
\begin{pf}
Obviously, $\mathbf{\Phi}(p(t)) \equiv 0$ implies $p(t)  \equiv 0$ for $p(t) \in  \overline{\mathbf{S}}(2,1,\mathrm{T})$.
Thus, $\mathbf{\Phi}$ is injective.
\end{pf}

From Lemma \ref{injectiveuni1111111111111111}, the mapping we defined in Definition \ref{mapping1} is an injective mapping.
To ensure the mapping is a bijective mapping, for each basis function of  $\overline{\mathbf{S}}(0,-1,\mathrm{G})$,
we need to construct a basis function of $\overline{\mathbf{S}}(2,1,\mathrm{T})$.

\subsection{Construction of the basis functions for $\overline{\mathbf{S}}(2,1,\mathrm{T})$}\label{constructuni11111111111}
In this subsection, %we will use the value of a basis function in $\overline{\mathbf{S}}(0,-1,\mathrm{G})$
%to construct a basis function in $\overline{\mathbf{S}}(2,1,\mathrm{T})$.
we first initialize the B-ordinates of the quadratic basis function in $\overline{\mathbf{S}}(2,1,\mathrm{T})$ via the value of a basis function in $\overline{\mathbf{S}}(0,-1,\mathrm{G})$,
and then give an algorithm to calculate the B-ordinates of the quadratic basis function.

\begin{figure}[htp!]
\captionsetup{font={small}}
  \centering
  \subfloat[\footnotesize{$q(t)$}]{\includegraphics[width=3.5cm]{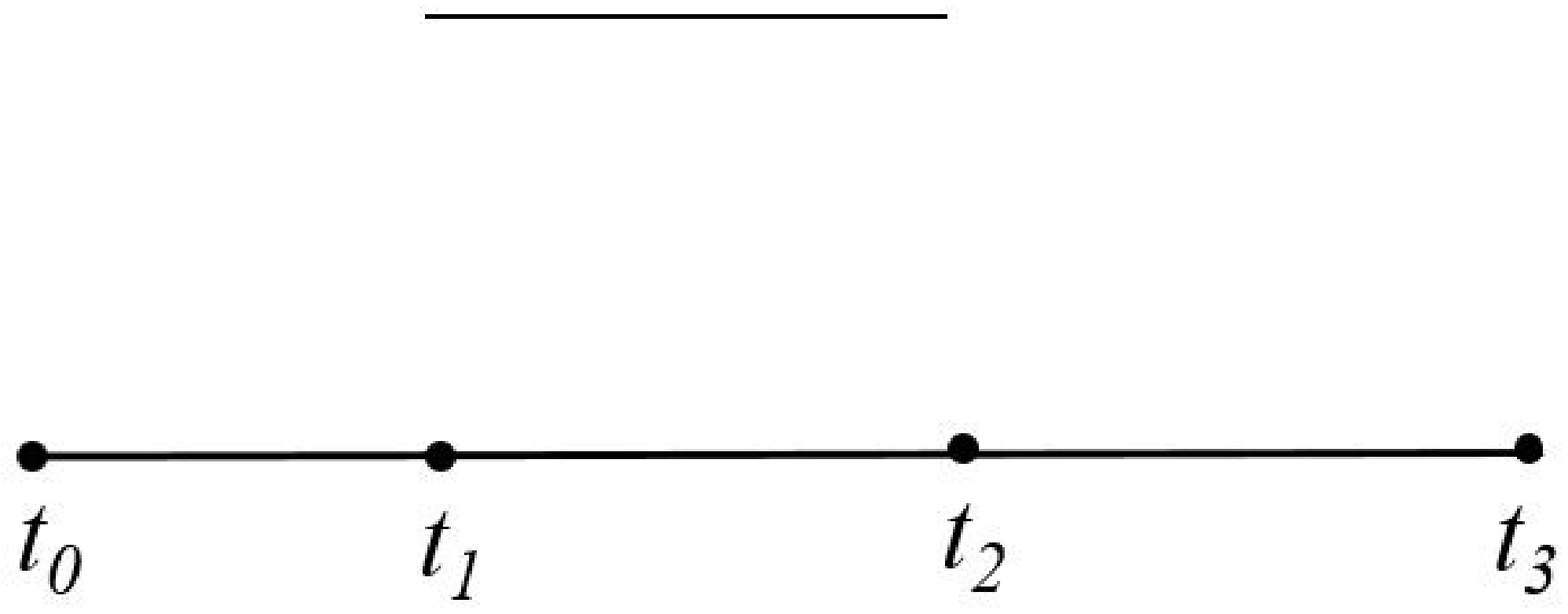}}\hspace{1cm}
  %\subfloat[B-ordinates]{\includegraphics[width=4.5cm]{bordinate.eps}}
   \subfloat[\footnotesize{$p(t)$}]{\includegraphics[width=3.5cm]{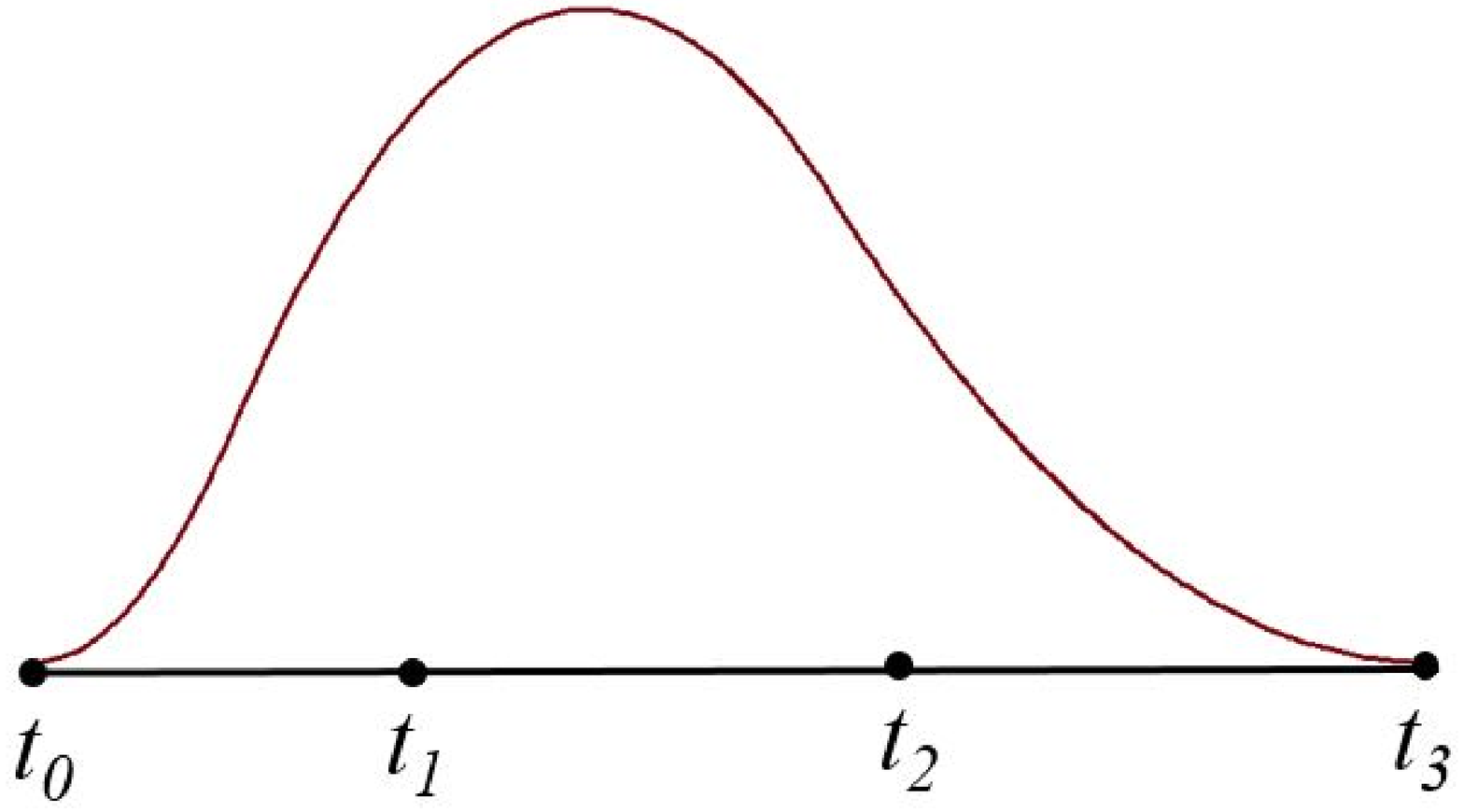}}
  \caption{The inverse mapping.}\label{UEval}
\end{figure}
\begin{figure}[!ht]
\captionsetup{font={small}}
\begin{center}
\psset{unit=0.65cm,linewidth=0.5pt}
\begin{tabular}{c@{\hspace*{1cm}}c@{\hspace*{1cm}}c}
\pspicture(0,0)(8,2)
\psline(0,0)(8,0)
\rput(0,0){\footnotesize${\bullet}$}\rput(0,0.4){\footnotesize${b^0_0}$}
\rput(1,0){\footnotesize${\green{\bullet}}$}\rput(1,0.4){\footnotesize${b^0_1}$}
\rput(2,0){\footnotesize${\bullet}$}\rput(1.75,0.4){\footnotesize${b^0_2}$}
\rput(2,0){\footnotesize${\bullet}$}\rput(2.25,0.4){\footnotesize${b^1_0}$}
\rput(3.5,0){\footnotesize${\green{\bullet}}$}\rput(3.5,0.4){\footnotesize${b^1_1}$}
\rput(5,0){\footnotesize${\bullet}$}\rput(4.75,0.4){\footnotesize${b^1_2}$}
\rput(5,0){\footnotesize${\bullet}$}\rput(5.25,0.4){\footnotesize${b^2_0}$}
\rput(6.5,0){\footnotesize${\green{\bullet}}$}\rput(6.5,0.4){\footnotesize${b^2_1}$}
\rput(8,0){\footnotesize${\bullet}$}\rput(8,0.4){\footnotesize${b^2_2}$}
\endpspicture \\
\end{tabular}
\caption{The B-ordinates $b^i_j,i=0,1,2;j=0,1,2$ of $p(t)$. \label{BB-PT}}
\end{center}
\end{figure}

Given the basis function of $\overline{\mathbf{S}}(0,-1,\mathrm{G})$ in Fig. \ref{UEval}(a):
\begin{equation}
\label{uivrateq}
 q(t)=
\begin{cases}
 1, &   [t_1,t_2]\\
 0, &   \text{other intervals}.
\end{cases}
\end{equation}
With Equation (\ref{uivrateq}), we initialize the B-ordinates of $p(t) \in \overline{\mathbf{S}}(2,1,\mathrm{T})$  as:
%\begin{equation}
%\label{Fbase11111}
%\text{B-ordinate}=
%\begin{cases}
%1,  & \frac{t_1+t_2}{2}\\
%0, & \text{other centre positions}
%\end{cases}.
%\end{equation}

\begin{equation}
\label{Fbase11111}
b^i_1=
\begin{cases}
1,  & i=1\\
0, & i \neq 1
\end{cases},
\end{equation}
which are the B-ordinates on ``$\green{{\bullet}}$" in Fig. \ref{BB-PT}.
Obviously, the support of $p(t)$ is $[t_0,t_3]$.
To calculate all of the B-ordinates $b^i_j,i=0,1,2;j=0,2$
for the polynomial function $p(t) \in \overline{\mathbf{S}}(2,1,\mathrm{T})$ we give some conclusions
as follows:
\begin{prop}
\label{univC1}
We use Fig.\ref{BB-PT} to illustrate some conclusions as follows:

1. $p(t)$ is $C^1$ continuous on $t_1$ if and only if
 $(t_1,b^0_2)$  occupies on the linear function that determinated by $(\frac{t_0+t_1}{2},b^0_1)$ and $(\frac{t_1+t_2}{2},b^1_1)$.

2. $p(t)$ is $C^1$ continuous on $t_2$ if and only if
$(t_2,b^1_2)$  occupies on the linear function that determinated by  $(\frac{t_1+t_2}{2},b^1_1)$ and $(\frac{t_2+t_3}{2},b^2_1)$.
\end{prop}
\begin{pf}
1. As $p(t)$ is $C^1$ continuous on $t_1$, we obtain $b^1_0=b^0_2$ and
\begin{equation}
\frac{b^0_2-b^0_1}{t_1-t_0}= \frac{b^1_1-b^0_2}{t_2-t_1},
\label{C111111}
\end{equation}
with Equation (\ref{C111111}), we obtain
\begin{equation}
\frac{b^0_2-b^0_1}{b^1_1-b^0_1}=\frac{t_1-\frac{t_1+t_0}{2}}{\frac{t_2+t_1}{2}-\frac{t_1+t_0}{2}},
\label{useC1}
\end{equation}
with Equation (\ref{useC1}),
the point$(t_1,b^0_2)$ is on the linear function that is determined by the points
$(\frac{t_0+t_1}{2},b^0_1)$ and $(\frac{t_1+t_2}{2},b^1_1)$.

The reverse proving process can be derived naturally.

2. Similar to 1, the proposition is correct.
\end{pf}

\begin{algorithm}[H]
\caption{Calculate  the B-ordinates of  $p(t) \in \mathbf{S}(2,1,\mathrm{T})$ \label{thedanlinear}}% 算法名字
\LinesNumbered %要求显示行号
\KwIn{The support $[t_0,t_{1},t_{2},t_{3}]$; $b^0_1,b^1_1,b^2_1$.}% 输入参数
\KwOut{All of the B-ordinates of  $p(t)$ on $[t_0,t_{3}]$ in Fig. \ref{BB-PT}.}%输出
%The linear function values on $t_i$ and $t_{i+3}$ are zero\;
%The B-ordinates on $t_i$ and $t_{i+3}$ are zero\;
%Obtain the linear function on $(t_{1},b^0_2)$ by $(\frac{t_0+t_{1}}{2},0)$ and $(\frac{t_{1}+t_{2}}{2},1)$\;
Using the two points $(\frac{t_0+t_{1}}{2},0)$ and $(\frac{t_{1}+t_{2}}{2},1)$ to calculate the linear function  on $(t_{1},b^0_2)$\;
Calculate the B-ordinate on $t_{1}$ as $b^0_2= \frac{t_{1}-t_0}{t_{2}-t_0}$\;
%Obtain the linear function on $(t_{2},b^{2}_0)$ by  $(\frac{t_{1}+t_{2}}{2},1)$  and $(\frac{t_{2}+t_{3}}{2},0)$\;
Using the two points $(\frac{t_{1}+t_{2}}{2},1)$  and $(\frac{t_{2}+t_{3}}{2},0)$ to calculate the linear function on $(t_{2},b^{2}_0)$\;
Calculate the B-ordinate on $t_{2}$ as $b^{2}_0= \frac{t_{2}-t_{1}}{t_{1}-t_{3}}+1$\;
Obtain the B-ordinates on $[t_{1},t_{2}]$ as $\{b^0_2, 1,b^2_0\}$\;
Using the $C^1$ continuous condition to calculate that the B-ordinates on $t_0$ and$ \frac{t_0+t_1}{2}$ are 0\;
Obtain the B-ordinates on $[t_0,t_{1}]$ as $\{0,0,b^0_2\}$\;
Using the $C^1$ continuous condition to calculate that the B-ordinates on $\frac{t_2+t_3}{2}$ and $t_3$ are 0\;
Obtain the B-ordinates on $[t_{2},t_{3}]$ as $\{b^{2}_0,0,0\}$\;
\end{algorithm}

Then, given the B-ordinates by Equation (\ref{Fbase11111}), we can calculate the B-ordinates of
$p(t) \in \overline{\mathbf{S}}(2,1,\mathrm{T})$ via Algorithm \ref{thedanlinear}.
We show $p(t)$ in Fig. \ref{UEval}(b), $p(t)$ is $C^1$ continuous, and $p(t)=\mathbf{\Phi}^{-1}( q(t))$.
%By the Algorithm \ref{thedanlinear}, we obtain the B-ordinates of $p(t)=\mathbf{\Phi}^{-1}( q(t))$, and $p(t)$ is $C^1$ continue.

\subsection{The isomorphic univariate spaces and properties}
In this subsection, we prove that the mapping is bijective,
$\overline{\mathbf{S}}(2,1,\mathrm{T})$ is isomorphic to $\overline{\mathbf{S}}(0,-1,\mathrm{G})$,
the basis functions of $\overline{\mathbf{S}}(2,1,\mathrm{T})$ hold the properties of are linearly independence, partition of unity and completeness.
\begin{thm}
\label{mybijectivemapping1111}
The mapping defined in definition \ref{mapping1} holds the property of bijectivity.
\end{thm}
\begin{pf}
For each basis function of $\overline{\mathbf{S}}(0,-1,\mathrm{G})$,
we can obtain a quadratic function of
$\overline{\mathbf{S}}(2,1,\mathrm{T})$ via Algorithm \ref{thedanlinear},
the mapping is surjective.
As the mapping is injective,
the mapping holds the property of bijectivity.
\end{pf}

As the mapping between $\overline{\mathbf{S}}(2,1,\mathrm{T})$ and $\overline{\mathbf{S}}(0,-1,\mathrm{G})$ is bijective.
We obtain the following corollary naturally.

\begin{coro}
\label{mybijectivemapping22222}
$\overline{\mathbf{S}}(2,1,\mathrm{T})$ is isomorphic to $\overline{\mathbf{S}}(0,-1,\mathrm{G})$.
\end{coro}

%Then, $\mathbf{S}(0,-1,\mathrm{G})$ is isomorphic to $\mathbf{S}(2,1,\mathrm{T})$.
\begin{thm}
\label{mybijectivemapping33333}
The basis functions of $\mathbf{S}(2,1,\mathrm{T})$, which are constructed in \ref{constructuni11111111111},
hold the properties of linearly independence, partition of unity and completeness on $[t_0,t_n]$.
\end{thm}
\begin{pf}
Assume that the basis functions of $\overline{\mathbf{S}}(0,-1,\mathrm{G}^{\varepsilon})$ are $q_i(t)$,
the basis functions of $\overline{\mathbf{S}}(2,1,\mathrm{T}^{\varepsilon})$ are $p_i(t), i=1,...,N$,
where $\mathrm{G}^{\varepsilon}$ is the C-knots of $\mathrm{T}^{\varepsilon}$.
We obtain that the mapping between $\overline{\mathbf{S}}(2,1,\mathrm{T}^{\varepsilon})$  and $\mathbf{S}(0,-1,\mathrm{G}^{\varepsilon})$ is bijective,
and $\overline{\mathbf{S}}(2,1,\mathrm{T}^{\varepsilon})$  is isomorphic to $\overline{\mathbf{S}}(0,-1,\mathrm{G}^{\varepsilon})$.

As the spaces are linear spaces and $\mathbf{\Phi}^{-1}( q_i(t))= p_i(t)$, we obtain
\[\sum^{N}_{i=1} p_i(t) = \sum^{N}_{i=1} \mathbf{\Phi}^{-1}( q_i(t) ) =\mathbf{\Phi}^{-1} (\sum^{N}_{i=1} q_i(t)).\]
As
$\sum^{N}_{i=1}( q_i(t) )= 1, t \in [t_{-1}, t_{n+1}],$ the B-ordinate on the centre position of each interior interval on $\mathrm{T}^{\varepsilon}$ are $1$.
By Algorithm \ref{thedanlinear}, the B-ordinates on each interval of $\mathrm{T}$ is $1$.
With Equation (\ref{usp}) and Equation (\ref{uspdim}), the basis functions of $\mathbf{S}(2,1,\mathrm{T})$ have partition of unity on $[t_0,t_n]$.

As the basis functions of $\overline{\mathbf{S}}(0,-1,\mathrm{G}^{\varepsilon})$ are  linearly independent and  complete on $[t_0,t_n]$,
and $\overline{\mathbf{S}}(2,1,\mathrm{T}^{\varepsilon})$  is isomorphic to $\overline{\mathbf{S}}(0,-1,\mathrm{G}^{\varepsilon})$,
the polynomial functions of $\overline{\mathbf{S}}(2,1,\mathrm{T}^{\varepsilon})$ are linearly independent and  complete on $[t_0,t_n]$.
With Equation (\ref{usp}) and Equation (\ref{uspdim}), the basis functions of $\mathbf{S}(2,1,\mathrm{T})$, which are constructed in \ref{constructuni11111111111}, are linearly independent and  complete on $[t_0,t_n]$.
The theorem is proved.
\end{pf}

Till now, we construct a bijective mapping between $\overline{\mathbf{S}}(2,1,\mathrm{T})$  and $\overline{\mathbf{S}}(0,-1,\mathrm{G})$,
the two spaces are isomorphic to each other,
some important properties are the same for the basis functions of the two spaces.
For $\mathbf{S}(2,2,1,1,\mathscr{T})$, we want to obtain similar conclusions.
We denote $\mathbf{S}(d,d,d-1,d-1,\mathscr{T})$ as  $\mathbf{S}^d(\mathscr{T})$ for convenience.
By Equation \ref{relationshipbetweenTG}, we first discuss the spline space $\mathbf{S}^2(\mathscr{T})$ with homogeneous boundary conditions,
which is denoted by $\overline{\mathbf{S}}^2(\mathscr{T})$.
\section{The mapping between $\overline{\mathbf{S}}^2(\mathscr{T})$ and $\overline{\mathbf{S}}^0(\mathscr{G})$}\label{mappingformulea111111}
In this section, we will introduce the mapping between $\overline{\mathbf{S}}^2(\mathscr{T})$ and $\overline{\mathbf{S}}^0(\mathscr{G})$.

\subsection{Some notations and CVR graphs}

Before we give the mapping,
we introduce some notations for a hierarchical T-mesh in Table \ref{marksandnotations},
we also give some abbreviations in brackets for convenience.

\begin{table*}
\captionsetup{font={small}}
\resizebox{\textwidth}{!}{
\begin{tabular*}{16.5cm}{lclclc}
\hline
Notations   &&  Definitions\\
\hline
P-cell($\mathcal{PC}$) && An interior-cell that four corner vertices are crossing-vertices.  \\
T-cell &&  An interior-cell that at least one of four corner vertices is a T-junction. \\
T-connected && Two T-cells are T-connected if they connect at a T-junction.\\
T-connection($\mathcal{TC}$) && The union of all T-connected T-cells.\\
P-domain($\mathcal{PD}$) && The domain on which a pure-cell occupies.\\
T-connection-domain($\mathcal{TCD}$) && The domain on which a T-connection occupies.\\
T-rectangle-domain($\mathcal{TRD}$) && The minimal rectangular domain that covers a T-connection.\\
Domain($\mathcal{D}$) && The minimal rectangular domain that covers a T-connection.\\
Domain-center && The centre point of the domain.\\
One-neighbour-cell && The lowest level cells adjacent to the T-connection.\\
The level of the T-connection && The level of the one-neighbour-cells corresponds to the T-connection.\\
\hline
\end{tabular*}}
\caption{Some notations for hierarchical T-meshes \label{marksandnotations}}
\end{table*}

%1. For an interior cell,
%if the four corner vertices are crossing-vertices, the cell is referred to as a \textbf{pure-cell} (P-cell for short);
%otherwise, $C$ is  referred to as a \textbf{T-cell}.
%
%2. Two T-cells are \textbf{T-connected} if at least one of their common corners is a T-junction.
%A \textbf{T-connection} is the set that consists of all T-connected T-cells.
%%Each T-cell of the T-connection is T-connected with at least one of other T-cells in the set.
%
%3. The domain on which a pure cell occupies is called a \textbf{P-domain}.
%The domain on which a T-connection occupies is called a  \textbf{T-connection-domain}.
%The minimal rectangular domain that covers a T-connection is called a \textbf{T-rectangle-domain}.
%For the domain $\mathcal{D}: [x_0,x_1] $ $\times$ $ [y_0,y_1]$,
%the centre position is $(\frac{x_0+x_1}{2}, \frac{y_0+y_1}{2})$, which is denoted as the \textbf{domain-centre} of $\mathcal{D}$.
%
%4. For a T-connection, the lowest level cells adjacent to the T-connection
%are called the \textbf{one-neighbour-cell} of the T-connection.
%We denote the level of the one-neighbour-cell as the \textbf{level} of the T-connection.
\begin{figure}[!htb]
\captionsetup{font={small}}
\begin{center}
\psset{unit=0.65cm,linewidth=0.5pt}
\begin{tabular}{c@{\hspace*{1cm}}c@{\hspace*{1cm}}c}

\begin{pspicture}(0,0)(4,4)
\psframe*[linecolor=lightgray](2,1)(3,2)
\psframe*[linecolor=green](1,2)(1.5,3)\psframe*[linecolor=green](1.5,2.5)(2,3)

\psline(0,0)(4,0)(4,4)(0,4)(0,0)
\psline(0,1)(4,1)\psline(0,2)(4,2)\psline(0,3)(4,3)
\psline(1,0)(1,4)\psline(2,0)(2,4)\psline(3,0)(3,4)
\psline(1,1.5)(2,1.5)\psline(1,2.5)(3,2.5)
\psline(1.5,1)(1.5,3)\psline(2.5,2)(2.5,3)
\rput(1.75,2.25){\scriptsize${0}$}
\rput(1.25,2.25){\scriptsize${1}$}
\rput(1.25,2.75){\scriptsize${2}$}
\rput(1.75,2.75){\scriptsize${3}$}
\rput(0.5,2.5){\scriptsize${4}$}
\rput(2.5,1.5){\scriptsize${5}$}
\psline[linecolor=red,linewidth=1pt](0.97,1.97)(2.03,1.97)(2.03,3.03)(0.97,3.03)(0.97,1.97)
\rput(1.5,2.5){\scriptsize${\bullet}$}
%\rput(2.25,2.25){\scriptsize${6}$}\rput(2.75,2.25){\scriptsize${7}$}\rput(2.75,2.75){\scriptsize${8}$}
%\rput(2.25,2.75){\scriptsize${9}$}
%\rput(1.5,3.5){\scriptsize${\mathcal{TC}_0}$}\rput(1.5,3){\scriptsize${\downarrow}$}
\end{pspicture} &
%\begin{pspicture}(0,0)(4,4)
%\psline(0,0)(0,4)(4,4)(4,0)(0,0)
%\psline(0,1)(4,1)\psline(1,1.5)(2,1.5)\psline(0,2)(4,2)\psline(1,2.5)(3,2.5)\psline(0,3)(4,3)
%\psline(1,0)(1,4)\psline(1.5,1)(1.5,3)\psline(2,0)(2,4)\psline(2.5,2)(2.5,3)\psline(3,0)(3,4)
%%\psline[linecolor=blue](1.03,2.03)(1.47,2.03)(1.47,2.53)(1.97,2.53)(1.97,2.97)(1.03,2.97)(1.03,2.03)
%\end{pspicture}&

\begin{pspicture}(0,0)(4,4)
\psline(1,1)(3,1)(3,3)(1,3)(1,1)
\psline(1,2)(3,2)\psline(2,1)(2,3)
\psline(1.5,1.5)(1.5,2.5)(2.5,2.5)
\rput(1.75,2.2){\scriptsize{0}}
\rput(1.3,2.5){\scriptsize{1}}
%\rput(2.6,2.6){\scriptsize{2}}
%\psline[linecolor=red](0.97,1.97)(2.03,1.97)(2.03,3.03)(0.97,3.03)(0.97,1.97)
\end{pspicture}  \\
(a) \scriptsize{$\mathscr{T}$}  &  (b) \scriptsize{$\mathscr{G}$}
\end{tabular}
\caption{Notations and CVR graph.\label{littlecvr}}
\end{center}
\end{figure}
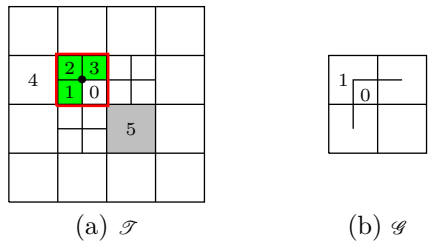

We use Fig. \ref{littlecvr}(a) to introduce the notations in Table \ref{marksandnotations}.
In Fig. \ref{littlecvr}(a), cell $0$ and cell $5$ are  P-cells, while cell $1$, cell $2$ and cell $3$ are T-cells.
Cell $1$ and cell $2$ are T-connected, cell $2$ and cell $3$ are T-connected.
The T-connection, which can be denoted as  $\mathcal{TC}_0$, is the union that consists of cell $1$, cell $2$ and cell $3$.
As cell $5$ is a P-cell, the gray domain is a P-domain.
The green domain is the T-connection-domain of $\mathcal{TC}_0$,
the domain inside the red square is the T-rectangle-domain of $\mathcal{TC}_0$,
the domain-centre of the T-rectangle-domain is denoted as $``{\bullet}"$ in Fig. \ref{littlecvr}(a).
Cell $4$ is the one-neighbour-cell of $\mathcal{TC}_0$,
the level of cell $4$ is the level of $\mathcal{TC}_0$.

In \cite{Deng13}, Definition \ref{CVR2222222222} is introduced to propose a topological explanation to the dimension formula of $\mathbf{S}(2,2,1,1,\mathscr{T})$.
\begin{definition}\cite{Deng13}
\label{CVR2222222222}
Given a hierarchical T-mesh $\mathscr{T}$, we can construct a graph
$\mathscr{G}$ by retaining the crossing-vertices and the line
segments with two end points that are crossing-vertices and removing
the other vertices and the edges in $\mathscr{T}$. $\mathscr{G}$
is called the \textbf{crossing-vertex-relationship graph} (CVR graph
for short) of $\mathscr{T}$.
\end{definition}

We introduce some notations of CVR graph for the mapping in Table \ref{marksandnotationsCVR},
we also give some abbreviations in brackets for convenience.

\begin{table*}
\captionsetup{font={small}}
\resizebox{\textwidth}{!}{
\begin{tabular*}{16.5cm}{lclclc}
\hline
Notations   &&  Definitions\\
\hline
g-cell   ($\mathcal{GC}$)  && A grid element in CVR graph.\\
P-g-cell ($\mathcal{PGC}$) &&  A g-cell corresponds to a P-cell in $\mathscr{T}$.\\
T-g-cell ($\mathcal{TGC}$) &&  A g-cell corresponds to a T-connection-domain in $\mathscr{T}$.  \\
\hline
\end{tabular*}}
\caption{Cells for CVR graphs \label{marksandnotationsCVR}}
\end{table*}
%The rest part of the T-connection that all of the edges with T-junctions are removed
We also use Fig. \ref{littlecvr} to illustrate the notations in Table \ref{marksandnotationsCVR}.
Fig. \ref{littlecvr}(b) shows the  CVR graph $\mathscr{G}$ of the hierarchical T-mesh $\mathscr{T}$ in Fig. \ref{littlecvr}(a).
The P-cell 0 in \ref{littlecvr}(a) corresponds to
the P-g-cell 0 in \ref{littlecvr}(b).
In Fig. \ref{littlecvr}(a), for the T-connection $\mathcal{TC}_0$,
the  T-connection-domain of $\mathcal{TC}_0$ corresponds to
the T-g-cell 1 in Fig. \ref{littlecvr}(b).

From the relationship between the cells of a hierarchical T-mesh and its CVR graph, we consider the mapping between $\overline{\mathbf{S}}^2(\mathscr{T})$ and $\overline{\mathbf{S}}^0(\mathscr{G})$.

\subsection{The mapping formulae between $\overline{\mathbf{S}}^2(\mathscr{T})$ and $\overline{\mathbf{S}}^0(\mathscr{G})$}

By the notations in Table \ref{marksandnotations} and Table \ref{marksandnotationsCVR}, we use the B-ordinates to define a functional, and then use the functional to construct
the mapping between $\overline{\mathbf{S}}^2(\mathscr{T})$ and $\overline{\mathbf{S}}^0(\mathscr{G})$.

%Let $f(x,y) \in \overline{\mathbf{S}}^2(\mathscr{T})$
%be a polynomials
%defined over the cell
%$C:[x_0,x_1] \times [y_0,y_1]$,
%it can be expressed as:
%
%\begin{equation}
%f(x,y)= \sum_{j=0}^{2} \sum_{k=0}^{2} b_{j,k}B^2_j(u)B^2_k(v),u=\left(\frac{x-x_0}{x_1-x_0}\right),v=\left(\frac{y-y_0}{y_1-y_0}\right).
%\label{Tbordi111}
%\end{equation}
For the Bernstein polynomials
$B^2_j(u)B^2_k(v),j,k=0,1,2$ on $[x_0,x_1] \times [y_0,y_1]$, where  $u=\left(\frac{x-x_0}{x_1-x_0}\right),v=\left(\frac{y-y_0}{y_1-y_0}\right)$, we obtain
\begin{equation}
B^2_1(u)B^2_1(v)=1-\sum_{j=0}^{2} B^2_j(u)B^2_0(v)- \sum_{j=0,j \neq 1}^{2} B^2_j(u)B^2_1(v) - \sum_{j=0}^{2} B^2_j(u)B^2_2(v).
\label{sumisone}
\end{equation}
As $B^2_j(u),j=0,2$ possess the factors $(x-x_0)^2$ or  $(x-x_1)^2$,
and $B^2_k(v),k=0,2$ possess the factors $(y-y_0)^2$ or  $(y-y_1)^2$.
We can give the functional as follows:
 \begin{definition}
 \label{theTfunctional}
Given  $f(x,y) \in \overline{\mathbf{S}}^2(\mathscr{T})$, and $\mathcal{D}:=[x_0,x_1] \times [y_0,y_1]$ is a rectangular domain.
Let $u=\frac{x-x_0}{x_1-x_0}, v=\frac{y-y_0}{y_1-y_0}$, $f(x,y)$ can be expressed as:
\[f(x,y)= \sum_{j=0}^{2} \sum_{k=0}^{2} b_{j,k}B^2_j(u)B^2_k(v),\]
we obtain
\[f(x,y)=   b_{1,1} + \sum_{j=0}^{2}( b_{j,0}-b_{1,1})B^2_j(u)B^2_0(v) + \sum_{j=0,j \neq 1}^{2}( b_{j,1}-b_{1,1})B^2_j(u)B^2_1(v)+ \sum_{j=0}^{2}( b_{j,2}-b_{1,1})B^2_j(u)B^2_2(v).\]
We define a mapping functional $\mathbf{\varphi}$, \[\mathbf{\varphi} (f(x,y):\mathcal{D})=b_{1,1}|_\mathcal{D},\]
where $b_{1,1}|_D$ denotes a piecewise constant function whose value is $b_{1,1}$ on $\mathcal{D}$.
\end{definition}

%\begin{table}
%\captionsetup{font={small}}
%\caption{Some notations and marks\label{marks}}
%\resizebox{\textwidth}{!}{
%\begin{tabular*}{15cm}{lclclc}
%\hline
%P-cell   ($\mathcal{PC}$) && A interior-cell that four corner vertices are crossing-vertices. \\
%P-domain ($\mathcal{PD}$) && The domain on which a pure-cell occupies.\\
%T-connection  ($\mathcal{TC}$) && The set that consists of all T-connected T-cells.\\
%T-connection-domain  ($\mathcal{TCD}$) && The domain on which a T-connection occupies.\\
%T-rectangle-domain ($\mathcal{TRD}$) && The minimal rectangular domain that covers a T-connection.\\
%Domain ($D$)&& P-domain and T-rectangle-domain\\
%g-cell ($\mathcal{GC}$) && A grid element in CVR graph\\
%P-g-cell ($\mathcal{PGC}$) &&Corresponds to a P-cell in $\mathscr{T}$.&&\\
%T-g-cell ($\mathcal{TGC}$) && The rest part of the T-connection that all of the edges \\
%       &&   with T-junctions are removed from the T-connection.\\
%\hline
%\end{tabular*}}
%\end{table}

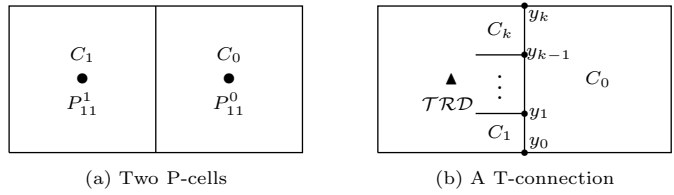
\begin{figure}[!htb]
\captionsetup{font={small}}
\begin{center}
\psset{unit=0.65cm,linewidth=0.5pt}
\begin{tabular}{c@{\hspace*{1cm}}c@{\hspace*{1cm}}c}

\begin{pspicture}(0,0)(6,3)
\psline(0,0)(6,0)(6,3)(0,3)(0,0)
\psline(3,0)(3,3)
\rput(1.5,1.5){${\bullet}$}\rput(4.5,1.5){${\bullet}$}
\rput(1.5,2){\scriptsize${C_1}$}\rput(4.5,2){\scriptsize${C_0}$}
\rput(1.5,1){\scriptsize${P^1_{11}}$}\rput(4.5,1){\scriptsize${P^0_{11}}$}
\end{pspicture} &
\begin{pspicture}(0,0)(6,3)
\psline(0,0)(6,0)(6,3)(0,3)(0,0)
\psline(3,0)(3,3)
\psline(2,0.8)(3,0.8)
\psline(2,2)(3,2)
\rput(2.5,1.5){$\vdots$}
\rput(2.5,0.4){\scriptsize${C_1}$}
\rput(2.5,2.5){\scriptsize${C_k}$}
\rput(3,0){\tiny${\bullet}$}\rput(3,0.8){\tiny${\bullet}$}\rput(3,2){\tiny${\bullet}$}\rput(3,3){\tiny${\bullet}$}
\rput(3.3,0.2){\scriptsize${y_0}$}\rput(3.3,0.8){\scriptsize${y_1}$}\rput(3.5,2){\scriptsize${y_{k-1}}$}\rput(3.3,2.8){\scriptsize${y_k}$}
\rput(4.5,1.5){\scriptsize${C_0}$}
\rput(1.5,1.5){\scriptsize${\blacktriangle}$}
\rput(1.5,1){\scriptsize${\mathcal{TRD}}$ }
\end{pspicture}  \\
\scriptsize{(a)} \scriptsize{Two P-cells} & \scriptsize{(b)} \scriptsize{A T-connection}
\end{tabular}
\caption{Two P-cells and a T-connection.\label{myC1}}
\end{center}
\end{figure}

%\begin{figure}[htp!]
%  \centering
%  \subfloat[Two P-cells]{\includegraphics[width=4cm]{c1continue.eps}}\hspace{1cm}
%  \subfloat[A T-connection]{\includegraphics[width=4cm]{c1mapping.eps}}
%  \caption{Two P-cells and a T-connection}\label{myC1}
%\end{figure}

%Similar to proposition \ref{univC1}, in Fig. \ref{myC1}, $(P^2_{11})$
In Fig. \ref{myC1}(a), $C_0$ and $C_1$ are two aligned P-cells in $\mathscr{T}$.
$P^i_{11}$ is the center domain point of $C_i, i=0,1$.
Let $f_i(x,y)$ be the polynomial with bi-degree $(2, 2)$
defined over $C_i$,
the B-ordinate on $P^i_{11}$  is denoted as $b^i_{11}$,
the domain that covers $C_i$ is denoted as $\mathcal{PD}_i, i=0,1$.
Applying the functional $\varphi$ in Definition \ref{theTfunctional},
$\mathbf{\varphi} (f_i(x,y):{\mathcal{PD}_i})=b^i_{1,1}|_{\mathcal{PD}_i}, i=0,1.$

In Fig. \ref{myC1}(b), $C_0$ is the one-neighbor-cell of $\mathcal{TC}$  in $\mathscr{T}$,
%the T-connection-domain of $\mathcal{TC}$ is denoted as $\mathcal{TCD}$,
the T-rectangle-domain of $\mathcal{TC}$ is denoted as $\mathcal{TRD}$.
Let $f_i(x,y)$ be the polynomial with bi-degree $(2, 2)$
defined over $C_i,i=0,...,k$,

$f_1(x,y)=f_0(x,y)+(x-x_1)^2v(y),$

$f_2(x,y)=f_0(x,y)+(x-x_1)^2v(y)+c_1(x-x_1)^2(y-y_1)^2,$

$\vdots$

$f_k(x,y)=f_0(x,y)+(x-x_1)^2v(y)+c_1(x-x_1)^2(y-y_1)^2+...+c_{k-1}(x-x_1)^2(y-y_{k_1})^2, v(y) \in \mathbb{P}_{2}(y), c_i \in \mathbb{R}.$

As $f_i(x,y),i=1,...,k$ possess the cofactor $(x-x_1)^2$, $\mathbf{\varphi} (f_i(x,y) : {\mathcal{TRD}})=\mathbf{\varphi} (f_0(x,y): {\mathcal{TRD}})$.

%For $\mathscr{T}$,
%if we want to construct the mapping similar to the univariate mapping,
%we need to consider the mapping result for each P-g-cell and T-g-cell in $\mathscr{G}$.
%As each P-g-cell in $\mathscr{G}$ is a P-cell in $\mathscr{T}$.
%For each P-cell in $\mathscr{T}$, we can use the polynomials on itself.
%As each  T-g-cell in $\mathscr{G}$ is generated by  a T-connection
%$\mathcal{TC}$ in $\mathscr{T}$,
%we can use the polynomials on the one-neighbour-cell of $\mathcal{TC}$.

Given a hierarchical T-mesh $\mathscr{T}$, $\mathscr{G}$ denotes the CVR graph of $\mathscr{T}$.
For $f(x,y) \in \overline{\mathbf{S}}^2(\mathscr{T})$,
we denote the support of $f(x,y)$ as $Sup(f)$.
From Table \ref{marksandnotations} and Table \ref{marksandnotationsCVR},
$\mathcal{PC}$ is denoted as a P-cell of $Sup(f)$,
the P-domain of $\mathcal{PC}$ is denoted as $\mathcal{PD}$,
the P-g-cell corresponds to $\mathcal{PC}$ in $\mathscr{G}$ is denoted as $\mathcal{PGC}$;
%and the polynomial on $\mathcal{PC}$ is denoted as $f(x,y)_P.$
$\mathcal{TC}$  is denoted as a T-connection of $Sup(f)$,
the T-rectangle-domain of $\mathcal{TC}$ is denoted as $\mathcal{TRD}$.
%,
%the one-neighbour-cell of $\mathcal{TC}$ is denoted as $C_T$,
%the polynomial on $C_T$ is denoted as $f(x,y)_T$,
And the T-g-cell corresponds to $\mathcal{TC}$ in $\mathscr{G}$ is denoted as $\mathcal{TGC}$.
We can define the mapping between $\overline{\mathbf{S}}^2(\mathscr{T})$
and $\overline{\mathbf{S}}^0(\mathscr{G})$  as follows:

\begin{definition}
\label{uniquemappingto}

The mapping formulae between $\overline{\mathbf{S}}^2(\mathscr{T})$ and $\overline{\mathbf{S}}^0(\mathscr{G})$ is defined as :
\begin{equation} \label{mapT}
\mathbf{\Phi}: \overline{\mathbf{S}}^2(\mathscr{T}) \rightarrow   \overline{\mathbf{S}}^0(\mathscr{G}),
\end{equation}
\begin{equation}
\begin{cases}
\mathbf{\varphi} (f(x,y)|_{\mathcal{PC}} : {\mathcal{PD}}) \to g(x,y)|_{\mathcal{PGC}}\\
\mathbf{\varphi} (f(x,y)|_{\mathcal{TC}} : {\mathcal{TRD}}) \to g(x,y)|_{\mathcal{TGC}}.
%\mathbf{\Phi}(f(x,y))|_{\mathcal{PGC}} = \mathbf{\varphi} (f(x,y)_P)|_{\mathcal{PD}} \\
%\mathbf{\Phi}(f(x,y))|_{\mathcal{TGC}} = \mathbf{\varphi} (f(x,y)_T)|_{\mathcal{TRD}}
\end{cases}
\end{equation}
Where $f(x,y) \in \overline{\mathbf{S}}^2(\mathscr{T})$, $f(x,y)|_{\mathcal{PC}}$ denotes the expression of $f(x,y)$ on $\mathcal{PC}$,
$g(x,y) \in \overline{\mathbf{S}}^0(\mathscr{G})$, and $g(x,y)|_{\mathcal{PGC}}$ denotes the expression of $g(x,y)$ on $\mathcal{PGC}$, which corresponds to $\mathcal{PC}$;
 $f(x,y)|_{\mathcal{TC}}$ denotes the expression of $f(x,y)$ on the one-neighbour-cell of $\mathcal{TC}$, and $g(x,y)|_{\mathcal{TGC}}$ denotes the expression of $g(x,y)$ on $\mathcal{TGC}$,
 which corresponds to $\mathcal{TC}$.
\end{definition}
\subsection{The injectivity property of the mapping}

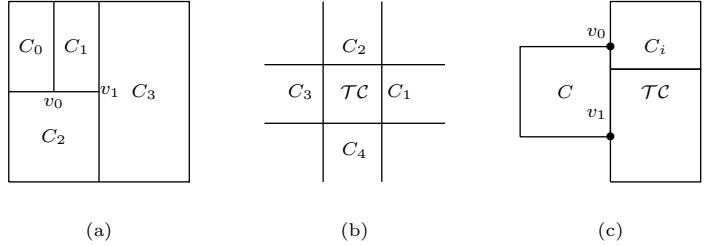
\begin{figure}[!ht]
\captionsetup{font={small}}
\begin{center}
\psset{unit=0.6cm,linewidth=0.5pt}
\begin{tabular}{c@{\hspace*{1cm}}c@{\hspace*{1cm}}c}
\begin{pspicture}(0,-0.5)(4,4.5)
\psline(0,0)(4,0)(4,4)(0,4)(0,0)
\psline(2,0)(2,4)\psline(0,2)(2,2)
\psline(1,2)(1,4)
\rput(0.5,3){\scriptsize${C_0}$}\rput(1.5,3){\scriptsize${C_1}$}
\rput(1,1){\scriptsize${C_2}$}\rput(3,2){\scriptsize${C_3}$}
\rput(1,1.75){\scriptsize${v_0}$}
\rput(2.25,2){\scriptsize${v_1}$}
\end{pspicture} &
\begin{pspicture}(0,-0.5)(4,4.5)
\psline(0,1.3)(4,1.3)\psline(0,2.6)(4,2.6)
\psline(1.3,0)(1.3,4)\psline(2.6,0)(2.6,4)
\rput(2,2){\scriptsize{$\mathcal{TC}$}}
\rput(3,2){\scriptsize{$C_1$}}\rput(2,3){\scriptsize{$C_2$}}
\rput(0.8,2){\scriptsize{$C_3$}}\rput(2,0.7){\scriptsize{$C_4$}}
\end{pspicture} &
\begin{pspicture}(0,-0.5)(4,4.5)
\psline(0,1)(2,1)(2,3)(0,3)(0,1)
\psline(2,0)(4,0)(4,2.5)(2,2.5)(2,0)
\psline(2,2.5)(4,2.5)(4,4)(2,4)(2,2.5)
\rput(1.7,3.3){\scriptsize{$v_0$}}\rput(2,3){\scriptsize{$\bullet$}}
\rput(1.7,1.5){\scriptsize{$v_1$}}\rput(2,1){\scriptsize{$\bullet$}}
\rput(3,2){\scriptsize{$\mathcal{TC}$}}
%\rput(1.6,3){$P_1$}\rput(2,3){$\bullet$}
%\rput(1.6,1.8){$P_2$}\rput(2,1.5){$\bullet$}
\rput(3,3){\scriptsize{$C_i$}} \rput(1,2){\scriptsize{$C$}}
\end{pspicture}\\
\scriptsize{(a)}  & \scriptsize{(b)} & \scriptsize{(c)}
\end{tabular}
\caption{Figures for Lemma \ref{T-comp-one-neibor}. \label{WD}}
\end{center}
\end{figure}

\begin{lem}
\label{T-comp-one-neibor}
Given a hierarchical T-mesh $\mathscr{T}$,
for each T-connection $\mathcal{TC} \in \mathscr{T}$,
the T-rectangle-domain of $\mathcal{TC}$ is denoted as $\mathcal{TRD}$.
%and the T-g-cell corresponds to $\mathcal{TC}$ in $\mathscr{G}$ is denoted as $\mathcal{TGC}$.
At least one one-neighbour-cell of $\mathcal{TC}$ exists,
and for all the one-neighbour-cells of $\mathcal{TC}$,
apply the functional $\mathbf{\varphi}$ with the polynomial of each one-neighbour-cell on $\mathcal{TRD}$,
the results are the same.
\end{lem}
\begin{pf}
Without loss of generality, we use Fig. \ref{WD} to illustrate the lemma.

1.  At least one one-neighbour-cell of $\mathcal{TC}$ exists.

 In Fig. \ref{WD}(a),
 we denote the level of $C_i$ as $l_i$ ,where $ i=0,1,2,3$.
 As $v_0$ is a T-junction, we get
 $l_2 \leqslant l_0$ and $l_2 \leqslant l_1$.
 As $v_1$ is a T-junction, we get
$l_3 \leqslant l_2$.
Then, $C_3$ is the cell with the lowest level of $C_0,C_1,C_2,C_3$.
In a similar manner, one-neighbour-cell must exist.

2.  If at least two one-neighbour-cells of $\mathcal{TC}$ exist,
they are connected to $\mathcal{TC}$ and have the same level.

According to Fig. \ref{WD} (b),
the neighbour cells of $\mathcal{TC}$ are $C_{1}, C_{2}, C_{3}$ and $C_{4}$.
The maximum level of the cells in $\mathcal{TC}$ is $l_0$,
and the  level of $C_i$ is $l_i$, where $ i=1,2,3,4$.
As  the mesh is a  hierarchical T-mesh, we can assume  $l_1=l_3=l$, and by (1), $l<l_0$ .

(1) Assume $l_2 > l$, then $l_4 < l$.

We use  proofs by contradiction  to prove $l_4 < l$.
If $l_4 > l$, by the assumption, $l_2 > l$, then $\mathcal{TC}$ will be divided,
and we obtain a contradiction.
Thus, $l_4 < l$  and $C_4$ is the only one-neighbour-cell.

(2) Assume $l_2<l$ and $l_4<l$.

If $l_2<l_4$, $C_2$  is the only one-neighbour-cell.
If $l_2>l_4$, $C_4$  is the only one-neighbour-cell.
If $l_2=l_4$, $C_2$ and $C_4$  are connected to $\mathcal{TC}$.

Thus, if $\mathcal{TC}$ has at least two one-neighbour-cells,
the levels of the one-neighbour-cells are same.

3. The one-neighbour-cell is aligned with $\mathcal{TC}$ .

According to Fig. \ref{WD}  (c),  $v_0$ must be cross-vertexes;
otherwise, the one-neighbour-cell of $\mathcal{TC}$ is not the lowest level cell.

%As the one-neighbour cell is a cell with the lowest level that is aligned at $\mathcal{TC}$,
%a well-defined, unique mapping can be defined by the functional $\mathbf{\varphi}$,
From the above, the mapping is well defined and
the mapping result is unique.
The lemma is proved.
\end{pf}

\begin{thm}
\label{Injectivemapping}
The mapping defined in Equation \eqref{mapT} is injective.
\end{thm}
\begin{pf}
By Lemma \ref{T-comp-one-neibor}, the mapping result on each T-connection is unique,
and $\mathbf{\Phi}(f(x,y)) \equiv 0$ implies $f(x,y) \equiv 0$
for $f(x,y) \in \overline{\mathbf{S}}^2(\mathscr{T})$.
Thus, $\mathbf{\Phi}$ is injective.
\end{pf}

The mapping  we defined in Definition \ref{uniquemappingto} is injective.
To verify that the mapping is a bijective mapping, we need to confirm
the mapping is surjective.
In other words,
given a basis function $q(x,y) \in \overline{\mathbf{S}}^0(\mathscr{G})$,
we need to construct the corresponding basis function $p(x,y) \in \overline{\mathbf{S}}^2(\mathscr{T})$,
and $p(x,y)$ is the inverse image of $q(x,y)$.
\section{Construction of the  basis functions for $\overline{\mathbf{S}}^2(\mathscr{T})$}\label{basisfunctions}

In this section, we give a general method to construct the basis functions of
$\overline{\mathbf{S}}^2(\mathscr{T})$ when there is no limitation for level difference on  $\mathscr{T}$.
Each basis function  corresponds  to a basis function of $\overline{\mathbf{S}}^0(\mathscr{G})$.
First, we propose a new structure and introduce how to use T-structures for calculating the B-ordinates.
Second, for each basis function of $\overline{\mathbf{S}}^2(\mathscr{T})$, we initialize the weights on each domain of $\mathscr{T}$ via the basis function of $\overline{\mathbf{S}}^0(\mathscr{G})$,
we use the T-structures to calculate the B-ordinates for the basis function of $\overline{\mathbf{S}}^2(\mathscr{T})$.
Finally, we propose that our computation can be reduced by simplifying $\mathscr{T}$.
\subsection{T-structures and some conclusions for T-structures}\label{TSTSTSTS}

T-structures will play an important role in calculating the B-ordinates for each basis function of $\overline{\mathbf{S}}^2(\mathscr{T})$.
%We can use T-structures as the carriers for B-ordinates.
In this subsection, we introduce how to use T-structures for calculating the B-ordinates of a polynomial function on the T-structure.
%First, we give some  notations for T-structures and T-structure-branches.
%Second, we prove that each
%T-structure-branch corresponds to a  T-connection.
%Third, for the purpose of calculating B-ordinates of a T-connection by the T-structure-branch, we introduce the B-net method on
%T-structures.

\subsubsection{Some  notations for T-structures}
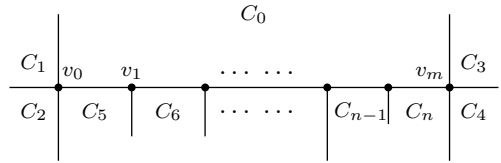
\begin{figure}[!ht]
\captionsetup{font={small}}
\begin{center}
\psset{unit=0.65cm,linewidth=0.5pt}
\begin{tabular}{ccc}
\begin{pspicture}(0,1)(12,3.5)
\psline(1,2)(11,2)\psline(2,0.5)(2,3.5)\psline(10,0.5)(10,3.5)
\psline(5,1)(5,2)\psline(7.5,0.5)(7.5,2)
\psline(3.5,1)(3.5,2)\psline(8.75,1.25)(8.75,2)
\psdots(2,2)\psdots(10,2)\psdots(5,2)\psdots(7.5,2)\psdots(3.5,2)\psdots(8.75,2)
\rput(6,3.5){\footnotesize${C_0}$}
\rput(1.5,2.5){\footnotesize${C_1}$}
\rput(1.5,1.5){\footnotesize${C_2}$}
\rput(10.5,2.5){\footnotesize${C_3}$}
\rput(10.5,1.5){\footnotesize${C_4}$}

\rput(2.75,1.5){\footnotesize${C_5}$}
\rput(4.25,1.5){\footnotesize${C_6}$}
\rput(8.2,1.5){\footnotesize${C_{n-1}}$}
\rput(9.4,1.5){\footnotesize${C_n}$}
\rput(2.3,2.3){\footnotesize${v_0}$}
\rput(9.6,2.3){\footnotesize${v_m}$}
\rput(3.5,2.3){\footnotesize${v_1}$}
\rput(5.6,2.3){$\hdots$}\rput(6.5,2.3){$\hdots$}
\rput(5.6,1.5){$\hdots$}\rput(6.5,1.5){$\hdots$}
\end{pspicture}\\
\end{tabular}
\caption{The T-structure $\mathcal{T}$. \label{notations of Tstr}}
\end{center}
\end{figure}
\begin{definition} \label{Tsdefinition}
In the  hierarchical T-mesh $\mathscr{T}$, a c-edge and all of the cells that
have at least one common vertex with the c-edge constitute a \textbf{T-structure} of $\mathscr{T}$, we denote the T-structure as $\mathcal{T}$ for convenience.
The c-edge is referred to as  the \textbf{mid-edge} of $\mathcal{T}$.
All of the vertices on the c-edge are referred to as the \textbf{interior-vertices} of $\mathcal{T}$.
The end-points of the c-edge are referred to as the \textbf{end-points}  of $\mathcal{T}$.
The lowest level cell that has the common edge with the c-edge is referred to as the \textbf{mother-cell} of $\mathcal{T}$.
The cells adjacent to the mid-edge except the mother-cell  are referred to as the \textbf{sub-cells} of $\mathcal{T}$.
The level of the mother-cell is denoted as the \textbf{level} of $\mathcal{T}$.
If the mid-edge is horizontal(vertical), the T-structure
is referred to as a \textbf{horizontal(vertical) T-structure}.
\end{definition}

In Fig. \ref{notations of Tstr},
the T-structure $\mathcal{T}$
consists of the c-edge $v_0v_m$ and the cells $C_0,...,C_n$.
$v_0v_m$ is the mid-edge of $\mathcal{T}$.
$v_0,...,v_m$ are the interior-vertices of $\mathcal{T}$.
$v_0$ and $v_m$ are the end-points of $\mathcal{T}$.
$C_0$ is the mother-cell of $\mathcal{T}$.
$C_5,...,C_n$ are the sub-cells of $\mathcal{T}$.
The level of the mother-cell $C_0$ is the level of the T-structure $\mathcal{T}$.
$\mathcal{T}$ is a horizontal T-structure.

\begin{figure}[!htb]
\captionsetup{font={small}}
\begin{center}
\psset{unit=0.65cm,linewidth=0.5pt}
\begin{tabular}{c@{\hspace*{1cm}}c}
\begin{pspicture}(0,0)(6,4)
%\psframe*[linecolor=lightgray](2.9,0.9)(3.1,3.2)
%\psframe*[linecolor=lightgray](1.9,1.8)(2.1,3.2)
%\psframe*[linecolor=lightgray](1.8,1.9)(3.2,2.1)
\psline(0,0)(6,0)(6,4)(0,4)(0,0)
\psline(0,2)(6,2)\psline(0,1)(2,1)\psline(0,3)(6,3)
\psline(2,0)(2,4)\psline(1,0)(1,4)\psline(3,2)(3,4)
\psline(4,0)(4,4)\psline(5,2)(5,4)
\psline(1.5,1)(1.5,3)\psline(1,2.5)(3,2.5)
\psline(1,1.5)(2,1.5)\psline(2.5,2)(2.5,3)
\psdots(2,2)\psdots(3,2)\psdots(4,2)\psdots(3,3)\psdots(2,3)
\rput(2.2,1.8){\footnotesize${v_0}$} \rput(3,1.8){\footnotesize${v_1}$} \rput(3.8,1.8){\footnotesize${v_2}$}
\rput(3.2,3.2){\footnotesize${v_3}$} \rput(2.2,3.2){\footnotesize${v_4}$}
\rput(3,1){\footnotesize${0}$}
\rput(1.75,1.75){\footnotesize${1}$}
\rput(1.75,2.25){\footnotesize${2}$} \rput(2.25,2.25){\footnotesize${3}$}\rput(2.75,2.25){\footnotesize${4}$}\rput(3.5,2.5){\footnotesize${5}$}
\rput(4.5,2.5){\footnotesize${6}$}\rput(5,1){\footnotesize${7}$}
\rput(2.75,2.75){\footnotesize${8}$}\rput(3.5,3.5){\footnotesize${9}$}\rput(2.5,3.5){\footnotesize${10}$}
\rput(2.25,2.75){\footnotesize${11}$}\rput(1.75,2.75){\footnotesize${12}$} \rput(1.5,3.5){\footnotesize${13}$}
%\rput(3.3,1.65){\footnotesize${\gets}$}
\end{pspicture} &
\begin{pspicture}(0,0)(6,4)
\psframe*[linecolor=lightgray](1.9,1.9)(4.1,2.1)
\psframe*[linecolor=lightgray](2.9,1.8)(3.1,3.2)
\psframe*[linecolor=lightgray](1.9,2.9)(3.2,3.1)
\psline(0,0)(6,0)(6,4)(0,4)(0,0)
\psline(0,2)(6,2)\psline(0,1)(2,1)\psline(0,3)(6,3)
\psline(2,0)(2,4)\psline(1,0)(1,4)\psline(3,2)(3,4)
\psline(4,0)(4,4)\psline(5,2)(5,4)
\psline(1.5,1)(1.5,3)\psline(1,2.5)(3,2.5)
\psline(1,1.5)(2,1.5)\psline(2.5,2)(2.5,3)
\psdots(2,2)\psdots(3,2)\psdots(4,2)\psdots(3,3)\psdots(2,3)
\rput(3,1.75){\footnotesize${\uparrow}$}\rput(3,1.4){\footnotesize${\mathcal{T}_0}$}
\rput(3.25,2.5){\footnotesize${\gets}$}\rput(3.6,2.5){\footnotesize${\mathcal{T}_1}$}
\rput(2.5,3.25){\footnotesize${\downarrow}$}\rput(2.5,3.6){\footnotesize${\mathcal{T}_3}$}
\end{pspicture} \\[2mm]
 \footnotesize{(a) T-structures} & \footnotesize{(b) A T-structure-branch}
\end{tabular}
\caption{T-structures on a hierarchical T-mesh .\label{Tsdef}}
\end{center}
\end{figure}
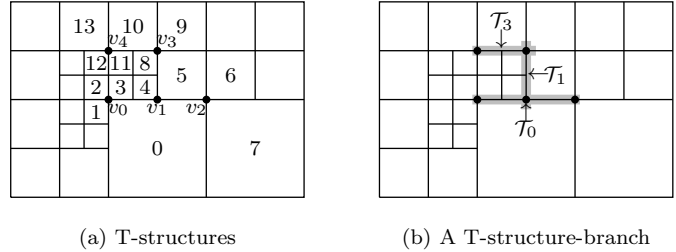

%From definition \ref{Tsdefinition}, we give some notations as follows:
\begin{definition}
\label{tstructurebranch11111}
One horizontal T-structure and one vertical T-structure are \textbf{connected} if they
have one common interior-vertex.
The union of all connected T-structures is referred to as a \textbf{T-structure-branch},
which is denoted as $\mathcal{TSB}$.
The minimal level of the  T-structures in a T-structure-branch
is denoted as the \textbf{level} of the T-structure-branch.
\end{definition}

 We show an example in Fig. \ref{Tsdef}:
$\mathcal{T}_0$ in Fig. \ref{Tsdef} (b) consists of the c-edge $v_0v_2$ and the surrounding cells $0,1,...,7$, which are shown in Fig. \ref{Tsdef} (a);
$\mathcal{T}_1$ in Fig. \ref{Tsdef} (b) consists of the c-edge $v_1v_3$ and the surrounding cells $0,4,5,8,9,{10}$, which are shown in Fig. \ref{Tsdef} (a);
and $\mathcal{T}_2$ in Fig. \ref{Tsdef} (b) consists of the c-edge $v_3v_4$ and the surrounding cells $5,8,9,...,{13}$, which are shown in Fig. \ref{Tsdef} (a).
$\mathcal{T}_0$ and $\mathcal{T}_1$ are connected at $v_1$, $\mathcal{T}_1$ and $\mathcal{T}_2$ are connected at $v_3$.
In Fig. \ref{Tsdef}(b),
$\mathcal{T}_0, \mathcal{T}_1$ and $\mathcal{T}_2$ constitute a T-structure-branch.
In Fig. \ref{Tsdef}(b), the level of $\mathcal{T}_0$ is denoted as the level of the T-structure-branch.

To make use of the T-structure-branches in Definition \ref{tstructurebranch11111}, we give a lemma to connect the T-structure-branches with T-connections.
\begin{lem}
\label{coverallcells}
Given the T-connection $\mathcal{TC} \in \mathscr{T}$,
$\mathcal{TC}$ will be covered by a T-structure-branch.
\end{lem}
\begin{pf}
Assume $\mathcal{T}_1, \mathcal{T}_2,...,\mathcal{T}_n$ are the T-structures
that cover the T-cells $\{C_1, C_2,..., C_m\}$.
\begin{enumerate}
\item [(1)] When $m =1$,
obviously, only one T-structure $T_1$ covers $\mathcal{TC}$.
The conclusion is right.

\item [(2)] When $m >1$, we prove it by reduction to absurdity.

Without loss of generality,
assume $\mathcal{T}_1$ is not connected with any T-structure of
$\mathcal{T}_2, \mathcal{T}_3,...,\mathcal{T}_n$.
Assume that the sub-cells of $\mathcal{T}_0$ are $I=\{C_{i_1},..., C_{i_k}\}$,
and $\{ i_1,...,i_k\}$ is a sub set of $\{1,...,m\}$.
We denote  $\{C_1, C_2,..., C_m\} \setminus I$ as the sub T-cell set of
$\{C_1, C_2,..., C_m\}$ except $\{C_{i_1},..., C_{i_k}\}$.
T-cells in $\{C_1, C_2,..., C_m\} \setminus I$
are not T-connected to the T-cells in $I$,
and $\mathcal{TC}$ will be divided into two,
it is a contradiction of the assumption.
Thus, T-structures $\mathcal{T}_1, \mathcal{T}_2,...,\mathcal{T}_n$ comprise a T-structure-branch.
\end{enumerate}
The lemma is proved.
\end{pf}

We also show an example by Fig. \ref{Tsdef}: In Fig. \ref{Tsdef}(a), the T-connection $\mathcal{TC}$  consists of cell 3, cell 4, cell 5, cell 8, cell 11;
in Fig. \ref{Tsdef}(b), the T-structure Branch $\mathcal{TSB}$  consists of $\mathcal{T}_0$,$\mathcal{T}_1$ and $\mathcal{T}_2$;
obviously, the cells of $\mathcal{TC}$ are covered by the cells belonging to $\mathcal{TSB}$.

\subsubsection{B-net method on T-structures}
%From Lemma \ref{coverallcells}, each T-connection is covered by a T-structure-branch.
In order to connect T-structures with B-ordinates, we introduce the B-net method on T-structures as follows.

\begin{figure}[!ht]
\captionsetup{font={small}}
\begin{center}
\psset{unit=0.65cm,linewidth=0.5pt}
\begin{tabular}{c@{\hspace*{1cm}}c}
\begin{pspicture}(0,0)(6,3.5)
\psline(0,1.5)(6,1.5)\psline(1,0)(1,3.5)\psline(5,0)(5,3.5)
\psline(2,0.5)(2,1.5)\psline(3,0.5)(3,1.5)\psline(4,0.5)(4,1.5)
\rput(3.5,1){$\hdots$}
\rput(0.8,1.3){\scriptsize{$\bullet$}}\rput(0.3,1.3){\scriptsize{$\bullet$}}\rput(0.8,0.8){\scriptsize{$\bullet$}}\rput(0.3,0.8){\scriptsize{$\bullet$}}
\rput(0.8,1.7){\scriptsize{$\bullet$}}\rput(0.3,1.7){\scriptsize{$\bullet$}}\rput(0.8,2.2){\scriptsize{$\bullet$}}\rput(0.3,2.2){\scriptsize{$\bullet$}}
\rput(1.2,1.7){\scriptsize{$\circ$}}\rput(3,1.7){\scriptsize{$\circ$}}\rput(4.8,1.7){\scriptsize{$\circ$}}
\rput(1.2,2.5){\scriptsize{$\circ$}}\rput(3,2.5){\scriptsize{$\circ$}}\rput(4.8,2.5){\scriptsize{$\circ$}}

\psdots[linecolor=green](1.2,1.3)\psdots[linecolor=green](1.2,0.9)
\psdots[linecolor=green](1.5,1.3)\psdots[linecolor=green](1.5,0.9)
\psdots[linecolor=green](1.8,1.3)\psdots[linecolor=green](1.8,0.9)

\psdots[linecolor=green](2.2,1.3)\psdots[linecolor=green](2.2,0.9)
\psdots[linecolor=green](2.5,1.3)\psdots[linecolor=green](2.5,0.9)
\psdots[linecolor=green](2.8,1.3)\psdots[linecolor=green](2.8,0.9)

\psdots[linecolor=green](4.2,1.3)\psdots[linecolor=green](4.2,0.9)
\psdots[linecolor=green](4.5,1.3)\psdots[linecolor=green](4.5,0.9)
\psdots[linecolor=green](4.8,1.3)\psdots[linecolor=green](4.8,0.9)

\rput(5.2,1.3){\scriptsize{$\bullet$}}\rput(5.7,1.3){\scriptsize{$\bullet$}}
\rput(5.2,0.8){\scriptsize{$\bullet$}}\rput(5.7,0.8){\scriptsize{$\bullet$}}
\rput(5.2,1.7){\scriptsize{$\bullet$}}\rput(5.7,1.7){\scriptsize{$\bullet$}}
\rput(5.2,2.2){\scriptsize{$\bullet$}}\rput(5.7,2.2){\scriptsize{$\bullet$}}
%\psline[linecolor=red](1.05,1.55)(3.1,1.55)(3.1,2.6)(1.05,2.6)(1.05,1.55)
\end{pspicture}&
\begin{pspicture}(0,0)(6,3.5)
\psline(0,1.5)(5,1.5)\psline(1,0)(1,3.5)\psline(5,0)(5,3.5)
\psline(2,0.5)(2,1.5)\psline(3,0.5)(3,1.5)\psline(4,0.5)(4,1.5)
\rput(3.5,1){$\hdots$}
\rput(0.8,1.3){\scriptsize{$\bullet$}}\rput(0.3,1.3){\scriptsize{$\bullet$}}\rput(0.8,0.8){\scriptsize{$\bullet$}}\rput(0.3,0.8){\scriptsize{$\bullet$}}
\rput(0.8,1.7){\scriptsize{$\bullet$}}\rput(0.3,1.7){\scriptsize{$\bullet$}}\rput(0.8,2.2){\scriptsize{$\bullet$}}\rput(0.3,2.2){\scriptsize{$\bullet$}}
\rput(1.2,1.7){\scriptsize{$\circ$}}\rput(3,1.7){\scriptsize{$\circ$}}\rput(4.8,1.7){\scriptsize{$\circ$}}
\rput(1.2,2.5){\scriptsize{$\circ$}}\rput(3,2.5){\scriptsize{$\circ$}}\rput(4.8,2.5){\scriptsize{$\circ$}}
\psdots[linecolor=green](1.2,1.3)\psdots[linecolor=green](1.2,0.9)
\psdots[linecolor=green](1.5,1.3)\psdots[linecolor=green](1.5,0.9)
\psdots[linecolor=green](1.8,1.3)\psdots[linecolor=green](1.8,0.9)

\psdots[linecolor=green](2.2,1.3)\psdots[linecolor=green](2.2,0.9)
\psdots[linecolor=green](2.5,1.3)\psdots[linecolor=green](2.5,0.9)
\psdots[linecolor=green](2.8,1.3)\psdots[linecolor=green](2.8,0.9)

\psdots[linecolor=green](4.2,1.3)\psdots[linecolor=green](4.2,0.9)
\psdots[linecolor=green](4.5,1.3)\psdots[linecolor=green](4.5,0.9)
\psdots[linecolor=green](4.8,1.3)\psdots[linecolor=green](4.8,0.9)

\rput(5.2,0.8){\scriptsize{$\bullet$}}\rput(5.7,0.8){\scriptsize{$\bullet$}}
\rput(5.2,1.5){\scriptsize{$\bullet$}}\rput(5.7,1.5){\scriptsize{$\bullet$}}
\rput(5.2,2.2){\scriptsize{$\bullet$}}\rput(5.7,2.2){\scriptsize{$\bullet$}}
\rput(3,3.3){\scriptsize{$\bullet$}}\rput(4.8,3.3){\scriptsize{$\bullet$}}
\rput(4.5,0.5){\scriptsize{$\bullet$}}\rput(4.8,0.5){\scriptsize{$\bullet$}}
\end{pspicture}\\[2mm]
\footnotesize{(a) The end points are two crossing-vertices} & \footnotesize{(b) One of the end-pint is a crossing-vertex}
\end{tabular}
\caption{The corresponding B-ordinates on T-structures \label{calculate T-structure}}
\end{center}
\end{figure}
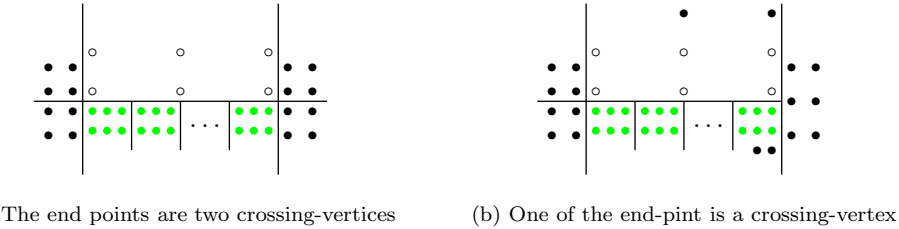

Let $p(x,y) \in \overline{\mathbf{S}}^2(\mathscr{T})$ be the polynomial
defined over the cells of the T-structure  $\mathcal{T} \in \mathscr{T}$.
We assume $\mathcal{T}$ as a horizontal T-structure in  Fig. \ref{calculate T-structure}(a),
we refer to the B-ordinates on ``$\green{\bullet}$", ``${\bullet}$" and ``$\circ$"
as the \textbf{corresponding B-ordinates} of $p(x,y)$ on $\mathcal{T}$.
%We refer to the B-ordinates in the red rectangle as the \textbf{ end-B-ordinates} of the left end-point,
%the end-B-ordinates  of the right end-point can be defined similarly.
For the T-structure in Fig. \ref{calculate T-structure}(b), the corresponding B-ordinates can be defined similarly.
For the vertical T-structures, we can also define the notations similarly.
\begin{lem}
\label{TSbordinates}
For a T-structure $\mathcal{T} \in \mathscr{T}$, let $p(x,y) \in \overline{\mathbf{S}}^2(\mathscr{T})$
be the polynomial defined over the cells of $\mathcal{T}$.
Then, when the two rows (columns) B-ordinates on the mother-cell that near the mid-edge are given,
the  two rows (columns) B-ordinates on sub-cells that near the mid-edge are determined.
\end{lem}

\begin{pf}
We prove the lemma for horizonal T-structures.
In Fig. \ref{calculate T-structure},
when the B-ordinates on ``$\circ$" of the mother-cell are given,
the B-ordinates on ``$\green{\bullet}$" of the sub-cells
can be calculated via B-net method.
If it is a vertical T-structure,
the lemma can be similarly proved.
\end{pf}

By Lemma \ref{TSbordinates}, if we want to obtain
the two rows (column) B-ordinates of sub-cells that are near the mid-edge,
we need to obtain the six B-ordinates of the mother-cell that are near the mid-edge.
\subsubsection{The corresponding B-ordinates on a crossing-vertex}
In this subsection, we introduce the B-ordinates associated with a crossing-vertex,
which can help us to obtain the six B-ordinates of the mother-cell that are near the mid-edge.

Let $p(x,y) \in \overline{\mathbf{S}}^2(\mathscr{T})$ be the polynomial
defined over the cells around a crossing-vertex $v^+ \in \mathscr{T}$, which are shown in Fig. \ref{Evalbilinear}(a).
We define the sixteen B-ordinates on the green domain points in Fig. \ref{Evalbilinear}(a)
as the \textbf{corresponding B-ordinates} of $p(x,y)$ on $v^+$.
In Fig. \ref{Evalbilinear},
the crossing-vertex $v^+$ is denoted as ``${\blacksquare}$",
%the edges around  $v^+$ are $E_i$,
the cells around  $v^+$ are denoted as $C_i$, where $i=1,2,3,4$,
and $C_1$ is denoted as the cell with the top level.
We use Fig. \ref{Evalbilinear} to give the lemma that describes the relationship between
the {corresponding B-ordinates} on a crossing-vertex and the bilinear function as follows:
\begin{figure}[!htb]
\captionsetup{font={small}}
\begin{center}
\psset{unit=0.65cm,linewidth=0.5pt}
\begin{tabular}{c@{\hspace*{1cm}}c@{\hspace*{1cm}}c}
\begin{pspicture}(-0.5,-0.5)(4,4)
\psline(0,2)(4,2)\psline(2,0)(2,4)
\psline(1,2)(1,1)(2,1)
\psline(2,0)(4,0)(4,2)
\psline(3.5,2)(3.5,3.5)(2,3.5)
\psline(0,2)(0,4)(2,4)

%c_1
\psdots[linewidth=0.02](1.1,1.1)\psdots[linewidth=0.02](1.5,1.1)\psdots[linewidth=0.02](1.9,1.1)
\psdots[linewidth=0.02](1.1,1.5)\psdots[linewidth=0.02,linecolor=green](1.5,1.5)\psdots[linewidth=0.02,linecolor=green](1.9,1.5)
\psdots[linewidth=0.02](1.1,1.9)\psdots[linewidth=0.02,linecolor=green](1.5,1.9)\psdots[linewidth=0.02,linecolor=green](1.9,1.9)

%c_2
\psdots[linewidth=0.02,linecolor=green](2.1,1.9)\psdots[linewidth=0.02,linecolor=green](3,1.9)\psdots[linewidth=0.02](3.9,1.9)
\psdots[linewidth=0.02,linecolor=green](2.1,1)\psdots[linewidth=0.02,linecolor=green](3,1)\psdots[linewidth=0.02](3.9,1)
\psdots[linewidth=0.02](2.1,0.1)\psdots[linewidth=0.02](3,0.1)\psdots[linewidth=0.02](3.9,0.1)
%c_3
\psdots[linewidth=0.02](2.1,3.4)\psdots[linewidth=0.02](2.75,3.4)\psdots[linewidth=0.02](3.4,3.4)
\psdots[linewidth=0.02,linecolor=green](2.1,2.75)\psdots[linewidth=0.02,linecolor=green](2.75,2.75)\psdots[linewidth=0.02](3.4,2.75)
\psdots[linewidth=0.02,linecolor=green](2.1,2.1)\psdots[linewidth=0.02,linecolor=green](2.75,2.1)\psdots[linewidth=0.02](3.4,2.1)
%c_4
\psdots[linewidth=0.02](0.1,2.1)\psdots[linewidth=0.02,linecolor=green](1,2.1)\psdots[linewidth=0.02,linecolor=green](1.9,2.1)
\psdots[linewidth=0.02](0.1,3)\psdots[linewidth=0.02,linecolor=green](1,3)\psdots[linewidth=0.02,linecolor=green](1.9,3)
\psdots[linewidth=0.02](0.1,3.9)\psdots[linewidth=0.02](1,3.9)\psdots[linewidth=0.02](1.9,3.9)
\rput(0.7,0.7){\scriptsize{$C_1$}}\rput(4.3,1.3){\scriptsize{$C_2$}}
\rput(4,4){\scriptsize{$C_3$}}\rput(-0.3,3.5){\scriptsize{$C_4$}}
%\rput(0,1.7){\scriptsize{$E_1$}}\rput(1.7,0){\scriptsize{$E_2$}}\rput(4,2.3){\scriptsize{$E_3$}}\rput(2.3,3.7){\scriptsize{$E_2$}}
\psdots[dotstyle=square*](2,2)
\end{pspicture} &
\begin{pspicture}(-0.5,-0.5)(4,4)
\psline(0,2)(4,2)\psline(2,0)(2,4)
\psline(1,2)(1,1)(2,1)
\psline(2,0)(4,0)(4,2)
\psline(3.5,2)(3.5,3.5)(2,3.5)
\psline(0,2)(0,4)(2,4)
\psdots[dotstyle=square*](2,2)
\psdots[linecolor=green](2.15,1.85)(3,1.85)(2.15,1)(3,1)
\psdots[linecolor=green](1,2.15)(1.85,2.15)(1,3)(1.85,3)
\psdots[linecolor=green](2.15,2.15)(2.15,2.75)(2.75,2.75)(2.75,2.15)
\rput(1.2,1.5){\scriptsize{$C_1$}}\rput(3.3,1.3){\scriptsize{$C_2$}}
\rput(3,3){\scriptsize{$C_3$}}\rput(0.5,3.5){\scriptsize{$C_4$}}
\rput(1.5,1.5){\tiny{$\bullet$}}\rput(1.5,1.85){\tiny{$\bullet$}}\rput(1.85,1.5){\tiny{$\bullet$}}\rput(1.85,1.85){\tiny{$\bullet$}}
\rput(1.5,2.5){\tiny{$\bullet$}}\rput(1.5,2.15){\tiny{$\bullet$}}\rput(1.85,2.15){\tiny{$\bullet$}}\rput(1.85,2.5){\tiny{$\bullet$}}
\rput(2.15,2.15){\tiny{$\bullet$}}\rput(2.15,2.5){\tiny{$\bullet$}}\rput(2.5,2.5){\tiny{$\bullet$}}\rput(2.5,2.15){\tiny{$\bullet$}}
\rput(2.15,1.85){\tiny{$\bullet$}}\rput(2.5,1.85){\tiny{$\bullet$}}\rput(2.5,1.5){\tiny{$\bullet$}}\rput(2.15,1.5){\tiny{$\bullet$}}
\end{pspicture} \\
\footnotesize{(a)}  & \footnotesize{(b)}
\end{tabular}
\caption{Corresponding B-ordinates around a crossing-vertex.\label{Evalbilinear}}
\end{center}
\end{figure}
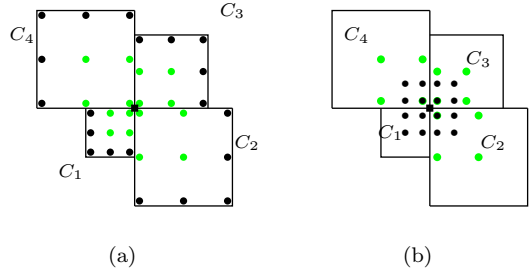
\begin{lem}
\label{evalbor}
For each crossing-vertex $v^+ \in \mathscr{T}$,
the corresponding B-ordinates of $p(x,y) \in \overline{\mathbf{S}}^2(\mathscr{T})$ on $v^+$ are on a bilinear function.
\end{lem}
\begin{pf}
We  also use Fig. \ref{Evalbilinear} to illustrate the proving process.
In Fig. \ref{Evalbilinear}(a),
the  domain points of $C_1$ around $v^+$ are
$P^1_{j,k}=(s^1_{j,k},t^1_{j,k})$, where $ j,k=1,2$,
which are denoted as ``$\green{\bullet}$" on $C_1$.
The bilinear function $f_1(s,t)=a_1st+b_1s+c_1t+d_1$, where $j,k=1,2$,
which satisfies $f_1(s^1_{j,k},t^1_{j,k})=b^1_{j,k}$, exists.
Similarly,  the domain points of $C_i$ around $v^+$
are denoted as ``{$\green\bullet$}" on $C_i$, where $i=2,3,4$, respectively,
and the corresponding bilinear functions are denoted as $f_i(s,t)= a_ist+b_is+c_it+d_i$,
where $ i=2,3,4$, respectively.

We use Fig. \ref{Evalbilinear}(b) to illustrate $f_1=f_2=f_3=f_4$.
Apply the $C^2$ continuous conditions to $C_i$, where $i=2,3,4$.
The B-ordinates on ``$\bullet$" and the B-ordinates on ``{$\green\bullet$}" belong to $C_i$
are on a common bilinear functions $f_i(s,t)$, where $i=2,3,4$, respectively.
In Fig. \ref{Evalbilinear}(b),
use the $C^1$ continuous condition of $C_1$ and $C_2$.
The B-ordinates on ``$\bullet$" of $C_1$ and $C_2$ are on a common bilinear function;
thus $f_1=f_2$.
Similarly, we obtain $f_1=f_3, f_2=f_3$ and $ f_3=f_4$;
then, $f_1=f_2=f_3=f_4$.
Thus, the lemma is proved.
\end{pf}

By Lemma \ref{evalbor}, if we want to calculate the corresponding B-ordinates for a crossing-vertex $v^+$,
we need to determine the \textbf{bilinear function} $f(s,t)= ast+bs+ct+d$ of $v^+$.
For $f(s,t)$,
to determine $a,b,c$ and $d$, we recall the notation of the \textbf{adaptive nodes} \cite{Dengbook} for a bilinear function.
\subsubsection{The adaptive nodes for a bilinear funtion}

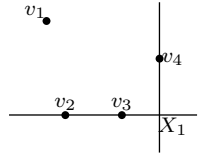
\begin{figure}[!htb]
\captionsetup{font={small}}
\begin{center}
\psset{unit=0.5cm,linewidth=0.5pt}
\begin{tabular}{c@{\hspace*{1cm}}c@{\hspace*{1cm}}c}
\begin{pspicture}(0,0.5)(5,4.5)
\psline(0,1.5)(5,1.5)\psline(4,0.5)(4,4.5)
\psdots(1,4)\rput(0.7,4.25){\footnotesize${v_1}$}
\psdots(1.5,1.5)\rput(1.5,1.8){\footnotesize${v_2}$}
\psdots(3,1.5)\rput(3,1.8){\footnotesize${v_3}$}
\psdots(4,3)\rput(4.35,3){\footnotesize${v_4}$}
\rput(4.35,1.2){\footnotesize${X_1}$}
\end{pspicture} \\
\end{tabular}
\caption{Adaptive nodes for a bilinear function.\label{adnodes}}
\end{center}
\end{figure}

We use Fig. \ref{adnodes} to illustrate how to obtain a group of adaptive nodes for a bilinear function.

\begin{enumerate}[Step 1]
\item Choose the point  $v_1(s_1,t_1) \in \mathbb{R}^2$.
\item Draw the cross  $X_1$ $\in \mathbb{R}^2$, and $v_1 \notin X_1$.
Choose two points $v_2(s_2,t_2)$ and $ v_3(s_3,t_3)$ on one edge of $X_1$,
and choose another point $v_4(s_4,t_4)$ on the other edge of $X_1$.
\end{enumerate}

$v_1, v_2, v_3$ and $ v_4$ are referred to as  adaptive nodes of the bilinear function $f(s,t)= ast+bs+ct+d$.
Given the corresponding $f(s_i,t_i),i=1,...,4$,
we can use the linear equations $f(s_i,t_i)= as_it_i+bs_i+ct_i+d, i=1,...,4$
to calculate the coefficients $a,b,c$ and $d$.

And then, we give the following lemma to illustrate the relationship between the corresponding B-ordinates and the adaptive nodes on each end-point of $\mathcal{T}$:

\begin{thm}
\label{bordinatesoftwocrossingverteices}
Assume that the two end-points of the T-structure $\mathcal{T} \in \mathscr{T}$ are crossing-vertices.
Let $p(x,y) \in \overline{\mathbf{S}}^2(\mathscr{T})$ be the polynomial defined over the cells of $\mathcal{T}$.
Given the values on a group of adaptive nodes for each end-point of $\mathcal{T}$,
the corresponding B-ordinates on $\mathcal{T}$ can be calculated by Lemma \ref{TSbordinates}.
\end{thm}
\begin{pf}
By Lemma  \ref{evalbor},
the  corresponding B-ordinates on each end-point are on a bilinear function.
For each end-point,
as the value on each adaptive node is given,
the coefficients of each bilinear function are calculated by four equations,
the corresponding B-ordinates on each end-points can be calculated by the corresponding bilinear functions,
the two rows (column) B-ordinates of the mother-cell that near the mid-edge are obtained,
and the corresponding B-ordinates on $\mathcal{T}$ can be calculated by Lemma \ref{TSbordinates},
and the corresponding B-ordinates satisfy the $C^1$ continuous conditions.
\end{pf}

\begin{figure}[!htb]
\captionsetup{font={small}}
\begin{center}
\psset{unit=0.65cm,linewidth=0.5pt}
\begin{tabular}{c@{\hspace*{1cm}}c@{\hspace*{1cm}}c}

\begin{pspicture}(0,-0.2)(6,3)
\psline(0,0)(6,0)(6,3)(0,3)(0,0)
\psline(3,0)(3,3)
\rput(1.5,1.5){${\bullet}$}\rput(2.85,1.5){${\bullet}$}\rput(4.5,1.5){${\bullet}$}\rput(3.15,1.5){${\bullet}$}
\rput(1.5,2){\scriptsize${C_1}$}\rput(4.5,2){\scriptsize${C_0}$}
\rput(1.5,1){\scriptsize${P^1_{11}}$}\rput(2.7,1){\scriptsize${P^1_{21}}$}
\rput(3.3,1){\scriptsize${P^0_{01}}$}\rput(4.5,1){\scriptsize${P^0_{11}}$}
\rput(0.1,-0.2){\scriptsize${x_0}$}\rput(3,-0.2){\scriptsize${x_1}$}\rput(6,-0.2){\scriptsize${x_2}$}
\rput(-0.3,0){\scriptsize${y_0}$}\rput(-0.2,3){\scriptsize${y_k}$}
\rput(3,3.2){\scriptsize${v_1}$}\rput(3.2,0.2){\scriptsize${v_2}$}
\rput(3,0){\tiny${\bullet}$}\rput(3,3){\tiny${\bullet}$}
\end{pspicture} &
\begin{pspicture}(0,0)(6,3)
\psline(0,0)(6,0)(6,3)(0,3)(0,0)
\psline(3,0)(3,3)
\psline(2,0.8)(3,0.8)
\psline(2,2)(3,2)
\rput(2.5,1.5){$\vdots$}
\rput(2.5,0.55){\scriptsize${C_1}$}
\rput(2.5,2.5){\scriptsize${C_k}$}
\rput(3,0){\tiny${\bullet}$}\rput(3,0.8){\tiny${\bullet}$}\rput(3,2){\tiny${\bullet}$}\rput(3,3){\tiny${\bullet}$}
\rput(3.3,0.2){\scriptsize${y_0}$}\rput(3.3,0.8){\scriptsize${y_1}$}\rput(3.5,2){\scriptsize${y_{k-1}}$}\rput(3.3,2.8){\scriptsize${y_k}$}
\rput(4.5,2){\scriptsize${C_0}$}
\rput(1.5,1.5){\scriptsize${\blacktriangle}$}
\rput(1.5,1){\scriptsize${\mathcal{TRD}}$ }
\rput(1.3,1.5){\scriptsize${P}$ }
\rput(0,-0.2){\scriptsize${x_0}$}\rput(3,-0.2){\scriptsize${x_1}$}\rput(6,-0.2){\scriptsize${x_2}$}
\rput(3,3.2){\scriptsize${v_1}$}\rput(2.75,0.2){\scriptsize${v_2}$}
\rput(3.15,1.5){${\bullet}$}\rput(4.5,1.5){${\bullet}$}
\rput(3.55,1.5){\scriptsize${P^0_{01}}$}\rput(4.9,1.5){\scriptsize${P^0_{11}}$}
\end{pspicture}  \\[2mm]
\footnotesize{(a)} \footnotesize{Two P-cells} & \footnotesize{(b)} \footnotesize{A T-connection and the corresponding one-neighbor-cell}
\end{tabular}
\caption{The nodes can be used in hierarchical T-mesh}\label{myC11111}
\end{center}
\end{figure}
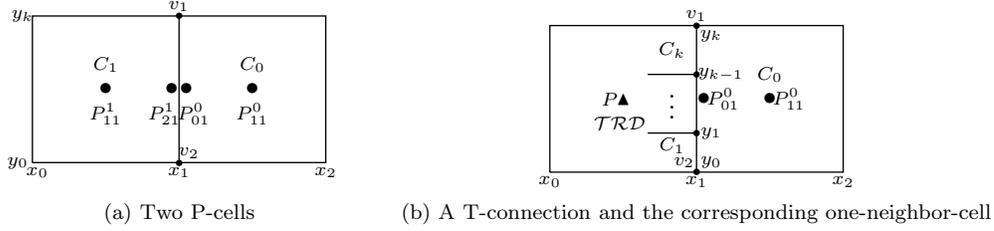

Till now, we obtain the conclusion that if we want to obtain the corresponding B-ordinates on a T-structure,
we need to obtain a group of adaptive nodes and the corresponding values on the nodes.
We use Fig. \ref{myC11111} to give the following proposition to discuss the values on the adaptive nodes for the end-points of a T-structure.

In Fig. \ref{myC11111}(a), $C_0$ and $C_1$ are two aligned P-cells in $\mathscr{T}$.
$P^0_{01}(x_1, \frac{y_0+y_k}{2})$ and $P^0_{11}(\frac{x_1+x_2}{2}, \frac{y_0+y_k}{2})$ are two domain-points of $C_0$.
$P^1_{11}(\frac{x_0+x_1}{2}, \frac{y_0+y_k}{2})$ and $P^1_{21}(x_1, \frac{y_0+y_k}{2})$  are two domain-points of $C_1$.
Let $p_i(x,y) \in \overline{\mathbf{S}}^2(\mathscr{T})$ be the polynomial
defined over $C_i, i=0,1$.
The B-ordinate of $p_0(x,y)$ on $P^0_{i1}$  is denoted as $b^0_{i1}, i=0,1$.
The B-ordinate of $p_1(x,y)$ on $P^1_{i1}$  is denoted as $b^1_{i1}, i=1,2$.

In Fig. \ref{myC11111}(b),
$C_0$ is the one-neighbor-cell of $\mathcal{TC}$  in $\mathscr{T}$,
the T-connection-domain of $\mathcal{TC}$ is denoted as $\mathcal{TCD}$,
the T-rectangle-domain of $\mathcal{TC}$ is denoted as $\mathcal{TRD}$.
Let $p_i(x,y) \in \overline{\mathbf{S}}^2(\mathscr{T})$ be the polynomial
defined over $C_i, i=0,..,k$,
$\mathbf{\varphi} (p_i(x,y):{\mathcal{TRD}})=\mathbf{\varphi} (p_0(x,y):{\mathcal{TRD}})=\omega|_{\mathcal{TCD}}$.
The center position of $\mathcal{TRD}$ is denoted as $P(\frac{x_0+x_1}{2}, \frac{y_0+y_k}{2})$.
$P^0_{01}(x_1, \frac{y_0+y_k}{2})$ and $P^0_{11}(\frac{x_1+x_2}{2}, \frac{y_0+y_k}{2})$ are denoted as two domain-points of $C_0$,
the B-ordinate on $P^0_{i1}$  is denoted as $b^0_{i1}, i=0,1$.
\begin{prop}
\label{CVRpointweight}
From the illustrations of Fig. \ref{myC11111}, we give the conclusions as follows:
\begin{enumerate}
\item In Fig. \ref{myC11111}(a), $p(x,y)$ is $C^1$ continuous on $v_1v_2$ if and only if
$(x_1, \frac{y_0+y_k}{2}, b^1_{21})$ is on the linear function that is determinated by
$(\frac{x_0+x_1}{2}, \frac{y_0+y_k}{2}, b^1_{11})$ and $(\frac{x_1+x_2}{2}, \frac{y_0+y_k}{2}, b^0_{11})$.
\item In Fig. \ref{myC11111}(b), $p(x,y)$ is $C^1$ continuous on $v_1v_2$ if and only if
$(x_1, \frac{y_0+y_1}{2}, b^0_{10})$ is on the linear function that is determinated by
$(\frac{x_0+x_1}{2}, \frac{y_0+y_k}{2},\omega)$ and $(\frac{x_1+x_2}{2}, \frac{y_0+y_k}{2}, b^0_{11})$.
\end{enumerate}
\end{prop}
\begin{pf}
\begin{enumerate}
\item Similar to Proposition \ref{univC1}, the proposition is true.
\item In  Fig. \ref{myC11111}(a), we denote $p(x,y) $ on $C_i$  as $p_i(x,y),i=0,1$.
As $p_1(x,y)=p_0(x,y)+(x-x_1)^2v(y)$, $b^1_{11}$  is the mapping result of $\varphi(p_0(x,y):{[x_0,x_1] \times [y_0,y_k]})$.
In Fig. \ref{myC11111}(b), $\omega$  is also the mapping result of $\varphi(p_0(x,y):{[x_0,x_1] \times [y_0,y_k]})$.
Thus, we have a similar conclusion as that $(x_1, \frac{y_0+y_1}{2}, b^0_{10})$ is on the linear function that is determined by
$(\frac{x_0+x_1}{2}, \frac{y_0+y_k}{2},\omega)$ and $(\frac{x_1+x_2}{2}, \frac{y_0+y_k}{2}, b^0_{11})$.
The reverse proving process can be derived naturally. The proposition is proved.
\end{enumerate}
\end{pf}

By Proposition \ref{CVRpointweight}, the \textbf{weight} on the domain-centre of $\mathcal{TC}$
and the B-ordinate on the center domain-point of the one-neighbor-cell $C \in \mathcal{TC}$ can be used as values in Theorem \ref{bordinatesoftwocrossingverteices}
to calculate  the coefficients of the bilinear functions for $v_1$ and $v_2$ in Fig. \ref{myC11111}(b).
We obtain the conclusion that we can use the weights on the  domain-centres and the B-ordinates on the domain-points
to calculate the corresponding B-ordinates of the polynomial $p(x,y) \in \overline{\mathbf{S}}^2(\mathscr{T})$.
We introduce the method for using T-structures to calculate the B-ordinates
for the basis functions of $\overline{\mathbf{S}}^2(\mathscr{T})$ as follows:

\subsection{Evaluate the B-ordinates of the basis functions of $\overline{\mathbf{S}}^2(\mathscr{T})$}
%First, we use the values of the basis functions of $\overline{\mathbf{S}}^0(\mathscr{G})$ to assign the domains of $\mathscr{T}$.
%In other words, we can use each  basis function of $\overline{\mathbf{S}}^0(\mathscr{G})$ to initialize the weights on each domain-center of $\mathscr{T}$, and then
In this subsection, we will evaluate the B-ordinates for the basis functions of $\overline{\mathbf{S}}^2(\mathscr{T})$.
%and then use T-structures to calculate the B-ordinates of a biquadratic polynomial function of $\overline{\mathbf{S}}^2(\mathscr{T})$.
First, we use each basis function of $\overline{\mathbf{S}}^0(\mathscr{G})$ to initialize the {weights} on each domain-center of $\mathscr{T}$,
and we obtain a domain $\mathbb{E}$ that covers $sup(p(x,y))$, $p(x,y)$ is denoted as the basis function in $\overline{\mathbf{S}}^2(\mathscr{T})$.
Second, we give an order for the T-structure-branches that corresponds to the T-connections in $\mathbb{E}$.
Third, we use the lowest level T-structure-branch to calculate the B-ordinates on each T-cell that belongs to the lowest level T-connection.
Finally, in a similar way to the lowest level T-connection, we calculate the B-ordinates on the T-cells of the rest T-connections.
\subsubsection{Initialize the weights on the domain-centres}\label{initializeweights11}
In this subsection, to make use of the CVR graph $\mathscr{G}$, we initialize the weights for the basis function $p(x,y) \in \overline{\mathbf{S}}^2(\mathscr{T})$ by a basis function $q(x,y) \in \overline{\mathbf{S}}^0(\mathscr{G})$.
%and obtain a

\begin{figure}[htbp!]
\captionsetup{font={small}}
\centering
\includegraphics[width=3.5cm]{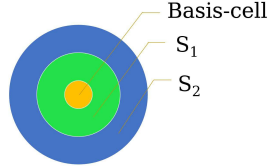}
\caption{The  sketch of $\mathbb{E}$}
\label{support}
\end{figure}

Given the basis function $q(x,y) \in \overline{\mathbf{S}}^0(\mathscr{G})$ as:
\begin{equation}
\label{cvrqxy}
q(x,y)=
\begin{cases}
 1, &   \mathcal{GC}\\
 0, &   \text{other g-cells}
\end{cases}.
\end{equation}
With Equation (\ref{cvrqxy}), we give the {weight} on each domain-centre as:
\begin{equation}
\label{CVRbase}
\omega =
\begin{cases}
 1, &   (\frac{x_0+x_1}{2}, \frac{y_0+y_1}{2})\\
 0, &   \text{other domain-centres}
\end{cases}.
\end{equation}
We use the weights in Equation (\ref{CVRbase}) to construct the basis function $p(x,y) \in \overline{\mathbf{S}}^2(\mathscr{T})$
and $p(x,y)$ is a piecewise quadratic polynomial on some cells of $\mathscr{T}$.
The weight on the domain-centre of $\mathcal{D}$ is $1$,
while the weights on the other domain-centres are 0,
we give the domain $\mathbb{E} \in \mathscr{T}$ that covers $Sup(p(x,y))$.
The sketch of $\mathbb{E}$ is shown in Fig. \ref{support}:
the \textbf{basis-cell} is the P-cell or the T-connection that corresponds to $\mathcal{D}$.
$S_1$ consists of the center-cell, the P-cells and T-connections whose domains that cover, are covered, or adjacent to $\mathcal{D}$.
$S_2$ consists of the P-cells and T-connections whose domains that  cover, are covered, or adjacent to $S_1$.
Delete each repeat element in $S_1 \cup S_2 $ , we obtain the domain $\mathbb{E}$
that covered by a series of T-connections and  P-cells.

\subsubsection{The order of T-structure-branches in $\mathbb{E}$}\label{OrderTsOrderTs}
By Lemma \ref{coverallcells}, each T-connection is covered by a T-structure-branch.
Assume that the T-connections  in $\mathbb{E}$ are $\mathcal{TC}_0, \mathcal{TC}_1,...,\mathcal{TC}_n$,
we denote $\mathcal{TSB}_i$ as the corresponding T-structure-branch of $\mathcal{TC}_i,i=0,...,n$.
For each $\mathcal{TSB}_i$, we give the Algorithm \ref{orderTS} to order the T-structures in $\mathcal{TSB}_i,i=0,...,n$.
And then, sort $\mathcal{TSB}_i, i=0,...,n$ in descending level order.

\begin{algorithm}[H]
\caption{The algorithm to sort the T-structures for each T-structure-Branch \label{orderTS}}% 算法名字
\LinesNumbered %要求显示行号
\KwIn{$\mathcal{TSB}: \{\mathcal{T}_0,...,\mathcal{T}_m\}$}%输入参数
\KwOut{The order sorted T-structure-branch}%输出
Give an empty T-structure-branch $\mathcal{TSB}_0$\;
$\mathcal{TSB}_0 \gets $ the T-structure (T-structures) with the lowest level in $\mathcal{TSB}$\;
Remove each T-structure of $\mathcal{TSB}_0$ from $\mathcal{TSB}$\;
\While{$|\mathcal{TSB}| > 0$}{
\For{$i=0,i < |\mathcal{TSB}_0|, i++$}
{
%Denote the end-point of $\mathcal{TSB}_0[i]$ as $v_1$ and $v_2$ respectively\;
Give an empty T-structure vector $V$\;
\For{$j=0;j<|\mathcal{TSB}|;j++$}
{
\If{$\mathcal{TSB}_0[i]$ is connected to $\mathcal{TSB}[j]$ }{
$V \gets \mathcal{TSB}[j]$\;
 }
}
 Remove each T-structure of $V$ from $\mathcal{TSB}$\;
 Sort T-structures of  $V$ in descending level order\;
 $\mathcal{TSB}_0 \gets $ each T-structure of $V$\;
}
}
Output $\mathcal{TSB}_0$\;
\end{algorithm}
\subsubsection{The B-ordinates of $p(x,y)$ on $\mathcal{TC}_0$}\label{T0T0T0T0T0}

%We will use the weights assigned by equation (\ref{CVRbase}) to calculate the B-ordinates for T-cells in  $\mathcal{TC}_0$,
%which is the lowest level T-structure-branch.
%Our calculating tools are the T-structures.
%If we can calculate the corresponding B-ordinates for each crossing end-points of the T-structure $\mathcal{T}$,
%the corresponding B-ordinates on $\mathcal{T}$ can be calculated.

In this subsection, we use T-structures to calculate the B-ordinates of $p(x,y)$ on $\mathcal{TC}_0$.
We denote the T-rectangle-domain of $\mathcal{TC}_0$ as $\mathcal{TRD}_0$.
By Lemma \ref{coverallcells}, $\mathcal{TC}_0$ is covered by $\mathcal{TSB}_0$, $\mathcal{TSB}_0$ is the lowest level T-structure-branch in $\mathbb{E}$.
$\mathcal{TC}_0$ is also the lowest level T-connection in $\mathbb{E}$.
The B-ordinates on the T-cells that belong to  $\mathcal{TC}_0$ can be calculated by
T-structures in $\mathcal{TSB}_0:{\mathcal{T}_0,...,\mathcal{T}_m}$.
$\mathcal{T}_0,...,\mathcal{T}_m$ are already sorted by Algorithm \ref{orderTS},
and $\mathcal{T}_0$ is the T-structure with the lowest level.

As $\mathcal{TC}_0$ is also the lowest level T-connection in $\mathbb{E}$,
we give the initialization as follows:
\begin{ini}
\label{initialization1}
As $\mathcal{TC}_0$ is also the lowest level T-connection in $\mathbb{E}$,
we give the initialization as follows:
\end{ini}

As $\mathcal{T}_0$ is the T-structure with the lowest level in $\mathcal{TSB}_0$,
we give the lemma as follows:
\begin{lem}
\label{twoendpointsarecrossing}
The two end-points of $\mathcal{T}_0$ are crossing-vertices.
\end{lem}
\begin{pf}
We prove the lemma by reduction to absurdity.
If one of the end-points is a T-junction,
then a lower level T-structure $\mathcal{T}'_0$, which
connects to $\mathcal{T}_0$ at a T-junction, exists.
Then, the level of $\mathcal{T}'_0$ is  lower than the level of $\mathcal{T}_0$ and $\mathcal{T}'_0$ belongs to $\mathcal{TSB}_0$.
It contradicts the order of the T-structures in $\mathcal{TSB}_0$.
Thus, the two end-points of $\mathcal{T}_0$ are crossing-vertices.
The lemma is proved.
\end{pf}

%
%Form Proposition \ref{CVRpointweight},
%%For a T-connection $\mathcal{TC}$,
%%the T-rectangle-domain of  $\mathcal{TC}$ is denoted as $\mathcal{TRD}$.
%%The weight on the center position of $\mathcal{TRD}$ is given.
%if a crossing-vertex $v^+$ occupies on the common c-edge that belongs to $\mathcal{TC}$
%and the one-neighbor-cell of $\mathcal{TC}$,
%given the {weight} on  each domain-center of $\mathscr{T}$, we can choose center position of $\mathcal{TRD}$
%as one of the adaptive nodes of the bilinear function on $v^+$.

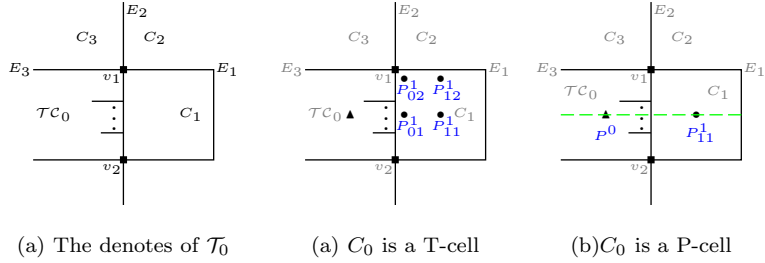
\begin{figure}[!ht]
\captionsetup{font={small}}
\begin{center}
\psset{unit=0.6cm,linewidth=0.5pt}
\begin{tabular}{c@{\hspace*{1cm}}c@{\hspace*{1cm}}c@{\hspace*{1cm}}c}
\begin{pspicture}(0,1)(4,5.5)
\psline(0,2)(4,2)(4,4)(0,4)
\psline(2,1)(2,5.5)\psline(1.5,2.6)(2,2.6)\psline(1.3,3.3)(2,3.3)
\rput(1.8,3.1){$\vdots$}
\psdots[dotstyle=square*](2,4)\rput(1.75,3.8){{\tiny${v_1}$}}
\psdots[dotstyle=square*](2,2)\rput(1.75,1.8){{\tiny${v_2}$}}
%\rput(3,3){\footnotesize${\bullet}$}
%\rput(3,3.8){\footnotesize${\bullet}$}\rput(2.2,3){\footnotesize${\bullet}$}\rput(2.2,3.8){\footnotesize${\bullet}$}
%\rput(1,3){\scriptsize${\blacktriangle}$}
\rput(0.45,3){{\tiny${\mathcal{TC}_0}$}}
\rput(3.5,3){{\tiny${C_1}$}}\rput(2.7,4.7){{\tiny${C_2}$}}\rput(1.2,4.7){{\tiny${C_3}$}}
%\rput(2.3,3.55){\blue{\tiny${P^1_{02}}$}}\rput(3.1,3.55){\blue{\tiny${P^1_{12}}$}}
%\rput(2.3,2.75){\blue{\tiny${P^1_{01}}$}}\rput(3.1,2.75){\blue{\tiny${P^1_{11}}$}}
\rput(-0.3,4){{\tiny${E_3}$}}\rput(4.3,4){{\tiny${E_1}$}}\rput(2.3,5.3){{\tiny${E_2}$}}
\end{pspicture}&
\begin{pspicture}(0,1)(4,5.5)
\psline(0,2)(4,2)(4,4)(0,4)
\psline(2,1)(2,5.5)\psline(1.5,2.6)(2,2.6)\psline(1.3,3.3)(2,3.3)
\rput(1.8,3.1){$\vdots$}
\psdots[dotstyle=square*](2,4)\rput(1.75,3.8){\gray{\tiny${v_1}$}}
\psdots[dotstyle=square*](2,2)\rput(1.75,1.8){\gray{\tiny${v_2}$}}
\rput(3,3){\tiny${\bullet}$}
\rput(3,3.8){\tiny${\bullet}$}\rput(2.2,3){\tiny${\bullet}$}\rput(2.2,3.8){\tiny${\bullet}$}
\rput(1,3){\tiny${\blacktriangle}$}
\rput(0.45,3){\gray{\tiny${\mathcal{TC}_0}$}}
\rput(3.55,3){\gray{\tiny${C_1}$}}\rput(2.7,4.7){\gray{\tiny${C_2}$}}\rput(1.2,4.7){\gray{\tiny${C_3}$}}
\rput(2.35,3.55){\blue{\tiny${P^1_{02}}$}}\rput(3.15,3.55){\blue{\tiny${P^1_{12}}$}}
\rput(2.35,2.75){\blue{\tiny${P^1_{01}}$}}\rput(3.15,2.75){\blue{\tiny${P^1_{11}}$}}
\rput(-0.3,4){\gray{\tiny${E_3}$}}\rput(4.3,4){\gray{\tiny${E_1}$}}\rput(2.3,5.3){\gray{\tiny${E_2}$}}
\end{pspicture}&
\begin{pspicture}(0,1)(4,5.5)
\psline(0,2)(4,2)(4,4)(0,4)
\psline(2,1)(2,5.5)\psline(1.5,2.6)(2,2.6)\psline(1.3,3.3)(2,3.3)
\rput(1.8,3.1){$\vdots$}
\psdots[dotstyle=square*](2,4)\rput(1.75,3.8){\gray{\tiny${v_1}$}}
\psdots[dotstyle=square*](2,2)\rput(1.75,1.8){\gray{\tiny${v_2}$}}
\rput(3,3){\tiny${\bullet}$}
%\rput(3,3.8){\footnotesize${\bullet}$}\rput(2.2,3){\footnotesize${\bullet}$}\rput(2.2,3.8){\footnotesize${\bullet}$}
\rput(1,3){\tiny${\blacktriangle}$}\rput(1,2.6){\blue{\tiny${P^0}$}}
\rput(0.45,3.5){\gray{\tiny${\mathcal{TC}_0}$}}
\rput(3.5,3.5){\gray{\tiny${C_1}$}}\rput(2.7,4.7){\gray{\tiny${C_2}$}}\rput(1.2,4.7){\gray{\tiny${C_3}$}}
\rput(3.1,2.6){\blue{\tiny${P^1_{11}}$}}
\rput(-0.3,4){\gray{\tiny${E_3}$}}\rput(4.3,4){\gray{\tiny${E_1}$}}\rput(2.3,5.3){\gray{\tiny${E_2}$}}
\psline[linestyle=dashed,dash=5pt 2pt, linecolor=green](0,3)(4,3)
\end{pspicture}\\[2mm]
\footnotesize{(a) The denotes of $\mathcal{T}_0$}  & \footnotesize{(a) $C_0$ is a T-cell}  & \footnotesize{(b)$C_0$ is a P-cell}
\end{tabular}
\caption{$\mathcal{T}_0$. \label{adapnodesofT}}
\end{center}
\end{figure}

By Lemma \ref{twoendpointsarecrossing}, the two end-points of $\mathcal{T}_0$ are crossing-vertices.
Without loss of generality, we use the vertical T-structure
in Fig. \ref{adapnodesofT}(a) to illustrate $\mathcal{T}_0$.
We denote the two end-points of $\mathcal{T}_0$ as $v_1$ and $v_2$ respectively,
denote the mother-cell of $\mathcal{T}_0$ as $C_1$.
$C_1$ is also the one-neighbor-cell of $\mathcal{TC}_0$.
We can calculate the corresponding B-ordinates of $\mathcal{T}_0$ via the bilinear function associated with each end-point of $\mathcal{T}_0$.
From Theorem \ref{bordinatesoftwocrossingverteices}, if the weights or B-ordinates
on the adaptive nodes of the bilinear function on each end-point of $\mathcal{T}_0$ are given,
we can use the bilinear function to calculate the corresponding B-ordinates for each end-point.
We give the theorem to obtain the corresponding B-ordinates on $\mathcal{T}_0$ as follows:

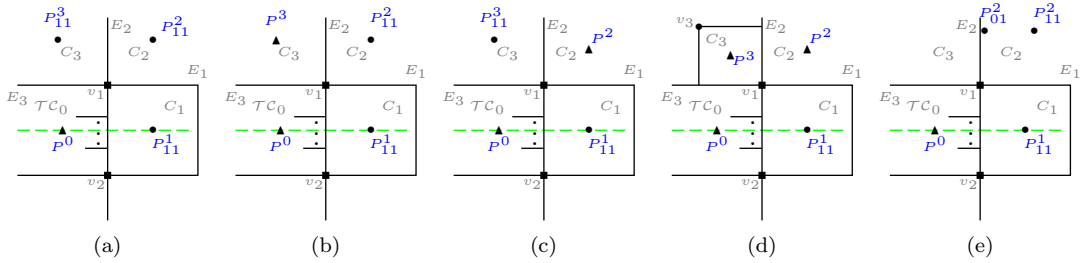
\begin{figure}[!ht]
\captionsetup{font={small}}
\begin{center}
\psset{unit=0.6cm,linewidth=0.5pt}
\begin{tabular}{c@{\hspace*{0.5cm}}c@{\hspace*{0.5cm}}c@{\hspace*{0.5cm}}c@{\hspace*{0.5cm}}c@{\hspace*{0.5cm}}c}
\begin{pspicture}(0,1)(4,5.5)
\psline[linestyle=dashed,dash=5pt 2pt, linecolor=green](0,3)(3.8,3)
%\psline[linestyle=dashed,dash=5pt 2pt, linecolor=green](3,1.2)(3,5.5)
\psline(0,2)(4,2)(4,4)(0,4)
\psline(2,1)(2,5.5)\psline(1.5,2.6)(2,2.6)\psline(1.3,3.3)(2,3.3)
\rput(1.8,3.1){$\vdots$}
\psdots[dotstyle=square*](2,4)\rput(1.75,3.8){\gray{\tiny${v_1}$}}
\psdots[dotstyle=square*](2,2)\rput(1.75,1.8){\gray{\tiny${v_2}$}}
\rput(3,3){\tiny${\bullet}$}
\rput(1,3){\tiny${\blacktriangle}$}
\rput(0.75,3.5){\gray{\tiny${\mathcal{TC}_0}$}}
\rput(3.5,3.5){\gray{\tiny${C_1}$}}
\rput(2.7,4.7){\gray{\tiny${C_2}$}}\rput(1.2,4.7){\gray{\tiny${C_3}$}}
\rput(0,3.7){\gray{\tiny${E_3}$}}\rput(4,4.3){\gray{\tiny${E_1}$}}\rput(2.3,5.3){\gray{\tiny${E_2}$}}
\rput(3.3,2.7){\blue{\tiny${P^1_{11}}$}}
\rput(3,5){\tiny${\bullet}$}\rput(3.5,5.3){\blue{\tiny${P^2_{11}}$}}
\rput(0.9,5){\tiny${\bullet}$}\rput(0.9,5.5){\blue{\tiny${P^3_{11}}$}}
\rput(1,2.7){\blue{\tiny${P^0}$}}
\end{pspicture}&

%\\
%\small{(a)}  & \small{(b)}
%\\[2mm]
\begin{pspicture}(0,1)(4,5.5)
\psline[linestyle=dashed,dash=5pt 2pt, linecolor=green](0,3)(3.8,3)
%\psline[linestyle=dashed,dash=5pt 2pt, linecolor=green](3,1.2)(3,5.5)
\psline(0,2)(4,2)(4,4)(0,4)
\psline(2,1)(2,5.5)\psline(1.5,2.6)(2,2.6)\psline(1.3,3.3)(2,3.3)
\rput(1.8,3.1){$\vdots$}
\psdots[dotstyle=square*](2,4)\rput(1.75,3.8){\gray{\tiny${v_1}$}}
\psdots[dotstyle=square*](2,2)\rput(1.75,1.8){\gray{\tiny${v_2}$}}
\rput(3,3){\tiny${\bullet}$}
\rput(1,3){\tiny${\blacktriangle}$}
\rput(0.75,3.5){\gray{\tiny${\mathcal{TC}_0}$}}
\rput(3.5,3.5){\gray{\tiny${C_1}$}}
\rput(2.7,4.7){\gray{\tiny${C_2}$}}\rput(1.2,4.7){\gray{\tiny${C_3}$}}
\rput(0,3.7){\gray{\tiny${E_3}$}}\rput(4,4.3){\gray{\tiny${E_1}$}}\rput(2.3,5.3){\gray{\tiny${E_2}$}}
\rput(1,2.7){\blue{\tiny${P^0}$}}
\rput(0.9,5){\tiny${\blacktriangle}$}\rput(0.9,5.5){\blue{\tiny${P^3}$}}
\rput(3.3,2.7){\blue{\tiny${P^1_{11}}$}}
\rput(3,5){\tiny${\bullet}$}\rput(3.3,5.5){\blue{\tiny${P^2_{11}}$}}
\end{pspicture}&
\begin{pspicture}(0,1)(4,5.5)

\psline(0,2)(4,2)(4,4)(0,4)
\psline[linestyle=dashed,dash=5pt 2pt, linecolor=green](0,3)(3.8,3)
\psline(2,1)(2,5.5)\psline(1.5,2.6)(2,2.6)\psline(1.3,3.3)(2,3.3)
\rput(1.8,3.1){$\vdots$}
\psdots[dotstyle=square*](2,4)\rput(1.75,3.8){\gray{\tiny${v_1}$}}
\psdots[dotstyle=square*](2,2)\rput(1.75,1.8){\gray{\tiny${v_2}$}}
\rput(3,3){\tiny${\bullet}$}
\rput(1,3){\tiny${\blacktriangle}$}
\rput(0.75,3.5){\gray{\tiny${\mathcal{TC}_0}$}}
\rput(3.5,3.5){\gray{\tiny${C_1}$}}
\rput(2.5,4.7){\gray{\tiny${C_2}$}}\rput(1.2,4.7){\gray{\tiny${C_3}$}}
\rput(0,3.7){\gray{\tiny${E_3}$}}\rput(4,4.3){\gray{\tiny${E_1}$}}\rput(2.3,5.3){\gray{\tiny${E_2}$}}
\rput(0.9,5){\tiny${\bullet}$}\rput(0.9,5.5){\blue{\tiny${P^3_{11}}$}}
\rput(1,2.7){\blue{\tiny${P^0}$}}
\rput(3.3,2.7){\blue{\tiny${P^1_{11}}$}}
\rput(3,4.8){\tiny${\blacktriangle}$}\rput(3.3,5.1){\blue{\tiny${P^2}$}}
\end{pspicture}&
\begin{pspicture}(0,1)(4,5.5)
\psline[linestyle=dashed,dash=5pt 2pt, linecolor=green](0,3)(3.8,3)
\psline(0,2)(4,2)(4,4)(0,4)
\psline(2,1)(2,5.5)\psline(1.5,2.6)(2,2.6)\psline(1.3,3.3)(2,3.3)
\rput(1.8,3.1){$\vdots$}
\psdots[dotstyle=square*](2,4)\rput(1.75,3.8){\gray{\tiny${v_1}$}}
\psdots[dotstyle=square*](2,2)\rput(1.75,1.8){\gray{\tiny${v_2}$}}
\rput(3,3){\tiny${\bullet}$}
\rput(1,3){\tiny${\blacktriangle}$}
\rput(0.75,3.5){\gray{\tiny${\mathcal{TC}_0}$}}
\rput(3.5,3.5){\gray{\tiny${C_1}$}}
\rput(2.5,4.7){\gray{\tiny${C_2}$}}\rput(1,5){\gray{\tiny${C_3}$}}
\rput(0,3.7){\gray{\tiny${E_3}$}}\rput(4,4.3){\gray{\tiny${E_1}$}}\rput(2.3,5.3){\gray{\tiny${E_2}$}}
\psline(0.6,4)(0.6,5.3)(2,5.3)
\rput(1.3,4.65){\tiny${\blacktriangle}$}\rput(1.6,4.6){\blue{\tiny${P^3}$}}
\rput(3,4.8){\tiny${\blacktriangle}$}\rput(3.3,5.1){\blue{\tiny${P^2}$}}
\rput(1,2.7){\blue{\tiny${P^0}$}}
\rput(3.3,2.7){\blue{\tiny${P^1_{11}}$}}
\rput(0.6,5.3){\tiny${\bullet}$}\rput(0.3,5.4){\gray{\tiny${v_3}$}}
\end{pspicture}&
\begin{pspicture}(0,1)(4,5.5)
\psline[linestyle=dashed,dash=5pt 2pt, linecolor=green](0,3)(3.8,3)
\psline(0,2)(4,2)(4,4)(0,4)
\psline(2,1)(2,5.5)\psline(1.5,2.6)(2,2.6)\psline(1.3,3.3)(2,3.3)
\rput(1.8,3.1){$\vdots$}
\psdots[dotstyle=square*](2,4)\rput(1.75,3.8){\gray{\tiny${v_1}$}}
\psdots[dotstyle=square*](2,2)\rput(1.75,1.8){\gray{\tiny${v_2}$}}
\rput(3,3){\tiny${\bullet}$}
\rput(1,3){\tiny${\blacktriangle}$}
\rput(0.75,3.5){\gray{\tiny${\mathcal{TC}_0}$}}
\rput(3.5,3.5){\gray{\tiny${C_1}$}}
\rput(2.7,4.7){\gray{\tiny${C_2}$}}\rput(1.2,4.7){\gray{\tiny${C_3}$}}
\rput(0,3.7){\gray{\tiny${E_3}$}}\rput(4,4.3){\gray{\tiny${E_1}$}}\rput(1.7,5.3){\gray{\tiny${E_2}$}}
\rput(3.3,2.7){\blue{\tiny${P^1_{11}}$}}
\rput(3.2,5.2){\tiny${\bullet}$}\rput(3.5,5.6){\blue{\tiny${P^2_{11}}$}}
\rput(2.1,5.2){\tiny${\bullet}$}\rput(2.3,5.6){\blue{\tiny${P^2_{01}}$}}
\rput(1,2.7){\blue{\tiny${P^0}$}}
\end{pspicture}\\
\footnotesize{(a)}  & \footnotesize{(b)}& \footnotesize{(c)} & \footnotesize{(d)} & \footnotesize{(e)}
\end{tabular}
\caption{Four adaptive nodes for a crossing-vertex of $\mathcal{T}_0$. \label{adapnodesEx}}
\end{center}
\end{figure}

\begin{thm}
\label{keyT}
For each end-point of $\mathcal{T}_0$, there exists a group of adaptive nodes that the weight or B-ordinate on each node is given,
the corresponding B-ordinates of $p(x,y)$ on $\mathcal{T}_0$ can be calculated by Theorem \ref{bordinatesoftwocrossingverteices}.
\end{thm}
\begin{pf}
Without loss of generality, we first discuss the adaptive nodes
for $v_1$,
%and we use the bilinear function to calculate the corresponding B-ordinates of $v_1$.
the adaptive nodes for $v_2$ can be obtained similar to $v_1$.
We denote
the level of $C_i$  as $l(C_i), i=1,2,3$ for
convenience.
If $C_i$ is a T-cell, without loss of generality, we assume the $C_i$ belongs to $\mathcal{TC}_i$, the level of $\mathcal{TC}_i$ is denoted as $l(\mathcal{TC}_i), i=1,2,3$ for
convenience.
By Definition \ref{tstructurebranch11111}, the level of $\mathcal{TC}_0$ is $l(C_1)$.

If $C_1$ is a T-cell, we use Fig \ref{adapnodesofT}(b) to illustrate the adaptive nodes.
Assume that $C_1$ is a sub-cell of the T-structure $\mathcal{T}'$.
As $C_1$ is the mother-cell of $\mathcal{T}_0$,
the level of $\mathcal{T}'$ is lower than $l(C_1)$.
If $\mathcal{T}'$ belongs to the T-structure-branch $\mathcal{TSB}'$,
the level of $\mathcal{TSB}'$ is lower than $l(C_1)$.
We denote  the T-connection corresponding to $\mathcal{TSB}'$ as $\mathcal{TC}'$,
the level of $\mathcal{TC}'$ is lower than $l(C_1)$.
By Initialization \ref{initialization1}, the nine B-ordinates of $p(x,y)$ on each T-cell belongs to $\mathcal{TC}'$ are $0$.
The four B-ordinates on ``$\bullet$" of $C_0$ in Fig \ref{adapnodesofT}(b) are given by the initialization,
we can choose $P^1_{02}(s^1_{02},t^1_{02}),P^1_{12}(s^1_{12},t^1_{12}),P^1_{01}(s^1_{01},t^1_{01})$ and $P^1_{11}(s^1_{11},t^1_{11})$
as the adaptive nodes of $v_1$.

If $C_1$ is a P-cell,
we use Fig. \ref{adapnodesofT}(c) to illustrate the two given adaptive nodes:
the weights on ``$\blacktriangle$" and ``$\bullet$" are given by Equation (\ref{CVRbase}).
As $C_1$ is a P-cell, the weight on ``$\bullet$" is also the B-ordinate on the centre domain point of $C_0$.
By Proposition \ref{CVRpointweight}, the points on ``$\blacktriangle$" and ``$\bullet$" can be used as two of the four adaptive nodes.
As $C_1$ is the one-neighbor-cell of $\mathcal{TC}_0$,
``$\blacktriangle$" and ``$\bullet$" are on the line that is parallel to $E_1$, which is shown as the green dashed line in Fig. \ref{adapnodesofT}(c).

In the next discussion, we need to obtain the whole group of adaptive nodes for $v_1$.
We use Fig. \ref{adapnodesEx} to illustrate the adaptive nodes,
the notations are the same as that in Fig . \ref{adapnodesofT}(c).
A careful analysis of the type of $C_i$ and $l(C_i), i=1,2,3$ will help us to obtain the four adaptive nodes.
We have the following two situations by discuss the type of $C_i,i=2,3$:
\begin{enumerate}
\item $C_2$ is a P-cell.
If $C_2$ is a P-cell, the vertices of $C_1$ and $C_2$ on $E$ are the same, and it is a crossing-vertex, then $l(C_2) = l(C_1)$.
As the B-ordinate of $p(x,y)$ on the centre domain-point of $C_2$ is given by Equation \ref{CVRbase},
the centre domain-point can be used as one node of the four adaptive nodes.
And then, we discuss the type of $C_3$  to obtain the adaptive nodes.
\item $C_2$ is T-cell. If $C_2$ is a T-cell, we assume that $C_2$ belongs to the T-connection $\mathcal{TC}_2$.
As $C_1$ is a P-cell that is adjacent to $\mathcal{TC}_2$,
the level of the one-neighbor-cell of $\mathcal{TC}_2$ is less than or equal to $l(C_1)$,
we obtain that $l(\mathcal{TC}_2)\leqslant l(C_1)$.
If $l(\mathcal{TC}_2)= l(C_1)$, $C_1$ is the one-neighbor-cell of $\mathcal{TC}_2$; otherwise, $C_1$ is not the one-neighbor-cell of $\mathcal{TC}_2$.
And then, we discuss the type of $C_3$ to obtain the adaptive nodes.
\end{enumerate}

Consider case 1.  If $C_2$ is a P-cell, the B-ordinate of $p(x,y)$ on the centre domain-point of $C_2$ is given by Equation \ref{CVRbase}.
We use Fig. \ref{adapnodesEx}(a) and (b) to discuss the adaptive nodes.
We denote the domain-point of $C_2$ as $P^2_{11}(s^2_{11},t^2_{11})$ in Fig. \ref{adapnodesEx}(a) and (b).
By Proposition \ref{CVRpointweight}, $P^2_{11}$  can be used as one of the four adaptive nodes.
$P^2_{11}$ is on a line that is perpendicular to the green dashed line, the two lines construct a cross, which can be denoted as $X_0$.
We discuss the whole group of adaptive nodes as follows:
\begin{enumerate}
\item [(1)]  $C_3$ is a p-cell.
If $C_3$ is a P-cell, the B-ordinate of $p(x,y)$ on the centre domain-point of $C_3$ is given by Equation \ref{CVRbase}.
We denote the domain-point of $C_3$ as $P^3_{11}(s^3_{11},t^3_{11})$ in Fig. \ref{adapnodesEx}(a).
By Proposition \ref{CVRpointweight}, $P^3_{11}$  can be used as one of the four adaptive nodes.
As $P^3_{11}$ is not on the cross $X_0$,
we can choose the four nodes as
$P^1_{11}(s^1_{11},t^1_{11})$, $P^2_{11}(s^2_{11},t^2_{11})$, $P^3_{11}(s^3_{11},t^3_{11})$
and $P^0(s^0,t^0)$ in Fig. \ref{adapnodesEx}(a).
\item [(2)] $C_3$ is a T-cell.
If $C_3$ is a T-cell,
we assume $C_3$ as a T-cell belongs to the T-connection $\mathcal{TC}_3$,
as $C_2$ is a P-cell adjacent to $\mathcal{TC}_3$, the level of the one-neighbor-cell of $\mathcal{TC}_3$ is not higher than $l(C_2)$,
we obtain that $l(\mathcal{TC}_3) \leqslant l(C_2)$.
\begin{enumerate}
\item  $l(\mathcal{TC}_3) < l(C_2)$.
If $l(\mathcal{TC}_3) < l(C_2)$, as $l(C_2)=l(C_1)$,  we obtain $l(\mathcal{TC}_3) < l(C_1)$.
By Initialization \ref{initialization1}, the nine B-ordinates of $p(x,y)$ on each T-cell of $\mathcal{TC}_3$ are given,
the B-ordinate of $p(x,y)$ on the centre domain-point of $C_3$ is given.
%We denote $P^3_{11}(s^3_{11},t^3_{11})$ in Fig. \ref{adapnodesEx}(a) as the  centre domain-point of $C_3$.
%By Proposition \ref{CVRpointweight}, $P^3_{11}$  can be used as one of the four adaptive nodes.
%As $P^3_{11}$ is not on the cross $X_0$,
Similar to (1) in case 1,
we can choose the four nodes as
$P^1_{11}(s^1_{11},t^1_{11})$, $P^2_{11}(s^2_{11},t^2_{11})$, $P^3_{11}(s^3_{11},t^3_{11})$
and $P^0(s^0,t^0)$ in Fig. \ref{adapnodesEx}(a).
\item $l(\mathcal{TC}_3) = l(C_2)$.
If $l(\mathcal{TC}_3) = l(C_2)$, $C_2$ is the one-neighbor-cell of $\mathcal{TC}_3$.
We denote the  domain-centre of  $\mathcal{TC}_3$ as $P^3(s^3,t^3)$ in Fig.\ref{adapnodesEx}(b).
As  the weight on $P^3$ is given by Equation \ref{CVRbase}. By Proposition \ref{CVRpointweight},
$P^3$ can be used as one of the four adaptive nodes.
As $P^3_{11}$ is not on the cross $X_0$,
we can choose the four nodes as
$P^3(s^3,t^3)$, $P^1_{11}(s^1_{11},t^1_{11})$, $P^2_{11}(s^2_{11},t^2_{11})$
and $P^0(s^0,t^0)$ in Fig. \ref{adapnodesEx}(b).
\end{enumerate}
\end{enumerate}
Consider case 2. If $C_2$ is a T-cell, $C_2$ is assumed as a cell belongs to $\mathcal{TC}_2$. As $l(\mathcal{TC}_2)\leqslant l(C_1)$, we discuss the adaptive nodes as follows:
\begin{enumerate}
\item [(1)]  $l(\mathcal{TC}_2) = l(C_1)$.
If $l(\mathcal{TC}_2) = l(C_1)$,
$C_1$ is the one-neighbor-cell of $\mathcal{TC}_2$.
We use Fig. \ref{adapnodesEx}(c)  and Fig. \ref{adapnodesEx}(d) to discuss the adaptive nodes.
We denote the domain-centre of $\mathcal{TC}_2$ as $P^2(s^2,t^2)$ in Fig. \ref{adapnodesEx}(c) Fig. \ref{adapnodesEx}(d).
The weight on the domain-centre  is given by Equation \ref{CVRbase}.
By Proposition \ref{CVRpointweight},
$P^2$ can be used as one of the four adaptive node,
and $P^2$ is on a line that is perpendicular to the green dashed line, the two lines construct a cross, which can be denoted as $X_1$.
\begin{enumerate}
\item $C_3$ is a P-cell.
If $C_3$ is a P-cell,  the B-ordinate on the centre domain-point  of $C_3$ is given by Equation \ref{CVRbase},
we denote the centre domain-point  of $C_3$ as $P^3_{11}(s^3_{11},t^3_{11})$ in Fig. \ref{adapnodesEx}(c).
By Proposition \ref{CVRpointweight}, $P^3_{11}$ can be used as one of the four nodes.
As $P^3_{11}$ is not on the cross $X_1$,
we can choose the four nodes as
$P^3_{11}(s^3_{11},t^3_{11})$, $P^1_{11}(s^1_{11},t^1_{11})$, $P^2(s^2,t^2)$
and $P^0(s^0,t^0)$ in Fig. \ref{adapnodesEx}(c).
\item $C_3$ is a T-cell. If $C_3$ is a T-cell,
we assume $C_3$ as a T-cell of the T-connection $\mathcal{TC}_3$.
Then, $l(\mathcal{TC}_3)$ is not equal to $l(C_1)$.
Otherwise, if $l(\mathcal{TC}_3)=l(C_1)$,  as the one-neighbor-cell of $l(\mathcal{TC}_3)$ is neither on $E_2$ nor on $E_3$.
There exist some cross vertices of $\mathcal{TC}_3$ on $E_2$ or $E_3$, $\mathcal{TC}_3$ is split into several parts, it is contrary to the assumption.
Thus, $l(\mathcal{TC}_3)<l(C_1)$ or $l(\mathcal{TC}_3)>l(C_1)$.
\begin{enumerate}
\item $l(\mathcal{TC}_3)<l(C_1)$.
If $l(\mathcal{TC}_3)<l(C_1)$, by Initialization \ref{initialization1}, the nine B-ordinates of $p(x,y)$ on each T-cell in $\mathcal{TC}_3$ are given,
the B-ordinate on the center domain-point of  $C_3$ is given.
%By Proposition \ref{CVRpointweight}, the centre domain-point of $C_3$,
%which is denoted as $P^3_{11}(s^3_{11},t^3_{11})$ in Fig. \ref{adapnodesEx}(c),
%can be used as one of the four adaptive nodes.
%As $P^3_{11}(s^3_{11},t^3_{11})$ is not on the cross $X_1$,
Similar to (1)(a) in case 2,
we can choose the four nodes as
$P^3_{11}(s^3_{11},t^3_{11})$, $P^1_{11}(s^1_{11},t^1_{11})$, $P^2(s^2,t^2)$
and $P^0(s^0,t^0)$ in Fig. \ref{adapnodesEx}(c).
\item $l(\mathcal{TC}_3)>l(C_1)$
If $l(\mathcal{TC}_3)>l(C_1)$, as $C_3$ is a cell divided by one cell of $\mathscr{T}^k,k>=l_1$,
$v_3$ is a crossing-vertex, and the one-neighbor-cell is either on $E_2$ or on $E_3$.
As the weight on the  domain-centre of $\mathcal{TC}_3$ is given by Equation \ref{CVRbase},
by Proposition \ref{CVRpointweight}, the domain-centre of $\mathcal{TC}_3$, which is denoted as $P^3(s^3,t^3)$ in Fig. \ref{adapnodesEx}(d), can be used as one of the four adaptive nodes.
As $P^3$ is not on the cross $X_1$,
we can choose the four nodes as
$P^3(s^3,t^3)$, $P^1_{11}(s^1_{11},t^1_{11})$, $P^2(s^2,t^2)$
and $P^0(s^0,t^0)$ in Fig. \ref{adapnodesEx}(d).
\end{enumerate}
\end{enumerate}
\item [(2)] $l(\mathcal{TC}_2) < l(C_1)$.
If $l(\mathcal{TC}_2)<l(C_1)$.
By Initialization \ref{initialization1}, the B-ordinates of $p(x,y)$ on each T-cell of $\mathcal{TC}_2$ are given.
By Proposition \ref{CVRpointweight}, the two domain-points,
which are denoted as $P^2_{01}(s^2_{01},t^2_{01})$ and $P^2_{11}(s^2_{11},t^2_{11})$ in Fig.\ref{adapnodesEx}(e),
can be used as two of the four adaptive nodes.
$P^2_{11}$ is on a line that is perpendicular to the green dashed line, the two lines construct a cross, which can be denoted as $X_2$.
As $P^2_{01}$ is not on $X_2$,
we can choose one group of adaptive nodes as
$P^2_{01}(s^2_{01},t^2_{01})$, $P^2_{11}(s^2_{11},t^2_{11})$, $P^0(s^0,t^0)$ and $P^1_{11}(s^1_{11},t^1_{11})$ in Fig. \ref{adapnodesEx}(e).

\end{enumerate}

From the discussion above, each group of nodes is adaptive.
As the weight or B-ordinate on each node is given by the discussion,
we can calculate the coefficients of the bilinear function on $v_1$.
Similar to the situation of $v_1$, we can calculate the coefficients of the bilinear function on $v_2$.
And then, using Theorem \ref{bordinatesoftwocrossingverteices}, we obtain the corresponding B-ordinates
that satisfy the $C^1$ continuous condition on each edge of $\mathcal{T}_0$, we obtain the corresponding B-ordinates of $p(x,y)$ on $\mathcal{T}_0$.
The theorem is proved.
\end{pf}
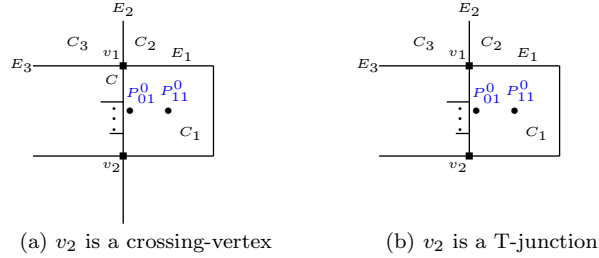
\begin{figure}[!ht]
\captionsetup{font={small}}
\begin{center}
\psset{unit=0.6cm,linewidth=0.5pt}
\begin{tabular}{c@{\hspace*{1cm}}c}
\begin{pspicture}(-0.5,0)(5.5,5.5)
\psline(0,1)(4,1)\psline(2,-0.5)(2,4)\psline(0,3)(4,3)\psline(4,3)(4,1)
\psdots[dotstyle=square*](2,1)
\psdots[dotstyle=square*](2,3)
\psline(1.7,1.5)(2,1.5)
\psline(1.5,2.2)(2,2.2)
\rput(1.8,2){$\vdots$}
\rput(1.75,0.75){\tiny{$v_2$}}
\rput(1.75,3.25){\tiny{$v_1$}}
\rput(-0.25,3){\tiny{$E_3$}}
\rput(2,4.25){\tiny{$E_2$}}
\rput(3.3,3.25){\tiny{$E_1$}}
\rput(3.5,1.5){\tiny{$C_1$}}
\rput(2.5,3.5){\tiny{$C_2$}}
\rput(1,3.5){\tiny{$C_3$}}
\rput(1.75,2.7){\tiny{$C$}}
\rput(3,2){\tiny${\bullet}$}
\rput(3.15,2.45){\blue{\tiny{$P^0_{11}$}}}
\rput(2.15,2){\tiny${\bullet}$}
\rput(2.4,2.4){\blue{\tiny{$P^0_{01}$}}}

\end{pspicture}&
\begin{pspicture}(-0.5,0)(5.5,5.5)
\psline(0,1)(4,1)\psline(2,1)(2,4)\psline(0,3)(4,3)\psline(4,3)(4,1)
\psdots[dotstyle=square*](2,1)
\psdots[dotstyle=square*](2,3)
\psline(1.7,1.5)(2,1.5)
\psline(1.5,2.2)(2,2.2)
\rput(1.8,2){$\vdots$}
\rput(1.75,0.75){\tiny{$v_2$}}
\rput(1.75,3.25){\tiny{$v_1$}}
\rput(-0.25,3){\tiny{$E_3$}}
\rput(2,4.25){\tiny{$E_2$}}
\rput(3.3,3.25){\tiny{$E_1$}}
\rput(3.5,1.5){\tiny{$C_1$}}
\rput(2.5,3.5){\tiny{$C_2$}}
\rput(1,3.5){\tiny{$C_3$}}
%\rput(1.75,2.7){\tiny{$C$}}
\rput(3,2){\tiny${\bullet}$}
\rput(3.15,2.45){\blue{\tiny{$P^0_{11}$}}}
\rput(2.15,2){\tiny${\bullet}$}
\rput(2.4,2.4){\blue{\tiny${P^0_{01}}$}}
\end{pspicture}\\[2mm]
\footnotesize{(a) $v_2$ is a crossing-vertex} & \footnotesize{(b) $v_2$ is a T-junction}
\end{tabular}
\caption{Other T-structures. \label{nokey-T}}
\end{center}
\end{figure}
\begin{coro}
\label{otherTS}
The corresponding B-ordinates of $p(x,y)$ on $\mathcal{T}_i,i=1,...,m$ can be calculated by Lemma \ref{TSbordinates}.
\end{coro}

\begin{pf}
We can obtain
all of the corresponding B-ordinates of $\mathcal{T}_0$ via the Theorem \ref{keyT}.
As the T-structures in $\mathcal{TSB}_0$ are sorted as $\mathcal{T}_0, \mathcal{T}_1,...\mathcal{T}_m$ via Algorithm \ref{orderTS},
$\mathcal{T}_1$ is connected to $\mathcal{T}_0$.
Without loss of generality,
we assume that $\mathcal{T}_1$ is connected to $\mathcal{T}_0$ at $v_2$, which is shown in Fig. \ref{nokey-T}.
$v_2$ is a crossing-vertex in Fig. \ref{nokey-T} (a) or T-junction in Fig. \ref{nokey-T} (b),
the corresponding B-ordinates of $v_2$ are obtained.
In Fig. \ref{nokey-T}, $C_0$ is denoted as the mother-cell of $\mathcal{T}_1$,
and the B-ordinates on $P^0_{01}$ and $P^0_{11}$ are given.
Replace $P^0$ in Fig. \ref{adapnodesofT} with $P^0_{01}$ in Fig. \ref{nokey-T},
we can obtain at least one group adaptive nodes for $v_1$ in Fig. \ref{nokey-T}.
The corresponding bilinear function for $v_1$ can be obtained naturally.
And then, we can calculate the corresponding B-ordinates of $\mathcal{T}_1$ via Lemma \ref{TSbordinates}.
By Algorithm \ref{orderTS}, $\mathcal{T}_i$ is connected to one of $\mathcal{T}_0,...,\mathcal{T}_{i-1}, i=1,...,m.$
In a similar way to $\mathcal{T}_1$, we can obtain the corresponding B-ordinates of $p(x,y)$ on $\mathcal{T}_i, i=2,...,m.$
The corollary is proved.
\end{pf}

We give algorithm \ref{shenduTS} to illustrate the process for calculating the corresponding B-ordinates on each T-structure in $\mathcal{TSB}_0$.

\begin{algorithm}[H]
\caption{Calculate  the B-ordinates for each T-cell in $\mathcal{TC}_0$ \label{shenduTS}}% 算法名字
\LinesNumbered %要求显示行号
\KwIn{$\mathcal{TSB}_0: \{\mathcal{T}_0,...,\mathcal{T}_m\}$}%输入参数
\KwOut{The B-ordinates of each T-cell in $\mathcal{TC}_0$}%输出
Calculate the corresponding B-ordinates of $\mathcal{TSB}_0[0]$\;
$V \gets \mathcal{TSB}_0[0]$\;
Remove $\mathcal{TSB}_0[0]$ from $\mathcal{TSB}_0$\;
\While{$|\mathcal{TSB}_0| > 0$}{
\For{$i=1,i < |\mathcal{TSB}_0|, i++$}
{
%Denote the end-point of $\mathcal{TSB}_0[i]$ as $v_1$ and $v_2$ respectively\;
Bool Connected=false\;
\For{$j=0;j<|V|;j++$}
{
\If{$\mathcal{TSB}_0[i]$ is connected to $V[j]$ at one end point of $\mathcal{TSB}_0[i]$ }{
Connected=true;
 }
}
\If{Connected}
{Calculate the corresponding B-ordinates on the other end-point of $\mathcal{TSB}_0[i]$\;
Using Lemma \ref{TSbordinates} to calculate the corresponding B-ordinates on $\mathcal{TSB}_0[i]$\;
Remove $\mathcal{TSB}_0[i]$ from $\mathcal{TSB}_0$\;
$V \gets \mathcal{TSB}_0[i]$\;}
}
}
Using the $C^1$ continuous conditions on the rest edges that the B-ordinates are not obtained\;
\end{algorithm}
\begin{thm}
\label{calculateTC0}
The B-ordinates of $p(x,y)$ on each T-cell belongs to $\mathcal{TC}_0$ can be calculated by Algorithm \ref{shenduTS}.
\end{thm}
\begin{pf}
By Lemma \ref{coverallcells}, the T-cells in $\mathcal{TC}_0$ are covered by $\mathcal{TSB}_0$,
The B-ordinate on the centre domain-point of each T-cell in $\mathcal{TC}_0$ is obtained.
As the corresponding B-ordinates on each T-structure in $\mathcal{TSB}_0$ are calculated,
we can calculate the rest B-ordinates on each T-cell of  $\mathcal{TC}_0$ via $C^1$ continuous conditions.
The theorem is proved.
\end{pf}
\subsubsection{The B-ordinates of $p(x,y)$ on $\mathcal{TC}_i, i=1,...,n$}

The B-ordinates on the T-cells of $\mathcal{TC}_0$ are calculated in Section \ref{T0T0T0T0T0}.
The B-ordinates on the T-cells of the rest T-connections
$\mathcal{TC}_i, i=1,...,n$ can be calculated similarly.

\begin{thm}
\label{thmeval}
For each T-connection in $\mathbb{E}$,
we can obtain the B-ordinates of $p(x,y)$ on the T-cells belong to the T-connection.
\end{thm}

\begin{pf}
%For the T-structure-branches $\mathcal{TSB}_i, i=0,...,n \in Sup(p(x,y))$,
%which are sorted in descending level order.
%The corresponding T-connection of  $\mathcal{TSB}_i$ is denoted as $\mathcal{TC}_i, i=0,...,n$.
As $\mathcal{TC}_i, i=0,...,n$ are in the same descending level order as  $\mathcal{TSB}_i, i=0,...,n$.
We can calculate all of the B-ordinates on $\mathcal{TC}_0$ via algorithm \ref{shenduTS}.
And then, $\mathcal{TC}_1$ is the T-connection whose level is the lowest.
We can calculate B-ordinates on $\mathcal{TC}_1$ in the same way as $\mathcal{TC}_0$.
By this analogy, we can calculate the B-ordinates on each T-connection $\mathcal{TC}_i, i=2,...,n$ by Algorithm \ref{shenduTS}.
The theorem is proved.
\end{pf}

Till now, we can calculate the B-ordinates of $p(x,y)$ on the the T-cells belong to $\mathcal{TC}_i, i=0,...,n$.
And we give the  following theorem to obtain the B-ordinates of $p(x,y)$:

\begin{thm}
\label{thmevaleeeeee}
For the polynomial function  $p(x,y) \in \overline{\mathbf{S}}^2(\mathscr{T})$,
the weights are denoted by Equation (\ref{CVRbase}), $\mathbb{E}$ denotes the domain that covers $sup(p(x,y))$.
Then, the B-ordinates on each cell in $\mathbb{E}$ can be calculated by Theorem \ref{thmeval} and $C^1$ continuous conditions.
\end{thm}
\begin{pf}
As the B-ordinate on the center domain-point of each T-cell in $\mathbb{E}$ is calculated by theorem \ref{thmeval},
and the B-ordinate on the center domain-point of each P-cell in $\mathbb{E}$ is given by Equation(\ref{CVRbase}).
The B-ordinates on each cell in $\mathbb{E}$ can be calculated via $C^1$ continuous conditions. The theorem is proved.
\end{pf}

For each cell $C \in \mathbb{E}$, if at least one B-ordinate of $C$ is nonzero, we save $C$ to the support of $p(x,y)$.
We can use Bernstein-B$\acute{e}$zier to express $p(x,y)$, which is $C^1$ continuous on the support.
Then, we obtained a  polynomial function with local support of  $\overline{\mathbf{S}}^2(\mathscr{T})$.
%corresponding to the basis function of $\overline{\mathbf{S}}^0(\mathscr{G})$.

\subsection{Simplify the hierarchical T-mesh}\label{Simplify1021}
%We can calculate the B-ordinates for each basis function for $\overline{\mathbf{S}}^2(\mathscr{T})$ in section \ref{evaluatepxy}.
%We can reduce the calculation by simplify a hierarchical T-mesh.
In particular, if the T-l-edge $E$ only contains $V_0$ vertices,
$\dim W[E]=(V_0-d-1)_+:=\max(0,V_0-d-1)$ holds for $\mathbf{S}^d(\mathscr{T})$,
$E$ is defined as a  \textbf{trivial l-edge}\cite{w13,Zeng15b} if $(V_0-d-1)_+=0$ .
For the T-l-edge $E$ with one interior crossing-vertex, we say $E$ is a trivial l-edge.
As $\dim W[E]=0$,
we can remove $E$ from $\mathscr{T}$, and the polynomial functions of the spline spaces will not change.

\begin{figure}[!htb]
\captionsetup{font={small}}
\begin{center}
\psset{unit=0.5cm,linewidth=0.5pt}
\begin{tabular}{c@{\hspace*{1cm}}c@{\hspace*{1cm}}c}
\begin{pspicture}(0,0)(4,4)
\psline(0,0)(4,0)(4,4)(0,4)(0,0)
\psline(0,2)(4,2)\psline(0,3)(4,3)\psline(0,1)(2,1)
\psline(2,0)(2,4)\psline(1,0)(1,4)\psline(3,2)(3,4)
\psline[linecolor=blue](1,2.5)(3,2.5)
\psline[linecolor=green](1.5,2)(1.5,3)
\psline[linecolor=green](2.5,2)(2.5,3)
\end{pspicture} &
\begin{pspicture}(0,0)(4,4)
\psline(0,0)(4,0)(4,4)(0,4)(0,0)
\psline(0,2)(4,2)\psline(0,3)(4,3)\psline(0,1)(2,1)
\psline(2,0)(2,4)\psline(1,0)(1,4)\psline(3,2)(3,4)
\psline[linecolor=blue](1,2.5)(3,2.5)
%\psline[linecolor=green](1.5,2)(1.5,3)
%\psline[linecolor=green](2.5,2)(2.5,3)
\end{pspicture} &
\begin{pspicture}(0,0)(4,4)
\psline(0,0)(4,0)(4,4)(0,4)(0,0)
\psline(0,2)(4,2)\psline(0,3)(4,3)\psline(0,1)(2,1)
\psline(2,0)(2,4)\psline(1,0)(1,4)\psline(3,2)(3,4)
\end{pspicture} \\[2mm]
\footnotesize{(a) $\mathscr{T}_0$} & \footnotesize{(b) $\mathscr{T}_1$} & \footnotesize{(c) $\mathscr{T}_2$}
\end{tabular}
\caption{Simplification of a T-mesh.\label{Simplify}}
\end{center}
\end{figure}
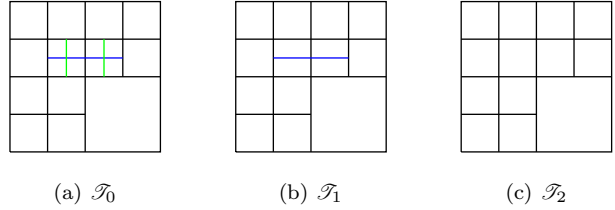

Fig. \ref{Simplify} shows the \textbf{simplification} of a hierarchical T-mesh.
As the  green lines are trivial l-edges in $\mathscr{T}_0$,
remove them, we obtain $\mathscr{T}_1$.
As the blue line is a trivial l-edges in $\mathscr{T}_1$,
remove it, we obtain $\mathscr{T}_2$.
$\mathscr{T}_2$ is denoted as the simplification of $\mathscr{T}_0$.
%After simplification, some overhanging edges in CVR graph are removed.

%At last, we give the algorithm to construct each basis function in $\overline{\mathbf{S}}^2(\mathscr{T}^k)$.
%For $\mathscr{T}^i$, we denote the CVR graph of $\mathscr{T}^i$ as $\mathscr{G}^i, i=0,...,k$.
%If a g-cell $\mathcal{GC}$ in $\mathscr{G}^{k}$  is different with any g-cell in $\mathscr{G}^{k-1}$, we denote the level of $\mathcal{GC}$ as $k$.
%The basis function $q(x,y) \in \overline{\mathbf{S}}^0(\mathscr{G})$ defined on $\mathcal{GC}$ is denoted as the basis function of level $k$.
%The basis function $p(x,y) \in \overline{\mathbf{S}}^2(\mathscr{T})$ corresponding to $q(x,y)$ is also denoted as the basis function of level $k$.
%Simplify $\mathscr{T}^k$, we can remove some the T-structures from $\mathscr{T}$.
%We can reduce computation load by simplify the $\mathscr{T}$.
It is natural that we can remove meshlines that do not contribute to the dimension of the spline space \cite{LR}.
The simplification
can reduce the number of T-structures, decrease the amount of calculation,
and remove some overhanging edges from CVR graph.
%We denote the simplification of $\mathscr{T}$ as $\mathscr{S}$.
%For each trivial l-edge $E$, $\dim \overline{\mathbf{S}}(2,1,E)= 0$.
%The basis functions of $\overline{\mathbf{S}}^2(\mathscr{T}^k)$
%are the same as the basis functions of  $\overline{\mathbf{S}}^2(\mathscr{T})$.
%We can reduce the computation load by constructing the basis functions in  $\overline{\mathbf{S}}^2(\mathscr{T})$.
We give Algorithm \ref{Evalabase} to construct a biquadratic polynomial function $p(x,y) \in \overline{\mathbf{S}}^2(\mathscr{T})$, where $\mathscr{T}$ is the simplified T-mesh.

\begin{algorithm}[H]
\caption{Calculate  the B-ordinates for $p(x,y)$ \label{Evalabase}}% 算法
\LinesNumbered %要求显示行号
\KwIn{$q(x,y) \in \overline{\mathbf{S}}^0(\mathscr{G})$}%输入参数
\KwOut{The B-ordinates of $p(x,y) \in \overline{\mathbf{S}}^2(\mathscr{T})$}%输出
Give the weights on each domain of $\mathscr{T}$  by Equation \ref{CVRbase}  \;
Obtain $\mathbb{E}:\{\mathcal{TC}_0,...,\mathcal{TC}_n \} \cup \{\mathcal{PC}_0,...,\mathcal{PC}_l\}$\;
Obtain each T-structure-branch  $\mathcal{TSB}_i$ corresponding to $\mathcal{TC}_i, i=0,...,n$\;
Sort the T-structures in $\mathcal{TSB}_i$ via algorithm \ref{orderTS}, $i=0,...,n$\;
Sort $\mathcal{TSB}_i, i=0,...,n$ in descending level order\;
Calculate the corresponding B-ordinates of $p(x,y)$ on $\mathcal{TSB}_i$ via Algorithm \ref{shenduTS}\;
Calculate the B-ordinates on the cells in $\mathbb{E}$ via $C^1$ continuous conditions\;
Save the support of $p(x,y)$ on $\mathscr{T}$\;
\end{algorithm}

%As the basis functions of $\overline{\mathbf{S}}^2(\mathscr{T})$
%are the same with  basis functions of $\overline{\mathbf{S}}^2(\mathscr{T}^k)$,
%And we obtain the basis function and we give the theorem as follows:
By Algorithm \ref{Evalabase}, we obtain the basis function of  $\overline{\mathbf{S}}^2(\mathscr{T})$
corresponding to the basis function of $\overline{\mathbf{S}}^0(\mathscr{G})$, and we give the theorem as:

\begin{thm}
\label{ThEval}
Each basis function of $\overline{\mathbf{S}}^0(\mathscr{G})$ corresponds to a biquadratic polynomial function of $\overline{\mathbf{S}}^2(\mathscr{T})$.
\end{thm}
\begin{pf}
Given the basis function $q(x,y) \in \overline{\mathbf{S}}^0(\mathscr{G})$,
denote the domain weights for $p(x,y) \in \overline{\mathbf{S}}^2(\mathscr{T})$ by Equation (\ref{CVRbase}),
we can calculate the B-ordinates for $p(x,y)$ by algorithm  \ref{Evalabase}.
The theorem is proved.
\end{pf}

So far, we construct the  biquadratic polynomial function of  $\overline{\mathbf{S}}^2(\mathscr{T})$ via a piecewise constant basis function of $\overline{\mathbf{S}}^0(\mathscr{G})$.
\section{The isomorphic bivariate spaces and properties}\label{Properties}
In this section, we discuss the bijective property of the mapping that constructed in Section \ref{mappingformulea111111}.
Some properties of the basis functions we constructed in Section \ref{basisfunctions} are also discussed.
\subsection{The bijective property of the mapping}
First, by Section \ref{mappingformulea111111} and Section \ref{basisfunctions}, we give a theorem about the mapping as follows:
\begin{thm}
\label{thm6.1}
For the hierarchical T-mesh $\mathscr{T}$, $\mathscr{G}$ denotes the CVR graph of $\mathscr{T}$.
The mapping between $\overline{\mathbf{S}}^2(\mathscr{T})$ and
$\overline{\mathbf{S}}^0(\mathscr{G})$ is bijective, and $\overline{\mathbf{S}}^2(\mathscr{T})$  is isomorphic to
$\overline{\mathbf{S}}^0(\mathscr{G})$.
\end{thm}
\begin{pf}
By Theorem \ref{Injectivemapping}, the mapping from
$\overline{\mathbf{S}}^2(\mathscr{T})$ to $\overline{\mathbf{S}}^0(\mathscr{G})$ is an injective mapping.
By Theorem \ref{ThEval},
each basis function of $\overline{\mathbf{S}}^0(\mathscr{G})$ corresponds to a polynomial function of $\overline{\mathbf{S}}^2(\mathscr{T})$,
the mapping is surjective.
Thus, the mapping is bijective,
and
$\overline{\mathbf{S}}^2(\mathscr{T})$  is isomorphic to
$\overline{\mathbf{S}}^0(\mathscr{G})$.
\end{pf}

%As the mapping in Section \ref{mappingformulea111111} is bijective, we given the corollary as follows:
%\begin{coro}
% The polynomial functions that constructed in Section \ref{basisfunctions} are the basis functions of $\overline{\mathbf{S}}^2(\mathscr{T})$.
%\end{coro}
As the properties of the piecewise constant basis functions over the CVR graph are simple and clear,
we use them to discuss the   properties of the  basis  function belongs to biquadratic spline space over the hierarchical T-mesh as follows:

\subsection{Properties}

%From Section \ref{basisfunctions}, some basis functions that constructed by our method are not B-spline
%or can be expressed by B-splines, but they hold some good properties similar to B-splines.
%The set of all basis functions of $\mathbf{S}^2(\mathscr{T})$ is denoted as $\mathcal{P}$.
%The set of all basis functions of $\mathbf{S}^0(\mathscr{G})$ is denoted as $\mathcal{Q}$.
In this subsection, we denote the extension of $\mathscr{T}$ as $\mathscr{T}^{\varepsilon}$, and we denote the CVR graph of $\mathscr{T}^{\varepsilon}$ as $\mathscr{G}^{\varepsilon}$.
We discuss the properties of basis functions of $\mathbf{S}^2(\mathscr{T})$ as follows:

\begin{thm}
The basis functions of $\mathbf{S}^2(\mathscr{T})$  hold the properties of linearly independence, completeness.
\end{thm}

\begin{pf}
Apply the mapping to $\overline{\mathbf{S}}^2(\mathscr{T}^{\varepsilon})$
and $\overline{\mathbf{S}}^2(\mathscr{G}^{\varepsilon})$.
By Theorem \ref{thm6.1}, the mapping between $\overline{\mathbf{S}}^2(\mathscr{T}^{\varepsilon})$
and $\overline{\mathbf{S}}^2(\mathscr{G}^{\varepsilon})$ is bijective, and $\overline{\mathbf{S}}^2(\mathscr{T}^{\varepsilon})$
is isomorphic to $\overline{\mathbf{S}}^0(\mathscr{G}^{\varepsilon})$.
As the basis functions of $\overline{\mathbf{S}}^0(\mathscr{G}^{\varepsilon})$ are linearly independent and complete,
the basis functions of $\overline{\mathbf{S}}^2(\mathscr{T}^{\varepsilon})$ are also linearly independent and complete.
By Theorem \ref{thm2.1},
and the basis functions  of $\mathbf{S}^2(\mathscr{T})$ are linearly independent and complete on $\mathscr{T}$.
\end{pf}

%As the mapping is bijective, for each basis function $p(x,y) \in \overline{\mathbf{S}}^2(\mathscr{T})$, there exists a basis function $q(x,y) in \overline{\mathbf{S}}^0(\mathscr{G})$
%satisfies  $\mathbf{\Phi}^{-1}(q)= p$. We give the theorem as follows:
\begin{thm}
All the basis functions of $\mathbf{S}^2(\mathscr{T})$ have the property of unit partition.
\end{thm}
\begin{pf}
Assume that the basis functions of $\overline{\mathbf{S}}^2(\mathscr{T}^{\varepsilon})$ are $p_i(x,y)$,
the basis functions of $\overline{\mathbf{S}}^0(\mathscr{G}^{\varepsilon})$ are $q_i(x,y)$,
and $\mathbf{\Phi}^{-1}( q_i )= p_i$ where $ i=1,2,...,N_{\mathscr{G}^{\varepsilon}}$.
As the space is a linear space, we obtain
\[ \sum^{N_{\mathscr{G}^{\varepsilon}}}_{i=1} p_i = \sum^{N_{\mathscr{G}^{\varepsilon}}}_{i=1} \mathbf{\Phi}^{-1}( q_i )
=\mathbf{\Phi}^{-1} (\sum^{N_{\mathscr{G}^{\varepsilon}}}_{i=1}( q_i )).\]
As $\sum^{N_{\mathscr{G}^{\varepsilon}}}_{i=1}( q_i )=1$ is true for each g-cell of  $\mathscr{G}^{\varepsilon}$,
the weight on each interior domain of $\mathscr{T}^{\varepsilon}$ is $1$.
By Algorithm \ref{Evalabase}, the B-ordinates on each cell of $\mathscr{T}$ are $1$.
%implies that
%The weight of each interior domain-centre on $\mathscr{T}^{\varepsilon}$ is $1$.
%By Algorithm \ref{Evalabase},
%the B-ordinates on each cell of $\mathscr{T}$ is $1$.

Thus, %\[\mathbf{\Phi}^{-1} (\sum^{N_{\mathscr{G}^{\varepsilon}}}_{i=1}( q_i ))=1,\]
$ \sum^{N_{\mathscr{G}^{\varepsilon}}}_{i=1} p_i = 1$ is true on $\mathscr{T}$,
by Theorem \ref{thm2.1}, the theorem is proved.
\end{pf}

Thus, the mapping is an isomorphism and
the basis functions that we construct for $\mathbf{S}^2(\mathscr{T})$ hold the properties of linearly independence, completeness and partition of unity.

\section{Surface fitting}\label{fitting}
Given an open surface triangulation with vertices $V_i, i=0,..,N$ in 3D space, the corresponding
parameter values $(x_i, y_i), i=0,..,N$ are obtained by the parametrization in \cite{Floater},
we denote the triangle on the  triangulation mesh as $\Delta$.
the parameter mesh is a triangle mesh  and  the parameter domain is $[0, 1] \times [0, 1]$.

To construct a spline to fit the given surface,
we need to compute all the basis functions $b_j(x,y),j=1,...,m$
and their corresponding control points $P_j,j=1,...,m$.
We denote the fitting spline $S(x,y)={\sum}^m_{j=1}P_jb_j(x,y)$.
To find the control points, we just need to solve an linear system

\[S(x_k,y_k)= {V'}_k, k=1,2,...,m,\]
%If the basis-cell of $b_k$ is a P-cell $C_i=[x_i,y_i] \times [x_{i+1}, y_{i+1}]$,
%$(x_k,y_k)=(\frac{x_i+x_{i+1}}{2}, \frac{y_i+y_{i+1}}{2})$.
%If the basis-cell of $b_k$  is a T-connection and the corresponding T-rectangle-domain
%is $[x_j,y_j] \times [x_{j+1}, y_{j+1}]$,
where $(x_k,y_k)=(\frac{x_{i_k}+x_{i_k+1}}{2}, \frac{y_{i_k}+y_{i_k+1}}{2})$ is the domain-centre of the domain $[x_{i_k},x_{i_k+1}] \times [y_{i_k}, y_{i_k+1}]$.
$(x_k,y_k) \in \Delta_l:((x_{k_1},y_{k_1}),(x_{k_2},y_{k_2}),(x_{k_3},y_{k_3}))$, $(x_k,y_k) = w_{k_1}(x_{k_1},y_{k_1}) +  w_{k_2}(x_{k_2},y_{k_2})+ w_{k_3}(x_{k_3},y_{k_3})$,
and ${V'}_k=w_{k_1}V_{k_1}+w_{k_2}V_{k_2}+w_{k_3}V_{k_3}$.

The surface fitting scheme repeats the following two steps until the fitting error in each cell
is less than the given tolerance $\varepsilon$.

1. Compute all the control points for all the basis functions.

2. Find all the cells whose errors are greater than the given error tolerance $\varepsilon$, then subdivide
these cells into four subcells to form a new mesh, simplify the new mesh, and construct basis functions for the new
mesh. The fitting error on cell $C$  is  $max_{(x,y) \in C} = \parallel V(x,y)-S(x,y)  \parallel$.

\begin{figure}[htp!]
\captionsetup{font={small}}
  \centering
  \subfloat[\footnotesize{An open surface patch}]{
  \includegraphics[width=3cm]{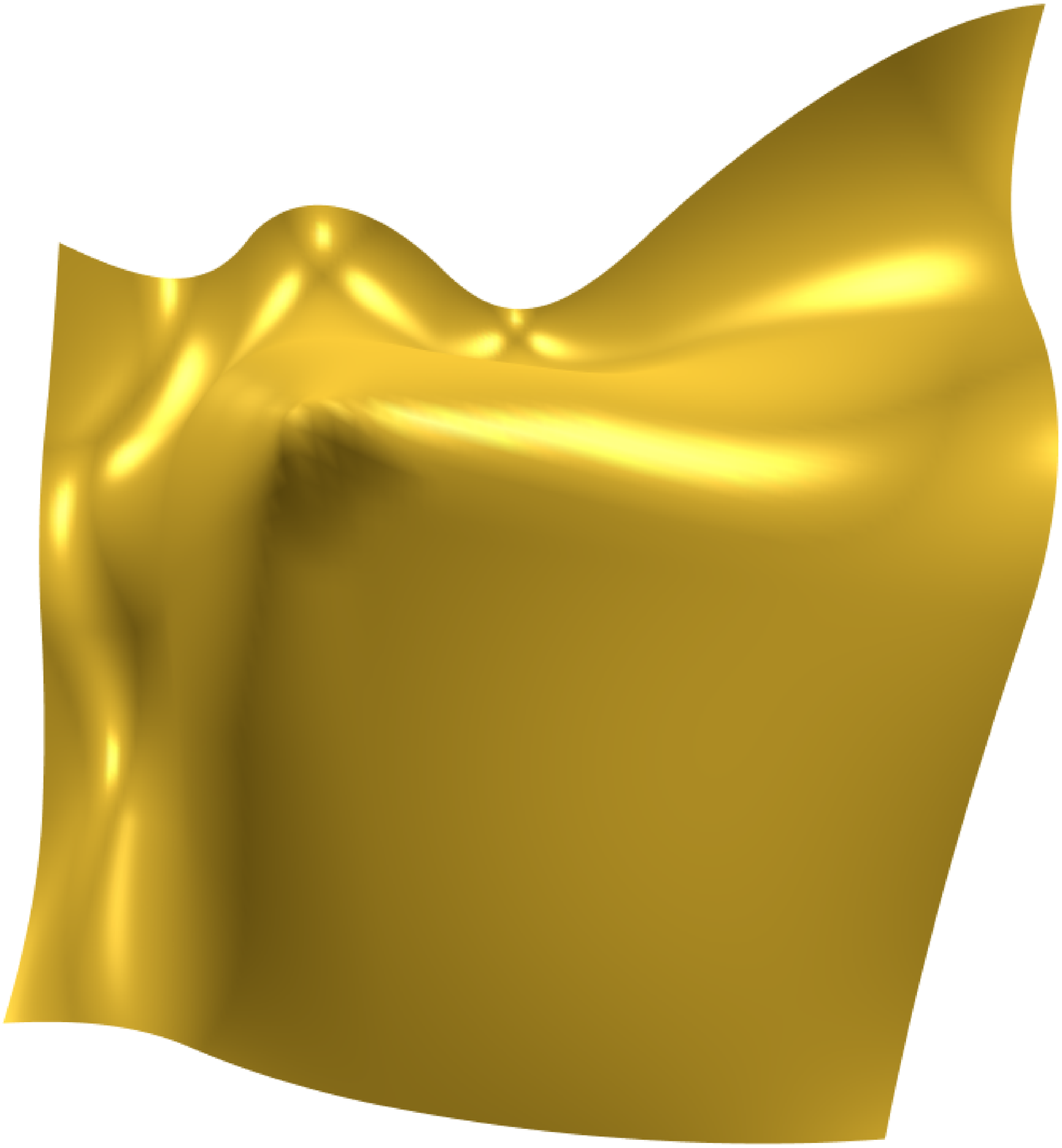}\hspace{0.5cm}
  \includegraphics[width=3cm]{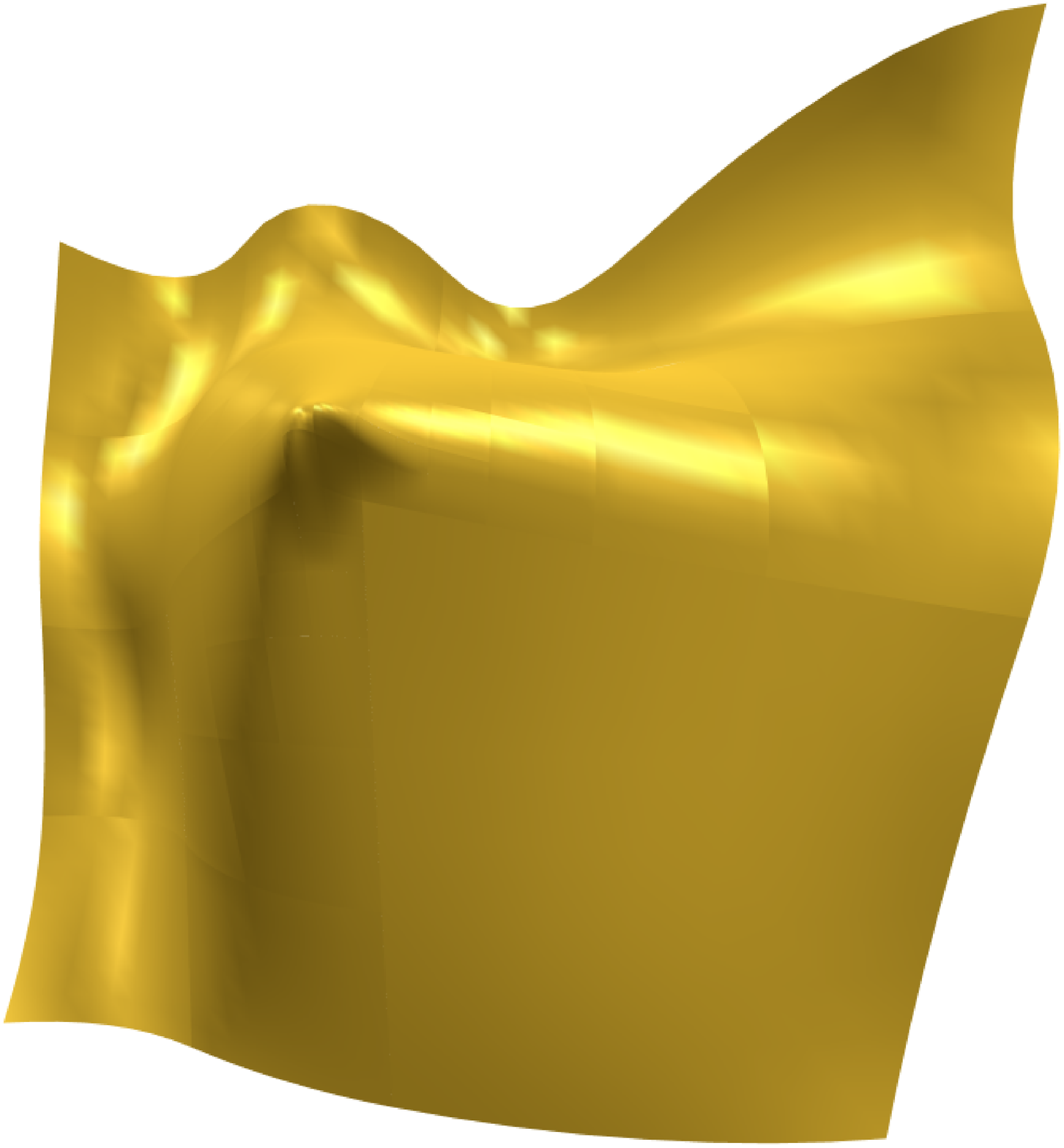}\hspace{0.5cm}
  \includegraphics[width=3cm]{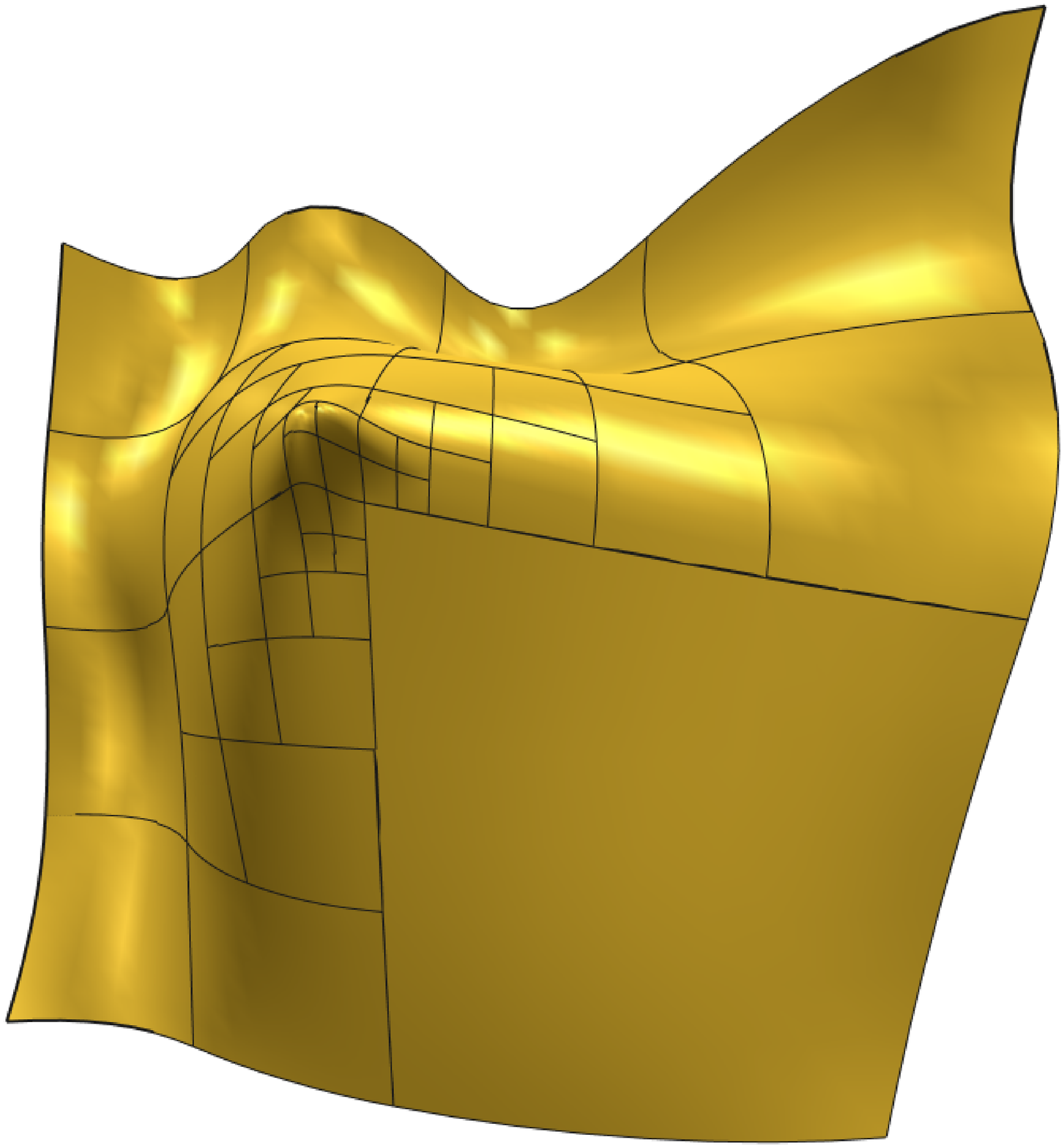}
  }\\[2mm]
  \subfloat[\footnotesize{Nefertiti face}]{
  \includegraphics[width=3cm]{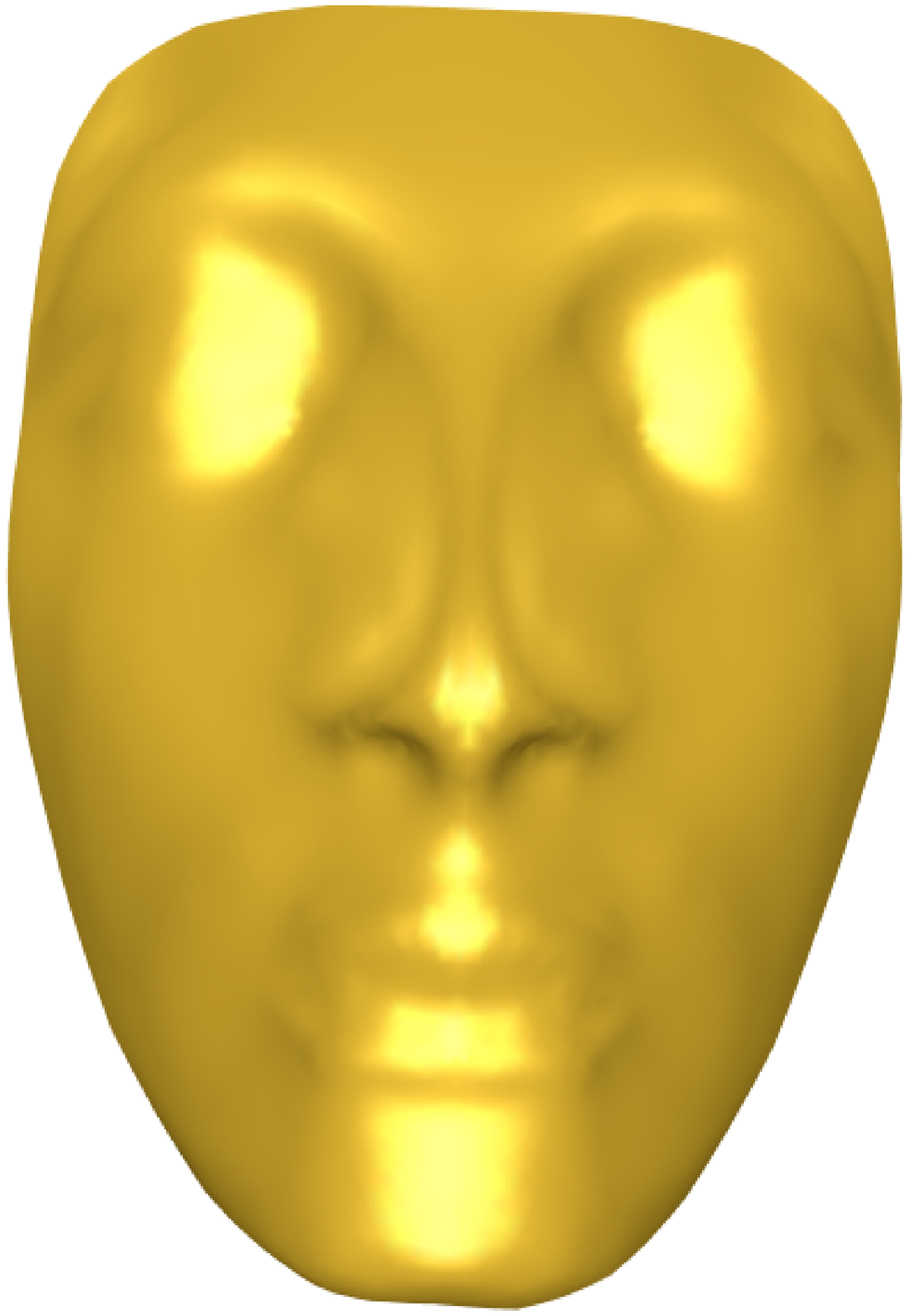}\hspace{0.5cm}
  \includegraphics[width=3cm]{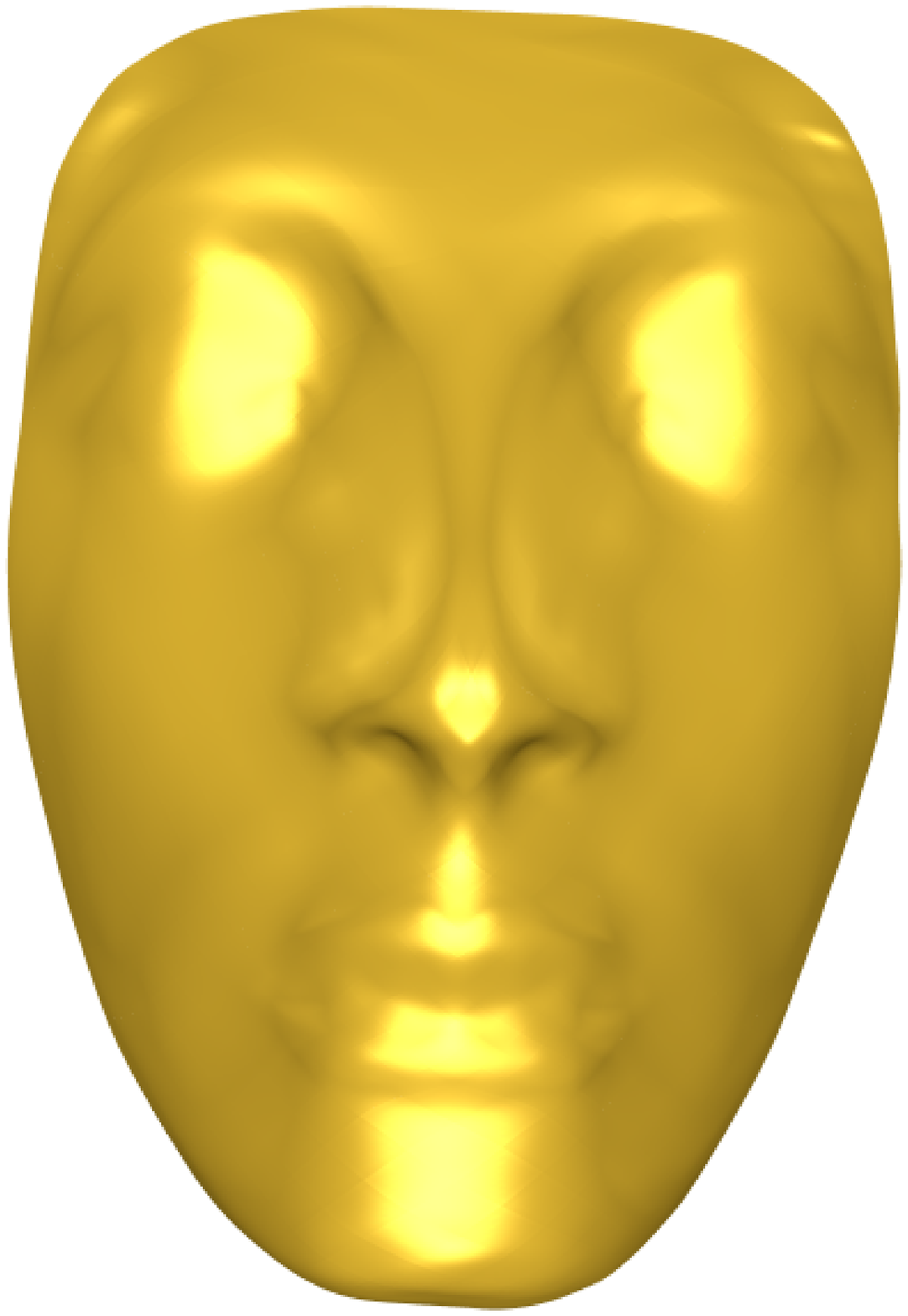}\hspace{0.5cm}
  \includegraphics[width=3cm]{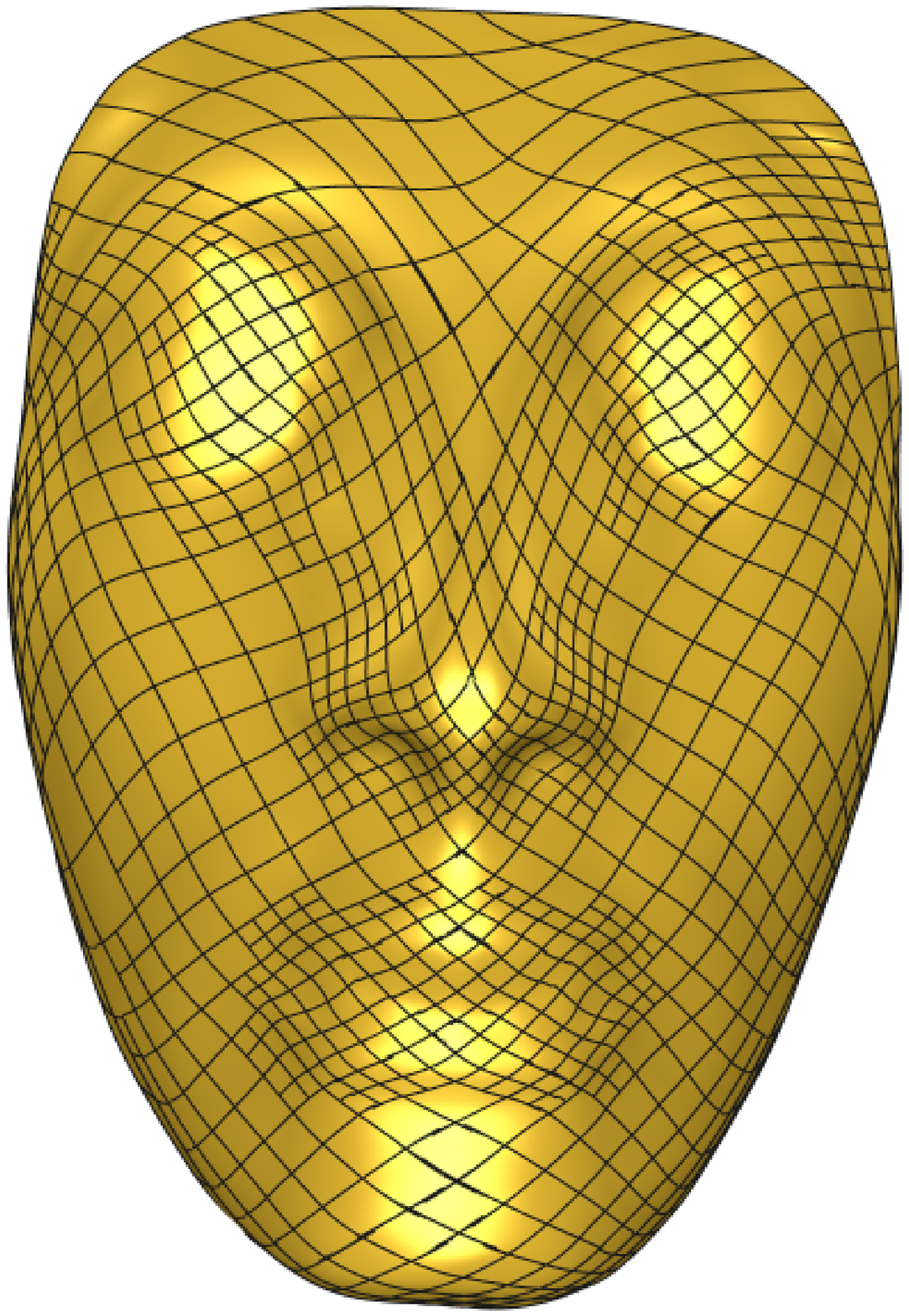}
    }\\[2mm]
    \subfloat[\footnotesize{Gargoyle}]{
  \includegraphics[width=3.5cm]{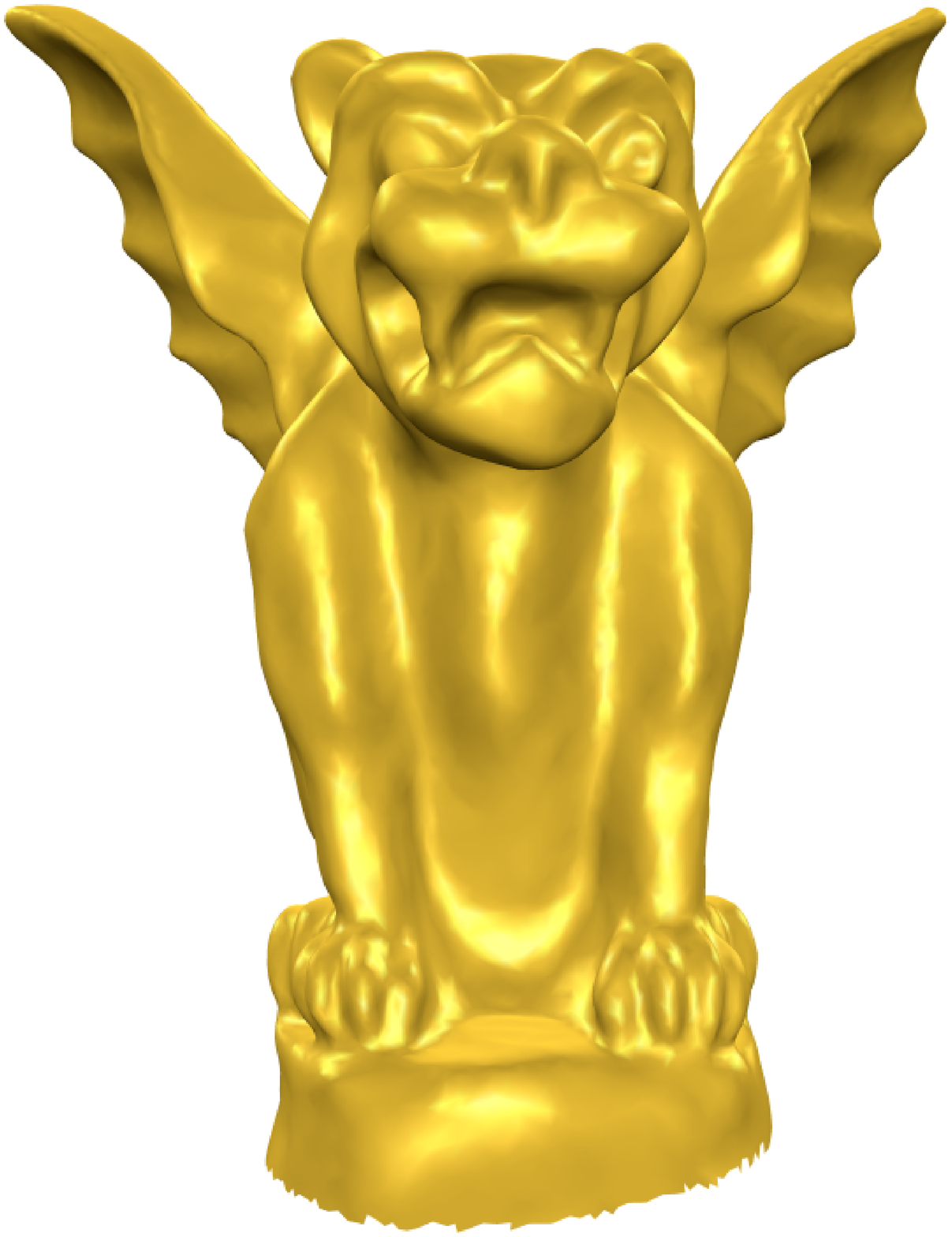}
  \includegraphics[width=3.5cm]{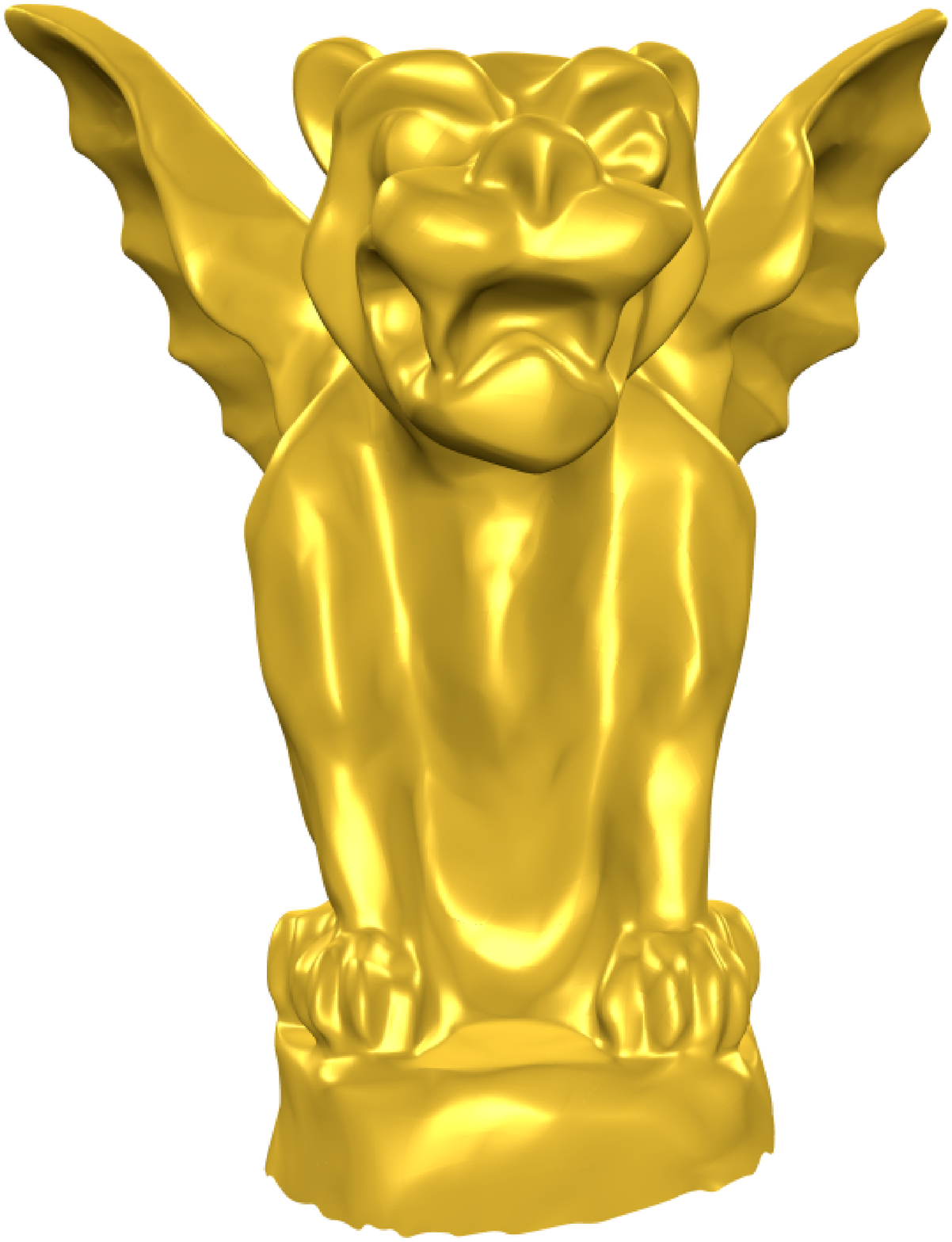}
  \includegraphics[width=3.5cm]{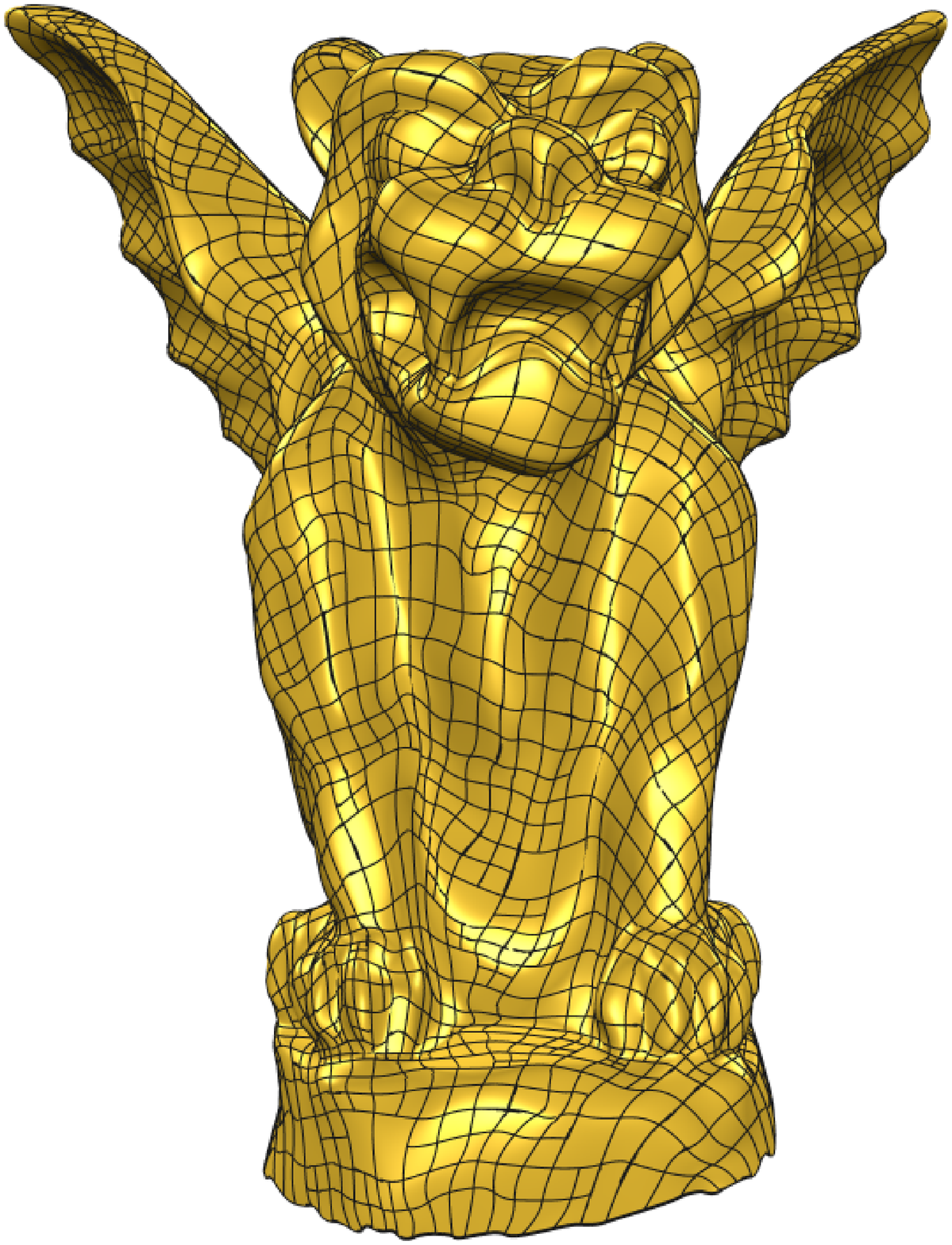}
    }\\[2mm]
     \subfloat[\footnotesize{Female head}]{
  \includegraphics[width=3.5cm]{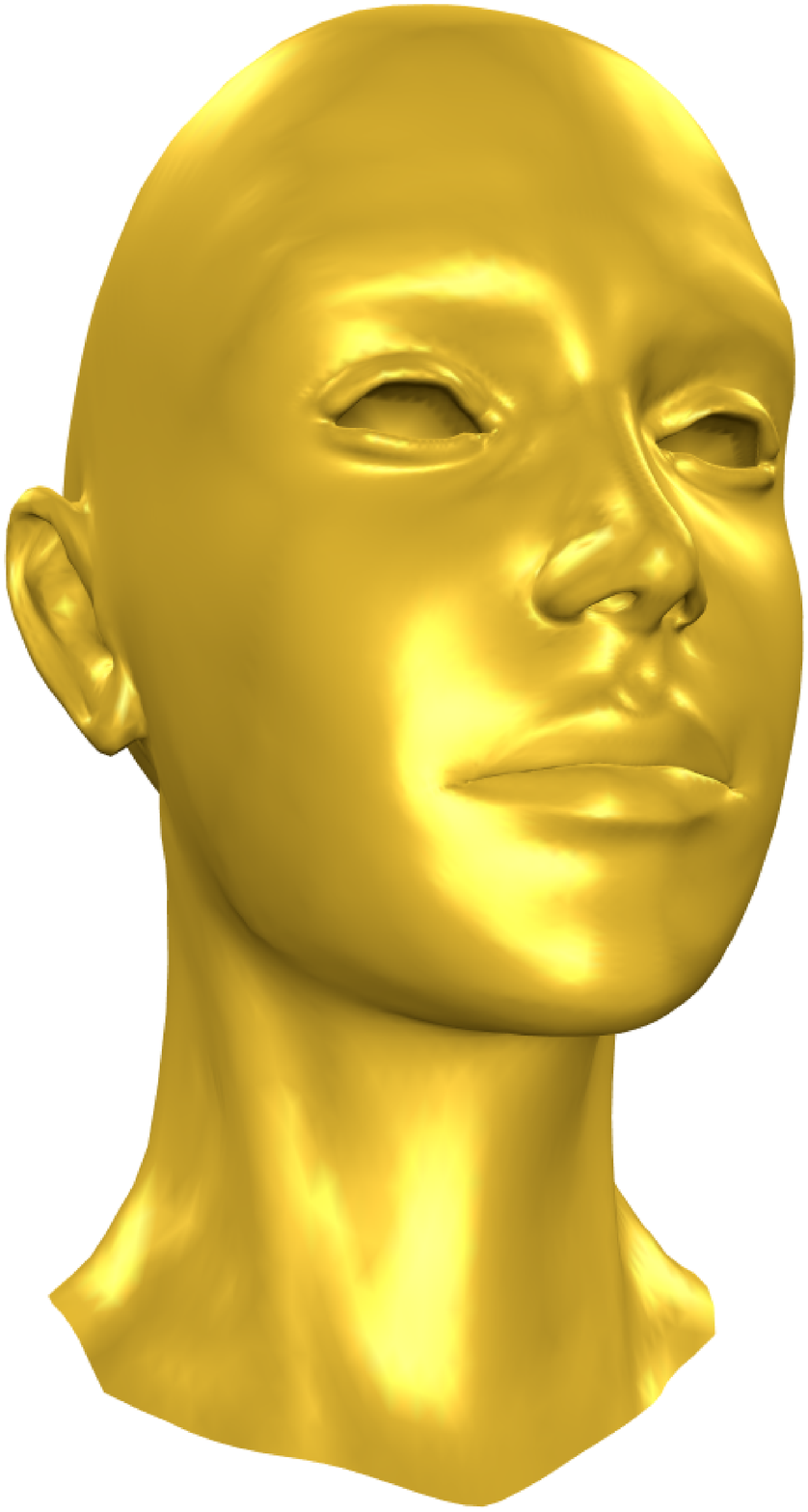}
  \includegraphics[width=3.5cm]{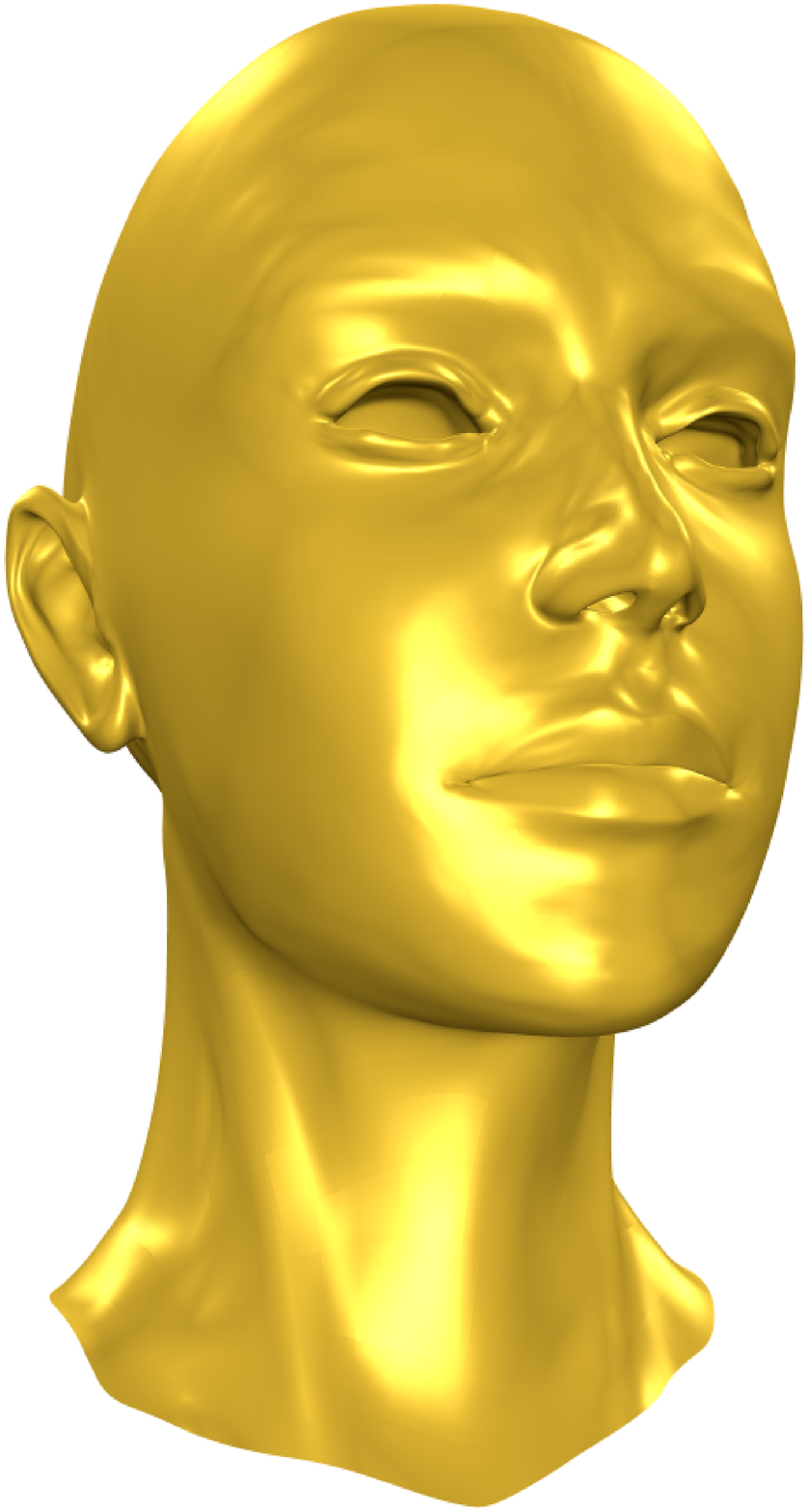}
  \includegraphics[width=3.5cm]{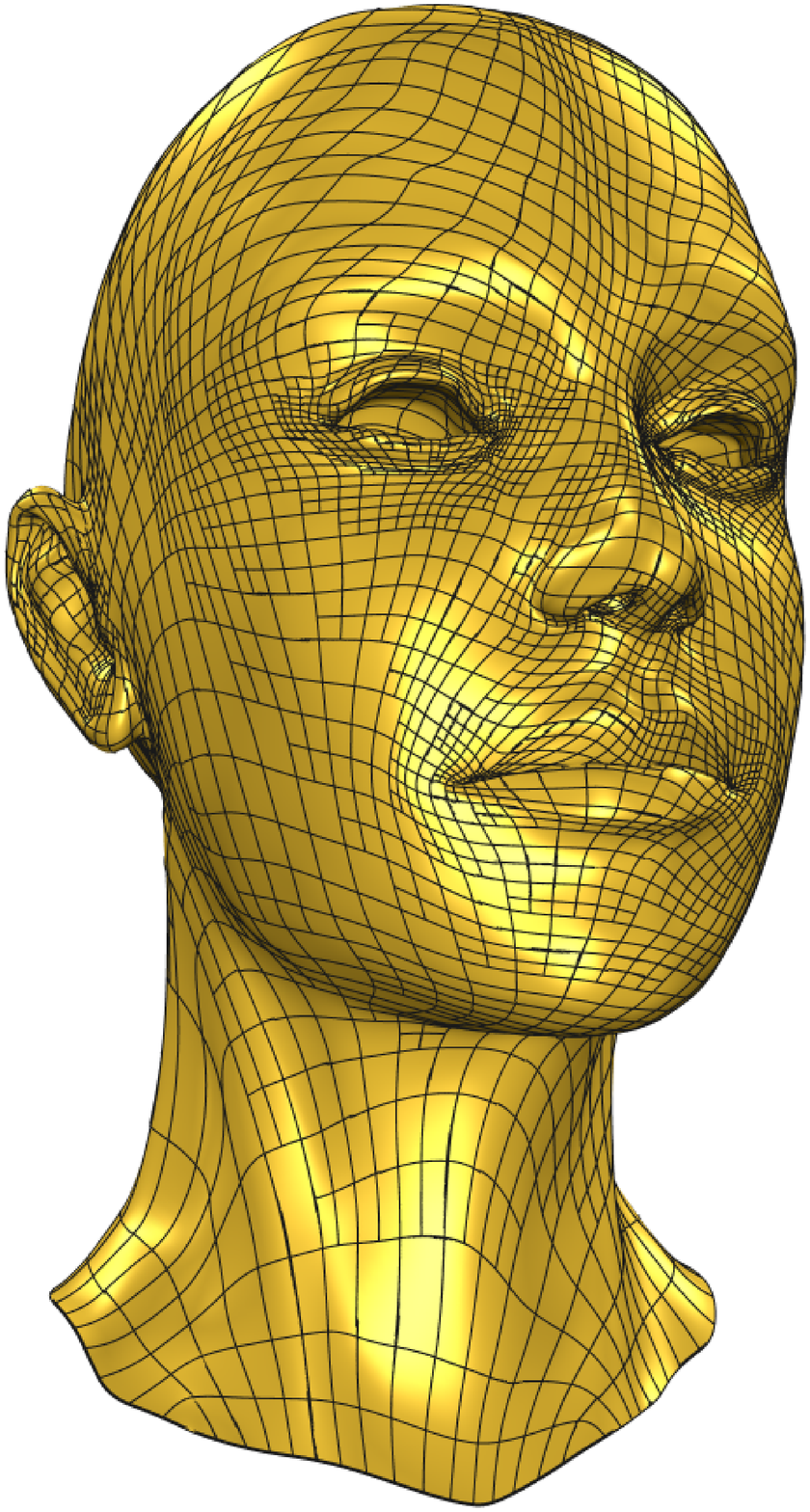}
    }
  \caption{Original meshes (left), result surfaces (middle), surfaces with T-meshes (right) }\label{Open}
\end{figure}

Four examples are provided to illustrate the above surface
fitting scheme in Fig. \ref{Open}.
The iteration number (n),
the dimensions of the spline spaces, the max error and the CPU time on 64 bit operating system are shown in Table \ref{erros}.

\begin{table}
\begin{tabular*}{14cm}{llllllllll}
\hline
model&&n  && dim && max error && t(s)\\
\hline
A surface patch && 4  && 33 && 2.4 $\times 10^{-5}$ && 0.293\\
Nefertiti face&&4  && 913 && 2.2 $\times 10^{-3}$ && 5.67\\
Gargoyle&&11  &&4908 && 7.8 $\times 10^{-3}$ && 101.25\\
Female head&& 13 && 4256 &&6.0 $\times 10^{-3}$ && 150.38\\
\hline
\end{tabular*}
\caption{Experiment data.\label{erros}}
\end{table}

\section{Conclusions and future works}\label{Conclusions}

We give bijective mapping between the biquadratic spline space over a  hierarchical T-mesh and the piecewise constant spline space over the corresponding CVR graph.
And we obtain the conclusion that the biquadratic spline space is isomorphic to the piecewise constant spline space.
By the bijective mapping, we proposed a novel method to discuss the dimensions of the biquadratic spline spaces
over hierarchical T-meshes.
We construct the basis functions of the biquadratic spline spaces via a novel structure, which is called a T-structure.
Our method is general when the level difference of the hierarchical T-meshes is more than one.
We overcome the limitations in \cite{Fang17},
and we need not subdivide the extra cells to maintain the level different is less or equal to $1$.
To reduce the computation, we give the simplifications of the hierarchical T-meshes.
Our method is easy operative,
and some numerical experiments are given to  show our method is effective.
By the bijective mapping, it is easy to prove that
the basis functions hold the properties of linearly independence, completeness and partition of unity.

As the mapping we construct is an isomorphism,
we will apply our basis functions to IGA and models with high genus in the future.
This mapping provides a new idea for us to study high-order spline spaces with low-order spline spaces.
We are also working to extend our work to high order spline spaces.
The 3-variate case is also a considerable question.
As some edges do not contribute to our dimension, improving our subdivision rules is also a considerable idea.
We will also consider improving our basis construction method to reduce the computational overload in the future.

\section*{Acknowledgements}
The authors are supported by the NSF of China (No. 11601114, No. 61772167 and No. 11771420).

\section*{References}
\bibliographystyle{plain}

\end{document}